\documentclass[10pt,letterpaper,aps,prd,twoside,showpacs,nofootinbib,%
	preprintnumbers,notitlepage]{revtex4-1}%
\usepackage[utf8]{inputenc}
\usepackage{fixltx2e}
\usepackage{amssymb}
\usepackage{array}
\usepackage[english]{babel}
\usepackage{bbm}
\usepackage{float}
\usepackage{hhline}
\usepackage[T1]{fontenc}
\usepackage{pb-diagram}
\usepackage{mathtools}
\usepackage{slashed}
\usepackage{stmaryrd}
\SetSymbolFont{stmry}{bold}{U}{stmry}{m}{n}
\usepackage{xspace}
\newcommand{\nc}{\newcommand}
\nc{\rnc}{\renewcommand}

\nc{\bosym}{\boldsymbol}
\nc{\enmat}{\ensuremath}

\nc{\TOM}{\TextOrMath}
\nc{\TOMt}[1]{\TOM {#1\xspace} {\text{#1}}}
\nc{\TOMm}[1]{\TOM {$#1$\xspace} {#1}}

\DeclareMathAlphabet{\mathpzc}{OT1}{pzc}{m}{it}

\nc{\noi}{\noindent}
\nc{\yesi}{\indent}

\nc{\sma}[1]{\TOM{{\small #1}}{\text{\small$#1$}}}
\nc{\smaa}[1]{\TOM{{\footnotesize #1}}{\text{\footnotesize$#1$}}}
\nc{\smaaa}[1]{\TOM{{\scriptsize #1}}{{\scriptstyle #1}}}
\nc{\smaaaa}[1]{\TOM{{\tiny #1}}{{\scriptscriptstyle #1}}}

\nc{\ssiz}{\fontsize{7pt}{9pt}\selectfont}
\nc{\sssiz}{\fontsize{5pt}{7pt}\selectfont}

\nc{\inv}[1][1]{^{-#1}}

\nc{\hs}[1]{^{\smaaaa{#1}}}
\nc{\hb}[1]{^{(#1)} }
\nc{\hbs}[1]{^{\smaaaa{(#1)}}}
\nc{\htx}[1]{^{\text{#1}}}
\nc{\htxs}[1]{^{\text{\tiny#1}}}
\nc{\hbtx}[1]{^{\text{(#1)}}}
\nc{\hbtxs}[1]{^{\text{\tiny(#1)}}}

\nc{\hint}[3]{^{\scriptscriptstyle[#1_{#2},#1_{#3}]}}

\nc{\ls}[1]{_{#1}}
\nc{\lb}[1]{_{(#1)} }
\nc{\lbs}[1]{_{\smaaaa{(#1)}}}
\nc{\ltx}[1]{_{\text{#1}}}
\nc{\ltxs}[1]{_{\text{\tiny#1}}}
\nc{\lbtx}[1]{_{\text{(#1)}}}
\nc{\lbtxs}[1]{_{\text{\tiny(#1)}}}

\nc{\intval}[3]{{[#1_{#2},#1_{#3}]}}%
\nc{\intvals}[3]{{\scriptscriptstyle\intval{#1}{#2}{#3} }}%
\nc{\lintval}[3]{_{\intvals{#1}{#2}{#3}}}
\nc{\hintval}[3]{^{\intval{#1}{#2}{#3}}}

\nc{\hvc}[2][]{^{\smaaaa{#1}\vc #2 }}
\nc{\hvcs}[2][]{^{\smaaaa{#1\vc #2 }}}

\nc{\lvc}[2][]{_{\smaaaa{#1}\vc #2 }}
\nc{\lvcs}[2][]{_{\smaaaa{#1\vc #2 }}}

\nc{\ovl}{\overline}
\nc{\unl}{\underline}
\nc{\unlh}[1]{\settowidth{\negphantomlength}{{#1}}%
	\unl{\hspace{\negphantomlength}}\negphantom{#1}#1}

\nc{\ovb}{\overbrace}
\nc{\unb}{\underbrace}
\nc{\ovs}[2]{\overset{#2}{#1}}			
\nc{\uns}[2]{\underset{#2}{#1}}		
\nc{\ovla}[1]{\overleftarrow{#1}}
\nc{\ovra}[1]{\overrightarrow{#1}}

\nc{\contraction}[1]{\overbracket{#1}}

\nc{\ti}[1]{{\TOMm{\tilde{#1}}}}
\nc{\tivc}[1]{\TOMm{\ti{\vc{#1}}}}

\nc{\Ti}[1]{\TOMm{\widetilde{#1}}}
\nc{\TTi}[1]{\TOMm{\!\widetilde{\,#1\,}\!}}	
\nc{\Tivc}[1]{\TOMm{\widetilde{\vc{#1}}}}

\nc{\Wi}[1]{\TOMm{\widehat{#1}}}
\nc{\WWi}[1]{\TOMm{\!\widehat{\,#1\,}\!}}	

\nc{\janusdel}{\partial}
\nc{\janusdelslashed}{\slashdel}
\declareslashed{}{\smaaa{\shortleftrightarrow}}{0}{1.6}{\janusdel}
\declareslashed{}{\smaaa{\shortleftrightarrow}}{0.07}{1.7}{\janusdelslashed}

\nc{\janus}{\slashed{\janusdel}}
\nc{\januslo}[1]{\janus\,\!_{\!#1}}
\nc{\janushi}[1]{\janus\,\!\hi{#1}}
\nc{\janusslashed}{\slashed{\janusdelslashed}}

\nc{\actson}{\,\!^{\!}\triangleright^{\!}\,\!}
\nc{\actsonn}{\!\triangleright\!}

\nc{\degree}{\TOMm{^\circ}}
\nc{\composed}{\circ}

\nc{\cartprod}{\mathchoice{\hspace{0.7pt}\sma\times\hspace{0.7pt}}
	{\hspace{0.7pt}\sma\times\hspace{0.7pt}}
	{\hspace{0.4pt}\times\hspace{0.4pt}}
	{\hspace{0.2pt}\times\hspace{0.2pt}} }

\nc{\cartprodn}{\! \cartprod \!}
\nc{\cartprods}{{\smaaa\cartprod}}
\nc{\cartprodns}{{\smaaa\cartprodn}}

\nc{\dirsum}{\mathchoice{\hspace{1pt}\sma\oplus\hspace{1pt}}
	{\hspace{1pt}\sma\oplus\hspace{1pt}}
	{\hspace{0.5pt}\oplus\hspace{0.5pt}}
	{\hspace{0.3pt}\oplus\hspace{0.3pt}} }

\nc{\tensored}{\mathchoice{\hspace{1pt}\sma\otimes\hspace{1pt}}
	{\hspace{1pt}\sma\otimes\hspace{1pt}}
	{\hspace{0.5pt}\otimes\hspace{0.5pt}}
	{\hspace{0.3pt}\otimes\hspace{0.3pt}} }

\nc{\cdotw}{\,\cdot\,}
\nc{\ldotsn}{{\scriptstyle\ldots}}

\nc{\bgl}{\left}
\nc{\biigl}{\Bigl}
\nc{\biiigl}{\biggl}
\nc{\biiiigl}{\Biggl}

\nc{\bgr}{\right}
\nc{\biigr}{\Bigr}
\nc{\biiigr}{\biggr}
\nc{\biiiigr}{\Biggr}

\nc{\bgm}{\middle}
\nc{\biigm}{\Bigm}
\nc{\biiigm}{\biggm}
\nc{\biiiigm}{\Biggm}

\nc{\bglrr}[1]{\bgl( #1 \bgr)}
\nc{\biglrr}[1]{\bigl( #1 \bigr)}
\nc{\biiglrr}[1]{\biigl( #1 \biigr)}
\nc{\biiiglrr}[1]{\biiigl( #1 \biiigr)}
\nc{\biiiiglrr}[1]{\biiiigl( #1 \biiiigr)}

\nc{\bglrs}[1]{\bgl[ #1 \bgr]}
\nc{\biglrs}[1]{\bigl[ #1 \bigr]}
\nc{\biiglrs}[1]{\biigl[ #1 \biigr]}
\nc{\biiiglrs}[1]{\biiigl[ #1 \biiigr]}
\nc{\biiiiglrs}[1]{\biiiigl[ #1 \biiiigr]}

\nc{\bglrc}[1]{\bgl\{ #1 \bgr\}}
\nc{\biglrc}[1]{\bigl\{ #1 \bigr\}}
\nc{\biiglrc}[1]{\biigl\{ #1 \biigr\}}
\nc{\biiiglrc}[1]{\biiigl\{ #1 \biiigr\}}
\nc{\biiiiglrc}[1]{\biiiigl\{ #1 \biiiigr\}}

\nc{\bglra}[1]{\bgl\langle #1 \bgr\rangle}
\nc{\biglra}[1]{\bigl\langle #1 \bigr\rangle}
\nc{\biiglra}[1]{\biigl\langle #1 \biigr\rangle}
\nc{\biiiglra}[1]{\biiigl\langle #1 \biiigr\rangle}
\nc{\biiiiglra}[1]{\biiiigl\langle #1 \biiiigr\rangle}

\nc{\bglrrn}[2]{\bgl( #1 \bgr)^{\!#2}}
\nc{\biglrrn}[2]{\bigl( #1 \bigr)^{\!#2}}
\nc{\biiglrrn}[2]{\biigl( #1 \biigr)^{\!#2}}
\nc{\biiiglrrn}[2]{\biiigl( #1 \biiigr)^{\!#2}}
\nc{\biiiiglrrn}[2]{\biiiigl( #1 \biiiigr)^{\!#2}}

\nc{\bglrsn}[2]{\bgl[ #1 \bgr]^{#2}}
\nc{\biglrsn}[2]{\bigl[ #1 \bigr]^{#2}}
\nc{\biiglrsn}[2]{\biigl[ #1 \biigr]^{#2}}
\nc{\biiiglrsn}[2]{\biiigl[ #1 \biiigr]^{#2}}
\nc{\biiiiglrsn}[2]{\biiiigl[ #1 \biiiigr]^{#2}}

\nc{\bglrcn}[2]{\bgl\{ #1 \bgr\}^{\!#2}}
\nc{\biglrcn}[2]{\bigl\{ #1 \bigr\}^{\!#2}}
\nc{\biiglrcn}[2]{\biigl\{ #1 \biigr\}^{\!#2}}
\nc{\biiiglrcn}[2]{\biiigl\{ #1 \biiigr\}^{\!#2}}
\nc{\biiiiglrcn}[2]{\biiiigl\{ #1 \biiiigr\}^{\!#2}}

\nc{\bglran}[2]{\bgl\langle #1 \bgr\rangle^{\!#2}}
\nc{\biglran}[2]{\bigl\langle #1 \bigr\rangle^{\!#2}}
\nc{\biiglran}[2]{\biigl\langle #1 \biigr\rangle^{\!#2}}
\nc{\biiiglran}[2]{\biiigl\langle #1 \biiigr\rangle^{\!#2}}
\nc{\biiiiglran}[2]{\biiiigl\langle #1 \biiiigr\rangle^{\!#2}}

\nc{\comcon}{\TOM{c.\hspace{-0.5pt}c.\hspace{-0.5pt}\xspace}%
			{\text{ c.c. }}}

\nc{\etc}{etc.\hspace{-0.5pt}\xspace}
\nc{\exgra}{e.\hspace{-0.5pt}g.\hspace{-0.5pt}\xspace}
\nc{\etal}{et al.\hspace{-0.5pt}\xspace}
\nc{\idest}{i.\hspace{-0.5pt}e.\hspace{-0.5pt}\xspace}
\nc{\resp}{resp.\hspace{-0.5pt}\xspace}

\declareslashed{}{\iota}{0.2}{0}{\pi}
\nc{\piu}{\TOMm{\slashed{\pi}}}

\nc{\iu}{\TOMt{i}}

\nc{\eu}{\TOMt{e}}

\nc{\ruut}[2][]{\sqrt[#1]{#2}}

\nc{\ruutabs}[2][]{\ruut[#1]{%
	\hspace{-1pt}\abs{\hspace{0.5pt}#2\hspace{0.5pt}}\hspace{-1pt}}%
	}
\nc{\ruutabsn}[2][]{\ruut[#1]{%
	\hspace{-1pt}\abs{#2}\hspace{-1pt}}%
	}

\nc{\ruuts}[2][]{\sma{\ruut[#1]{#2}}}
\nc{\ruutss}[2][]{\smaa{\ruut[#1]{#2}}}
\nc{\ruutsss}[2][]{\smaaa{\ruut[#1]{#2}}}
\nc{\ruutssss}[2][]{\smaaaa{\ruut[#1]{#2}}}

\nc{\ruutabss}[2][]{\sma{\ruutabs[#1]{#2}}}
\nc{\ruutabsss}[2][]{\smaa{\ruutabs[#1]{#2}}}
\nc{\ruutabssss}[2][]{\smaaa{\ruutabs[#1]{#2}}}
\nc{\ruutabsssss}[2][]{\smaaaa{\ruutabs[#1]{#2}}}

\nc{\ruutabssn}[2][]{\sma{\ruutabsn[#1]{#2}}}
\nc{\ruutabsssn}[2][]{\smaa{\ruutabsn[#1]{#2}}}
\nc{\ruutabssssn}[2][]{\smaaa{\ruutabsn[#1]{#2}}}
\nc{\ruutabsssssn}[2][]{\smaaaa{\ruutabsn[#1]{#2}}}

\nc{\fracs}[2]{\sma{\frac{#1}{#2}} }
\nc{\fracss}[2]{\smaa{\frac{#1}{#2}} }
\nc{\fracsss}[2]{\smaaa{\frac{#1}{#2}} }
\nc{\fracssss}[2]{\smaaaa{\frac{#1}{#2}} }

\nc{\fraccoco}[2]{\frac{\coco{#1}}{\coco{#2}}}
\nc{\fracscoco}[2]{\sma{\fraccoco{#1}{#2}} }
\nc{\fracsscoco}[2]{\smaa{\fraccoco{#1}{#2}} }
\nc{\fracssscoco}[2]{\smaaa{\fraccoco{#1}{#2}} }
\nc{\fracsssscoco}[2]{\smaaaa{\fraccoco{#1}{#2}} }

\nc{\fracwspace}{\hspace{0.2ex}}

\nc{\fracw}[2]{\TOMm{\, \frac{\fracwspace #1 \fracwspace}%
			{\fracwspace #2 \fracwspace} \,}%
			}
\nc{\fracws}[2]{\sma{\fracw{#1}{#2}} }
\nc{\fracwss}[2]{\smaa{\fracw{#1}{#2}} }
\nc{\fracwsss}[2]{\smaaa{\fracw{#1}{#2}} }
\nc{\fracwssss}[2]{\smaaaa{\fracw{#1}{#2}} }

\nc{\fracwcoco}[2]{\TOMm{\, \frac{\fracwspace \coco{#1} \fracwspace}%
			{\fracwspace \coco{#2} \fracwspace} \,}%
			}
\nc{\fracwscoco}[2]{\sma{\fracwcoco{#1}{#2}} }
\nc{\fracwsscoco}[2]{\smaa{\fracwcoco{#1}{#2}} }
\nc{\fracwssscoco}[2]{\smaaa{\fracwcoco{#1}{#2}} }
\nc{\fracwsssscoco}[2]{\smaaaa{\fracwcoco{#1}{#2}} }

\nc{\tfracw}[2]{\TOMm{\, \tfrac{\fracwspace #1 \fracwspace}%
			{\fracwspace #2 \fracwspace} \,}%
			}

\nc{\tfraccoco}[2]{\TOMm{\tfrac{\coco{#1}}{\coco{#2}}\,}}

\nc{\tfracwcoco}[2]{\TOMm{\, \tfrac{\fracwspace \coco{#1} \fracwspace}%
			{\fracwspace \coco{#2} \fracwspace} \,}%
			}

\nc{\dual}[1]{\TOMm{\,^* #1}}   
\nc{\dualspace}[1]{\TOMm{#1^*}}	

\nc{\coco}[1]{\TOMm{\ovl{#1}}}	
\nc{\cocow}[1]{\coco{\,#1\,}}
 	       	      
\nc{\hodge}{\TOMm{\star}}		

\nc{\pushfwd}[1]{\TOMm{#1 _{\smaaaa\rhd} }}	
\nc{\pushfwdat}[2]{\TOMm{#1 _{\smaaaa\rhd} ^{#2}}}
\nc{\pullback}[1]{\TOMm{#1 ^{\smaaaa\lhd} }}	
\nc{\pullbackat}[2]{\TOMm{#1 ^{\smaaaa\lhd} _{#2}}}

\nc{\Trans}[1]{\TOMm{#1^\top}}
\nc{\adjoint}[1]{\TOMm{#1^{\dagger}}}

\nc{\opspace}[1]{\text{Op}\Arg{#1}}
\nc{\linopspace}[1]{\text{Lin}\Arg{#1}}

\rnc{\emptyset}{\TOMm{\varnothing}}

\nc{\union}{\mathop{\cup}}
\nc{\unionn}{\mathop{\bigcup}}
\nc{\unionnl}[2]{\mathop{\bigcup} \limits_{#1}^{#2}}
\nc{\disunion}{\mathop{\sqcup}}
\nc{\disunionn}{\mathop{\bigsqcup}}
\nc{\disunionnl}[2]{\mathop{\bigsqcup} \limits_{#1}^{#2}}

\nc{\intsec}{\mathop{\cap}}
\nc{\intsecw}{\,\intsec\,}
\nc{\intsecc}{\mathop{\bigcap}}
\nc{\intseccw}{\,\intsecc\,}
\nc{\intsecclim}[2]{\mathop{\bigcap} \limits_{#1}^{#2}}
\nc{\intseccliml}[1]{\mathop{\bigcap} \limits_{#1}}

\nc{\without}{\setminus}

\nc{\Else}{\text{else}}

\nc{\const}{\text{const.}}
\nc{\deth}[1][]{\text{det}^{#1}\,}
\nc{\diag}{\text{diag}\,}
\nc{\sign}{\text{sign}\,}

\nc{\vcnabla}{\,\vc{\!\nabla\!}\,}
\nc{\grad}{\text{grad}\,}					
\nc{\gradvc}{\vc{\text{grad}}\;}
\rnc{\div}{\text{div}\;}						

\nc{\Id}[1][]{\TOMm{\text{Id} \ltx{#1} \,\!}}
\nc{\Idop}[1][]{\TOMm{\op{\text{Id}} \ltx{#1} \,\!}}
\nc{\One}[1][]{\TOMm{\mathbbm{1} \ltx{#1} \,\!}}
\nc{\Oneop}[1][]{\TOMm{\op{\mathbbm{1}} \ltx{#1} \,\!}}
\nc{\trace}{\text{tr }}
\nc{\rank}{\text{rank }}
\nc{\card}{\text{card }}
\rnc{\det}[1][]{\text{det}{#1}\;}
\nc{\Repart}{\mathbb{R}\text{e}\:}
\nc{\Impart}{\mathbb{I}\text{m}\:}

\nc{\interior}[1]{\TOMm{\text{int } #1}}
\nc{\closure}[1]{\TOMm{\ovl{#1}}}
\nc{\kernel}{\text{Ker}\;}
\nc{\image}{\text{Im}\;}
\nc{\supp}{\text{Supp}\;}
\nc{\domain}{\text{Dom}\;}

\nc{\eq}{\, = \,}
\nc{\eqn}{\! = \!}
\nc{\eqnn}{\!\! = \!\!}
\nc{\eqw}{\,\eq\,}
\nc{\neqq}{\, \neq \,}
\nc{\defeq}{\, := \,}
\nc{\eqdef}{\, =: \,}
\nc{\eqos}[2][]{\,\uns{\ovs{=}{#2}}{#1}\,}
\nc{\eqostx}[2][]{\eqos[\text{\tiny#1}]{\text{\tiny#2}} }
\nc{\eqosref}[2][]{\eqostx[\eqref{#2}]{\eqref{#1}}}
\nc{\eqx}{\eqos{!}}
\nc{\eqq}{\eqos{?}}

\nc{\approxn}{\! \approx \!}
\nc{\approxw}{\, \approx \,}

\nc{\equivw}{\,\equiv\,}
\nc{\equivn}{\!\equiv\!}
\nc{\nequiv}{\slashed{\equiv}}
\nc{\nequivw}{\,\nequiv\,}

\nc{\eqapprox}{\simeq}
\nc{\verysmall}{\smaa{\ll}}
\nc{\verylarge}{\smaa{\gg}}

\nc{\orthog}{\bot}
\nc{\orthcomp}[1][]{^{\orthog_{#1}} }

\nc{\simeqdif}{\ovs{\smaaa\simeq}{}}
\declareslashed{}{\text{\tiny dif}}{0}{0.9}{\simeqdif}
\nc{\diffeomorph}{\enmat{\,\slashed{\simeqdif}\,}}

\nc{\simeqiso}{\ovs{\smaaa\simeq}{}}
\declareslashed{}{\text{\tiny iso}}{0}{0.9}{\simeqiso}
\nc{\isomorph}{\enmat{\,\slashed{\simeqiso}\,}}

\nc{\Diffgroup}[2][]{\TOMm{\text{Diff}{^{^{\,}#1}}(#2)}}	

\nc{\smallplus}{{\smaaa+}}
\nc{\smallminus}{{\smaaa-}}
\declareslashed{}{{\smaaa\circ}}{0}{0}{\smallplus}
\declareslashed{}{{\smaaa\circ}}{0}{0}{\smallminus}
\nc{\preseta} {\phantom{i}}
\nc{\premseta}{\phantom{i}}
\declareslashed{}{\slashed{\smallplus}} {0}{-0.2}{\preseta}
\declareslashed{}{\slashed{\smallminus}}{0}{-0.2}{\premseta}
\nc{\seta} {\slashed{\preseta}}
\nc{\mseta}{\slashed{\premseta}}					

\nc{\presplus} {\phantom{i}}
\nc{\presminus}{\phantom{i}}
\declareslashed{}{\smallplus} {0}{-0.2}{\presplus}
\declareslashed{}{\smallminus}{0}{-0.2}{\presminus}
\nc{\splus} {\slashed{\presplus}}
\nc{\sminus}{\slashed{\presminus}}					

\nc{\pn}{\! + \!}
\nc{\mn}{\! - \!}
\nc{\pmn}{\! \pm \!}
\nc{\mpn}{\! \mp \!}
\nc{\pns}{\smaaa{\pn}}
\nc{\mns}{\smaaa{\mn}}
\nc{\pms}{{\smaaa\pm}}
\nc{\mps}{{\smaaa\mp}}
\nc{\len}{\! < \!}
\nc{\gen}{\! > \!}

\nc{\centerdots}{\phantom{i}}						
\declareslashed{}{{\smaaa\centerdot}}{0.4}{0.36}{\centerdots}
\nc{\dotpro}{\slashed{\centerdots}}
\nc{\dotprol}{\!\slashed{\centerdots}\,}

\nc{\limepszero}{\uns{\lim}{\epsilon \rightarrow 0} \,}
\nc{\limargs}[2]{\uns{\lim}{#1 \rightarrow #2} \,}
\nc{\limzero}[1]{\limargs{#1}{0}}
\nc{\liminfty}[1]{\limargs{#1}{\infty}}
\nc{\limpinfty}[1]{\limargs{#1}{+\infty}}
\nc{\limminfty}[1]{\limargs{#1}{-\infty}}

\nc{\maps}{\TOM{:\;}{:\;\;}}
\rnc{\to}{\TOM{$\,\rightarrow\,$}{\;\rightarrow\;}}
\nc{\lto}{\;\longrightarrow\;}
\nc{\toar}[2][]{\;\uns{\ovs{\rightarrow}{#2}}{#1}\;}
\nc{\ltoar}[2][]{\;\longrightarrow\negphantom{$\longrightarrow$\,}\hs{#2}\ls{#1}\;}
\nc{\ltoartx}[2][]{\;\longrightarrow\negphantom{$\longrightarrow$\,}\htxs{#2}\ltxs{#1}\;}
\nc{\toiso}{\ltoa{\text{iso}}}
\nc{\mappedto}{\;\mapsto\;}
\nc{\mappedtoar}[1]{\;\ovs{\mapsto}{#1}\;}
\nc{\lmappedtoar}[1]{\;\ovs{\longmapsto}{#1}\;}
\nc{\krarrow}{\;\rightsquigarrow\;}
\nc{\krarrowar}[1]{\;\ovs{\rightsquigarrow}{#1}\;}

\nc{\mapsn}{:\,}
\nc{\ton}{\rightarrow}
\nc{\lton}{\longrightarrow}
\nc{\toarn}[1]{\ovs{\rightarrow}{#1}}
\nc{\ltoarn}[1]{\ovs{\longrightarrow}{#1}}
\nc{\toison}{\ltoa{\text{iso}}}
\nc{\mappedton}{\mapsto}
\nc{\mappedtoarn}[1]{\ovs{\mapsto}{#1}}
\nc{\lmappedtoarn}[1]{\ovs{\longmapsto}{#1}}
\nc{\krarrown}{\rightsquigarrow}
\nc{\krarrowarn}[1]{\ovs{\rightsquigarrow}{#1}}

\rnc{\implies}{\; \Rightarrow\;}
\nc{\Implies}{\; \Longrightarrow \;}
\nc{\impliesar}[2][]{\; \uns{\ovs{\Rightarrow}{#2}}{#1} \;}
\nc{\Impliesar}[2][]{\; \uns{\ovs{\Longrightarrow}{#2}}{#1} \;}

\rnc{\iff}{\; \Leftrightarrow \;}
\nc{\Iff}{\; \Longleftrightarrow \;}
\nc{\iffar}[2][]{\; \uns{\ovs{\Leftrightarrow}{#2}}{#1} \;}
\nc{\Iffar}[2][]{\; \uns{\ovs{\Longleftrightarrow}{#2}}{#1} \;}
\nc{\iffartx}[2][]{\; \uns{\ovs{\Leftrightarrow}{\text{#2}}}{\text{#1}} \;}
\nc{\Iffartx}[2][]{\; \uns{\ovs{\Longleftrightarrow}{\text{#2}}}{\text{#1}} \;}

\nc{\included}[1]{ \; \ovs{\hookrightarrow}{#1} \;}

\nc{\difpower}[1]{^{^{_{#1}}}\,\!}
\nc{\dif}[1][]{\TOMm{\text{d}\difpower{#1} }}
\nc{\vol}[2][]{\TOMm{\dif[#1] #2 \;}}
\nc{\volvc}[2][]{\TOMm{\dif[#1] \vc{#2} \;}}
\nc{\fundif}[1]{\TOMm{\mc{D} #1 \;\,}}
\nc{\fundifar}[2]{\TOMm{\mc{D} #1 \ar{#2} \;\,}}
\nc{\intfundif}[1]{\int\!\!\fundif{#1}}
\nc{\intfundifar}[2]{\int\!\!\fundifar{#1}{#2}}
\nc{\fundifn}[1]{\TOMm{\mc{D} #1}}
\nc{\fundifarn}[2]{\TOMm{\mc{D} #1 \ar{#2}}}
\nc{\intfundifn}[1]{\int\!\!\fundifn{#1}}
\nc{\intfundifarn}[2]{\int\!\!\fundifarn{#1}{#2}}

\nc{\del}{\partial}
\nc{\dell}{\del_}
\nc{\delh}{\del\hi}
\nc{\delco}[2]{\del_{#1^{#2}}}
\nc{\vcdel}[1][]{\vc{\del\hs{\!}} \ls{\vc{#1}}\,\! }
\nc{\vcdelsq}[1][]{\vc{\del\hs{\!}}^{\,2} \ls{\vc{#1}}\,\!}
\nc{\tivcdel}[1][]{\,\tilde{\!\vcdel} \ls{\vc{#1}}\,\!}

\nc{\fuder}[2][]{\TOMm{\fracw{\delta #1}{\delta #2}}}							
\nc{\fuders}[2][]{\TOMm{\smaaa{\fuder[#1]{#2}}}}								
\nc{\fuderat}[3][]{\TOMm{\fracw{\delta #1}{\delta #2}\biiiigr|_{#3}}}						

\nc{\covder}{\nabla} 

\nc{\dirdeltaexponent}[1]{^{^{(\!#1\!)}\!\!}\,\!}		
\nc{\dirdeltaexponenttx}[1]{^{\smaaa{(\!#1\!)\!}}\,\!}

\nc{\krodelta}[3][]{\delta\dirdeltaexponent{#1}_{#2#3}}			
\nc{\krodeltahi}[3][]{\delta\dirdeltaexponent{#1}\hi{#2,#3}}				
\nc{\krotensor}[2]{\TOMm{\delta \hilo{#1}{#2}}}								
\nc{\dirdelta}[2][]{\TOM{$\delta\dirdeltaexponenttx{#1} \ar{#2}$\xspace}		
					{ \delta\dirdeltaexponent{#1}   \ar{#2}}}			
\nc{\dirdeltaab}[1]{\TOMm{\delta \ab{#1}\,}}								

\nc{\mb}[1]{\TOMm{\mathbf{#1} }}
\nc{\mbb}[1]{\TOMm{\mathbb{#1} }}
\nc{\mbbm}[1]{\TOMm{\mathbbm{#1} }}
\nc{\mc}[1]{\TOMm{\mathcal{#1} }}
\nc{\mf}[1]{\TOMm{\mathfrak{#1} }}

\nc{\lagdens}{\mc{L}}										
\nc{\lagdensar}[1]{\TOMm{\mc{L} \ar{#1} }}					
\nc{\lagfunc}{\TOMm{L}}										
\nc{\lagfuncar}[1]{\TOMm{L \ar{#1} }}						
\nc{\hamdens}{\mc{H}}										
\nc{\hamdensar}[1]{\TOMm{\mc{H} \ar{#1} }}					
\nc{\hamfunc}{\TOMm{H }}										
\nc{\hamfuncar}[1]{\TOMm{H \ar{#1} }}						

\nc{\emtens}{\TOMm{\mc{T}}}	

\nc{\Sactionar} [2][]{\TOMm{S \ls{#1} \ab{#2} }}
\nc{\SactionArg}[2][]{\TOMm{S \ls{#1} \Argb{#2} }}
\nc{\Saction}   [1][]{\Sactionar[#1]{}\,\!}							

\nc{\scatmat}[1][]{\TOMm{\mc{S}\ls{#1}}\,\!}						

\nc{\timeevol}[1][]{\mc{U}}	
\nc{\timeorder}{\opT}										

\nc{\hilb}[1][]{\mathpzc{H}_{\,#1}\,\!}					
\nc{\hilbdual}[1][]{\TOMm{\mathpzc{H}^* \ls{\,#1} }\,\!}
\nc{\hilbtx}[1]{\hilb[\text{\tiny#1}]}
\nc{\hilbdualtx}[1]{\hilbdual[\text{\tiny#1}]}
\nc{\hilbin}{\hilbtx{in}}
\nc{\hilbdualtin}{\hilbdualtx{in}}
\nc{\hilbout}{\hilbtx{out}}
\nc{\hilbdualtout}{\hilbdualtx{out}}

\nc{\configspace}{\mc{C}}

\nc{\rep}[2][]{\TOMm{\op{\mc{R}}\ls{#1} \Arg{#2} }}				

\nc{\laplacian}[1][]{\TOMm{\bigtriangleup \ls{#1}\,\! }}
\nc{\dalembertian}[1][]{\TOMm{\Box \ls{#1}\,\! }}
\nc{\beltrami}[1][]{\TOMm{\Box \ls{#1}\,\! }}

\nc{\Norm}[2][]{\mc N^{#1}_{#2}\,\!}
\nc{\tiNorm}[2][]{\ti{\mc N}^{#1}_{#2}\,\!}
\nc{\opNorm}[2][]{\op{\mc N}^{#1}_{#2}\,\!}

\nc{\reals}[1][]{\TOMm{\mathbb{R} \hs{#1} }}
\nc{\realspos}{\TOMm{\mathbb{R} \hs{+} }}
\nc{\realsposzero}{\TOMm{\mathbb{R} \hs{+}\ls{0} }}
\nc{\realproj}[1][]{\TOMm{\mathbb{RP} \hs{#1} }}
\nc{\imags}[1][]{\TOMm{\mathbb{I} \hs{#1} }}
\nc{\complex}[1][]{\TOMm{\mathbb{C} \hs{#1} }}
\nc{\comproj}[1][]{\TOMm{\mathbb{CP} \hs{#1} }}
\nc{\krecom}[1][]{\TOMm{\mathbb{K} \hs{#1} }}

\nc{\hreals}{^{\mathbb{R}}}
\nc{\lreals}{_{\mathbb{R}}}
\nc{\himag}{^{\mathbb{I}}}
\nc{\limag}{_{\mathbb{I}}}
\nc{\hcomplex}{^{\mathbb{C}}}
\nc{\lcomplex}{_{\mathbb{C}}}

\nc{\integers}[1][]{\TOMm{\mathbb{Z} \hs{#1} }}
\nc{\naturals}[1][]{\TOMm{\mathbb{N} \hs{#1} }}
\nc{\naturalszero} {\TOMm{\mathbb{N} \ls{0}  }}
\nc{\rationals}[1][]{\TOMm{\mathbb{Q}\hs{#1} }}

\nc{\Zleq}[1]{\integers\,\!\hs{\leq}\!\ar{#1}}
\nc{\Zgeq}[1]{\integers\,\!\hs{\geq}\!\ar{#1}}

\nc{\sphere}[1][]{\TOMm{\mathbb{S}^{#1} }}
\nc{\spherel}[2][]{\TOMm{\mathbb{S}^{#1}\ls{#2} }}
\nc{\ball}  [1][]{\TOMm{\mathbb{B}^{#1} }}
\nc{\balll}  [2][]{\TOMm{\mathbb{B}^{#1}\ls{#2} }}
\nc{\disc}  [1][]{\TOMm{\mathbb{D}^{#1} }}
\nc{\discl}  [2][]{\TOMm{\mathbb{D}^{#1}\ls{#2} }}

\nc{\detgdef}[1][g]{{\smaaa{ #1\defeq\det #1_{\mu\nu}}}}

\nc{\GLgroup}[2][]{\TOMm{\txG\txL{_{#1}} \Arg{#2}} }
\nc{\SLgroup}[2][]{\TOMm{\txS\txL{_{#1}} \Arg{#2}} }
\nc{\Ogroup}[2][]{\TOMm{\txO{_{#1}} \Arg{#2}} }
\nc{\SOgroup}[2][]{\TOMm{\txS\txO{_{#1}} \Arg{#2}} }
\nc{\Ugroup}[2][]{\TOMm{\txU{_{#1}} \Arg{2}} }
\nc{\SUgroup}[2][]{\TOMm{\txS\txU{_{#1}} \Arg{2}} }
\nc{\Spgroup}[2][]{\TOMm{\txS\txp{_{#1}} \Arg{#2}} }

\nc{\GLalg}[2][]{\TOMm{\textbf{gl}{_{#1}} \Arg{#2}} }
\nc{\SLalg}[2][]{\TOMm{\textbf{sl}{_{#1}} \Arg{#2}} }
\nc{\Oalg}[2][]{\TOMm{\textbf{o}{_{#1}} \Arg{#2}} }
\nc{\SOalg}[2][]{\TOMm{\textbf{so}{_{#1}} \Arg{#2}} }
\nc{\Ualg}[2][]{\TOMm{\textbf{u}{_{#1}} \Arg{2}} }
\nc{\SUalg}[2][]{\TOMm{\textbf{su}{_{#1}} \Arg{2}} }
\nc{\Spalg}[2][]{\TOMm{\textbf{sp}{_{#1}} \Arg{#2}} }

\nc{\existsw}{\exists\;}
\nc{\nexistsw}{\nexists\;}
\nc{\forallw}{\forall\;}

\nc{\forrallel}[2]{\forall \; #1 {\elof #2}}			
\nc{\forrallels}[2]{{\smaa{\forall \; #1 {\elofs #2}}}}
\nc{\forrall}[1]{\forrallel{#1}{}}
\nc{\forralls}[1]{\forrallels{#1}{}}
\nc{\forrallelhilo}[4]{^{\forrallels{#1}{#2}}_{\forrallels{#3}{#4}}}	
\nc{\forrallhilo}[2]{^{\forralls{#1}}_{\forralls{#2}}}	

\nc{\notni}{\slashed{\ni}}

\nc{\smallin}{\phantom{\in}}
\nc{\smallins}{\phantom{\in}}

\declareslashed{}{{\smaaa\in}}{0}{0.1}{\smallin}
\declareslashed{}{{\smaaaa\in}}{0}{0.02}{\smallins}

\nc{\elof}{\,\slashed{\smallin}\,}							
\nc{\elofs}{\,\slashed{\smallins}\,}

\nc{\Arg}[1]    {{\left( #1 \right)}}
\nc{\AArg}[1]   {{\bigl( #1 \bigr)}}
\nc{\AAArg}[1]  {{\biigl( #1 \biigr)}}
\nc{\AAAArg}[1] {{\biiigl( #1 \biiigr)}}
\nc{\AAAAArg}[1]{{\biiiigl( #1 \biiiigr)}}

\nc{\ar}[1]    {{\smaa{(#1)} }}
\nc{\ars}[1]   {{{\smaaa{(#1)}} }}
\nc{\arvc}[1]    {{{\smaa{(\vc #1)}} }}
\nc{\arvcp}[1]    {{{\smaa{(\vc #1')}} }}
\nc{\arvcm}[1]    {{{\smaa{(-\vc #1)}} }}

\nc{\Argb}[1]{{\left[#1\right]}}
\nc{\ab}[1]{{{\smaa{[#1]}} }}

\nc{\artx}{\ar{t,\vcx}}

\nc{\sumlim}   [2]{\sum  \limits_{#1}^{#2}}
\nc{\sumliml}  [1]{\sumlim{#1}{}}
\nc{\sumlims}  [2]{\sum  \limits_{\smaaa{#1}}^{\smaaa{#2}}}
\nc{\sumlimls} [1]{\sumlims{#1}{}}
\nc{\prodlim}  [2]{\prod \limits_{#1}^{#2}}
\nc{\prodliml} [1]{\prodlim{#1}{}}
\nc{\prodlims} [2]{\prod  \limits_{\smaaa{#1}}^{\smaaa{#2}}}
\nc{\prodlimls}[1]{\prodlims{#1}{}}

\newlength{\intsymlength}
\nc{\intlim} [2]{	\int\limits_{#1}^{#2}}
\nc{\intliml}[1]{\intlim{#1}{}}
\nc{\intvol}[2][]{\int\!\!\vol[#1]{#2}}

\nc{\intn}{\int\!\!}
			
\declareslashed{\mathop}{\int}{0.05}{0}{\sum}
\nc{\sumint}{\slashed{\sum}}
\nc{\sumintlim}[2]{\slashed{\sum} \limits_{#1}^{#2}}
\nc{\sumintliml}[1]{\sumintlim{#1}{}}
\nc{\sumintlims}[2]{\slashed{\sum} \limits_{\smaaa{#1}}^{\smaaa{#2}}}
\nc{\sumintlimls}[1]{\sumintlims{#1}{}}

\nc{\asslegendre}[3][]{\TOMm{{P^{#2}_{#3}} #1 }}
\nc{\asslegendrear}[4][]{\TOMm{\asslegendre[#1]{#2}{#3} \ar{#4}}}

\nc{\bessel}[2][]{\TOMm{J_{#2} #1 }}						
\nc{\besselar}[3][]{\TOMm{\bessel[#1]{#2} \ar{#3}}}			
\nc{\neumann}[2][]{\TOMm{N_{#2} #1}}
\nc{\neumannar}[3][]{\TOMm{\neumann[#1]{#2} \ar{#3}}}
\nc{\hankel}[2][]{\TOMm{H_{#2} #1}}
\nc{\hankelar}[3][]{\TOMm{\hankel[#1]{#2} \ar{#3}}}

\nc{\gegenbauer}[3][]{\TOMm{C\hb{#2}_{#3} #1}}
\nc{\gegenbauerar}[4][]{\TOMm{\gegenbauer[#1]{#2}{#3} \ar{#4}}}

\nc{\hypergeo}[5][]{\TOMm{F #1 \ar{#2,\,#3;\,#4;\;#5}}}

\nc{\jacobipoly}[4][]{\TOMm{{P^{(#2,#3)}_{#4}} #1 }}
\nc{\jacobipolyar}[5][]{\TOMm{\jacobipoly[#1]{#2}{#3}{#4} \ar{#5}}}

\nc{\pochhammer}[2]{(#1)_{#2}\,}
\nc{\doublepochhammer}[2]{(\!(#1)\!)_{#2}\,}

\nc{\spherbessel}[2][]{\TOMm{j_{#2} #1 }}
\nc{\spherbesselar}[3][]{\TOMm{\spherbessel[#1]{#2} \ar{#3}}}
\nc{\spherneumann}[2][]{\TOMm{n_{#2} #1 }}
\nc{\spherneumannar}[3][]{\TOMm{\spherneumann[#1]{#2} \ar{#3}}}
\nc{\spherhankel}[2][]{\TOMm{h_{#2} #1 }}
\nc{\spherhankelar}[3][]{\TOMm{\spherhankel[#1]{#2} \ar{#3}}}

\nc{\spherharmonic}[3][]{\TOMm{{Y^{#2}_{#3}} #1} }
\nc{\spherharmonicar}[4][]{\TOMm{\spherharmonic[#1]{#2}{#3} \ar{#4}}}

\nc{\sinn}[1]{\sin^{#1}\!}
\nc{\sinsq}{\sinn{2}}
\nc{\cosn}[1]{\cos^{#1}\!}
\nc{\cossq}{\cosn{2}}
\nc{\tann}[1]{\tan^{#1}\!}
\nc{\tansq}{\tann{2}}
\nc{\cotn}[1]{\cot^{#1}\!}
\nc{\cotsq}{\tann{2}}

\nc{\sinhn}[1]{\sinh^{#1}\!}
\nc{\sinhsq}{\sinhn{2}}
\nc{\coshn}[1]{\cosh^{#1}\!}
\nc{\coshsq}{\coshn{2}}
\nc{\tanhn}[1]{\tanh^{#1}\!}
\nc{\tanhsq}{\tanhn{2}}
\nc{\cothn}[1]{\coth^{#1}\!}
\nc{\cothsq}{\cothn{2}}

\nc{\abs}[1]{\TOMm{\left| #1 \right|}}
\nc{\abss}[1]{\bigl| #1 \bigr|}
\nc{\absss}[1]{\biigl| #1 \biigr|}
\nc{\abssss}[1]{\biiigl| #1 \biiigr|}
\nc{\absssss}[1]{\biiiigl| #1 \biiiigr|}

\nc{\absq}[1]{\TOMm{\left| #1 \right|^{2}}}
\nc{\abssq}[1]{\bigl| #1 \bigr|^{2}}
\nc{\absssq}[1]{\biigl| #1 \biigr|^{2}}
\nc{\abssssq}[1]{\biiigl| #1 \biiigr|^{2}}
\nc{\absssssq}[1]{\biiiigl| #1 \biiiigr|^{2}}

\nc{\norm}[2][]{\TOMm{\left| \! \left| \, #2 \, \right| \! \right|\ls{#1}} }
\nc{\normm}[2][]{\bigl| \! \bigl| \, #2 \, \bigr| \! \bigr|\ls{#1} }
\nc{\normmm}[2][]{\biigl| \! \biigl| \, #2 \, \biigr| \! \biigr|\ls{#1} }
\nc{\normmmm}[2][]{\biiigl| \! \biiigl| \, #2 \, \biiigr| \! \biiigr|\ls{#1} }
\nc{\normmmmm}[2][]{\biiiigl| \! \biiiigl| \, #2 \, \biiiigr| \! \biiiigr|\ls{#1} }

\nc{\normsq}[2][]{\TOMm{\left| \! \left| \, #2 \, \right| \! \right|^{2}\ls{#1} }}
\nc{\normmsq}[2][]{\bigl| \! \bigl| \, #2 \, \bigr| \! \bigr|^{2}\ls{#1} }
\nc{\normmmsq}[2][]{\biigl| \! \biigl| \, #2 \, \biigr| \! \biigr|^{}\ls{#1} }
\nc{\normmmmsq}[2][]{\biiigl| \! \biiigl| \, #2 \, \biiigr| \! \biiigr|^{2}\ls{#1} }
\nc{\normmmmmsq}[2][]{\biiiigl| \! \biiiigl| \, #2 \, \biiiigr| \! \biiiigr|^{2}\ls{#1} }

\nc{\floor}[1]{\left\lfloor #1 \right\rfloor}
\nc{\floorr}[1]{\bigl\lfloor #1 \bigr\rfloor}
\nc{\floorrr}[1]{\biigl\lfloor #1 \biigr\rfloor}
\nc{\floorrrr}[1]{\biiigl\lfloor #1 \biiigr\rfloor}
\nc{\floorrrrr}[1]{\biiiigl\lfloor #1 \biiiigr\rfloor}

\nc{\ceil}[1]{\left\lceil #1 \right\rceil}
\nc{\ceill}[1]{\bigl\lceil #1 \bigr\rceil}
\nc{\ceilll}[1]{\biigl\lceil #1 \biigr\rceil}
\nc{\ceillll}[1]{\biiigl\lceil #1 \biiigr\rceil}
\nc{\ceilllll}[1]{\biiiigl\lceil #1 \biiiigr\rceil}

\nc{\vc}[1]{\TOMm{\unl{#1}} }				

\nc{\multii}[1]{\unl{#1}}						
\nc{\lmultii}[1]{_{\multii{#1}}}					

\nc{\Op}[1]{\TOMm{\widehat{#1}} }			
\nc{\op}[1]{\TOMm{\hat{#1}} }					
\nc{\optx}[2][]{\TOMt{\^{#2}}^{#1}\,\!}	
\nc{\opmc}[1]{\op{\mc{#1}}}

\providecommand{\maincheck}[1]{}

\nc{\mlarge}[1]{\text{\large$#1$}}

\newlength{\negphantomlength}
\nc{\negphantom}[1]{\settowidth{\negphantomlength}{{#1}}\hspace{-\negphantomlength}}	

\nc{\igx}{\includegraphics}

\nc{\bal}[1]{\begin{align} #1 \end{align}}
\nc{\bals}[1]{\begin{align*} #1 \end{align*}}

\nc{\bsplit}[1]{\begin{split}#1\end{split}}				
\nc{\balsplit}[1]{\bal{\bsplit{#1}} }		

\nc{\bmult}[1]{\begin{multline}#1\end{multline}}
\nc{\bmults}[1]{\begin{multline*}#1\end{multline*}}

\nc{\bmat}[1]{\begin{matrix} #1 \end{matrix}}
\nc{\bpmat}[1]{\begin{pmatrix} #1 \end{pmatrix}}
\nc{\bdia}[1]{\begin{diagram} #1 \end{diagram}}
\nc{\bcas}[1]{\begin{cases} #1 \end{cases}}

\nc{\bflal}[1]{\begin{flalign} #1 \end{flalign}}
\nc{\bflals}[1]{\begin{flalign*} #1 \end{flalign*}}

\nc{\btab}[1]{\begin{table}[H]\centering #1 \end{table}}
\nc{\btabu}[2]{\begin{tabular}{#1} #2 \end{tabular}}
\nc{\btabuc}[1]{\btabu{c}{#1}}

\nc{\al}{\TOMm{\alpha}}
\nc{\be}{\TOMm{\beta}}
\nc{\ga}{\TOMm{\gamma}}
\nc{\de}{\TOMm{\delta}}
\nc{\ep}{\TOMm{\epsilon}}
\nc{\vep}{\TOMm{\varepsilon}}
\nc{\ph}{\TOMm{\phi}}
\nc{\vph}{\TOMm{\varphi}}
\nc{\ps}{\TOMm{\psi}}
\nc{\et}{\TOMm{\eta}}
\nc{\io}{\TOMm{\iota}}
\nc{\ka}{\TOMm{\kappa}}
\nc{\la}{\TOMm{\lambda}}
\nc{\muu}{\TOMm{\mu}}
\nc{\nuu}{\TOMm{\nu}}
\nc{\pii}{\TOMm{\pi}}
\nc{\ro}{\TOMm{\rho}}
\nc{\si}{\TOMm{\sigma}}
\nc{\ta}{\TOMm{\tau}}
\nc{\te}{\TOMm{\theta}}
\nc{\vte}{\TOMm{\vartheta}}
\nc{\om}{\TOMm{\omega}}
\nc{\ki}{\TOMm{\chi}}
\nc{\xii}{\TOMm{\xi}}
\nc{\ze}{\TOMm{\zeta}}

\nc{\Ga}{\TOMm{\Gamma}}
\nc{\De}{\TOMm{\Delta}}
\nc{\Ph}{\TOMm{\Phi}}
\nc{\Ps}{\TOMm{\Psi}}
\nc{\La}{\TOMm{\Lambda}}
\nc{\Pii}{\TOMm{\Pi}}
\nc{\Si}{\TOMm{\Sigma}}
\nc{\Om}{\TOMm{\Omega}}
\nc{\Te}{\TOMm{\Theta}}
\nc{\Up}{\TOMm{\Upsilon}}
\nc{\Xii}{\TOMm{\Xi}}

\nc{\upal}{\TOMm{\upalpha}}
\nc{\upbe}{\TOMm{\upbeta}}
\nc{\upga}{\TOMm{\upgamma}}
\nc{\upde}{\TOMm{\updelta}}
\nc{\upep}{\TOMm{\upepsilon}}
\nc{\upvep}{\TOMm{\upvarepsilon}}
\nc{\upph}{\TOMm{\upphi}}
\nc{\upvph}{\TOMm{\upvarphi}}
\nc{\upps}{\TOMm{\uppsi}}
\nc{\upet}{\TOMm{\upeta}}
\nc{\upio}{\TOMm{\upiota}}
\nc{\upka}{\TOMm{\upkappa}}
\nc{\upla}{\TOMm{\uplambda}}
\nc{\upmuu}{\TOMm{\upmu}}
\nc{\upnuu}{\TOMm{\upnu}}
\nc{\uppii}{\TOMm{\uppi}}
\nc{\upro}{\TOMm{\uprho}}
\nc{\upsi}{\TOMm{\upsigma}}
\nc{\upta}{\TOMm{\uptau}}
\nc{\upte}{\TOMm{\uptheta}}
\nc{\upvte}{\TOMm{\upvartheta}}
\nc{\upom}{\TOMm{\upomega}}
\nc{\upki}{\TOMm{\upchi}}
\nc{\upxii}{\TOMm{\upxi}}
\nc{\upze}{\TOMm{\upzeta}}

\nc{\alar}[2][]{\TOMm{\alpha #1 \ar{#2}}}
\nc{\bear}[2][]{\TOMm{\beta #1 \ar{#2}}}
\nc{\gaar}[2][]{\TOMm{\gamma #1 \ar{#2}}}
\nc{\dear}[2][]{\TOMm{\delta #1 \ar{#2}}}
\nc{\epar}[2][]{\TOMm{\epsilon #1 \ar{#2}}}
\nc{\vepar}[2][]{\TOMm{\varepsilon #1 \ar{#2}}}
\nc{\phar}[2][]{\TOMm{\phi #1 \ar{#2}}}
\nc{\vphar}[2][]{\TOMm{\varphi #1 \ar{#2}}}
\nc{\psar}[2][]{\TOMm{\psi #1 \ar{#2}}}
\nc{\etar}[2][]{\TOMm{\eta #1 \ar{#2}}}
\nc{\ioar}[2][]{\TOMm{\iota #1 \ar{#2}}}
\nc{\kaar}[2][]{\TOMm{\kappa #1 \ar{#2}}}
\nc{\laar}[2][]{\TOMm{\lambda #1 \ar{#2}}}
\nc{\muar}[2][]{\TOMm{\mu #1 \ar{#2}}}
\nc{\nuar}[2][]{\TOMm{\nu #1 \ar{#2}}}
\nc{\piar}[2][]{\TOMm{\pi #1 \ar{#2}}}
\nc{\roar}[2][]{\TOMm{\rho #1 \ar{#2}}}
\nc{\siar}[2][]{\TOMm{\sigma #1 \ar{#2}}}
\nc{\taar}[2][]{\TOMm{\tau #1 \ar{#2}}}
\nc{\tear}[2][]{\TOMm{\theta #1 \ar{#2}}}
\nc{\vtear}[2][]{\TOMm{\vartheta #1 \ar{#2}}}
\nc{\omar}[2][]{\TOMm{\omega #1 \ar{#2}}}
\nc{\kiar}[2][]{\TOMm{\chi #1 \ar{#2}}}
\nc{\xiar}[2][]{\TOMm{\xi #1 \ar{#2}}}
\nc{\zear}[2][]{\TOMm{\zeta #1 \ar{#2}}}

\nc{\Gaar}[2][]{\TOMm{\Gamma #1 \ar{#2}}}
\nc{\Dear}[2][]{\TOMm{\Delta #1 \ar{#2}}}
\nc{\Phar}[2][]{\TOMm{\Phi #1 \ar{#2}}}
\nc{\Psar}[2][]{\TOMm{\Psi #1 \ar{#2}}}
\nc{\Laar}[2][]{\TOMm{\Lambda #1 \ar{#2}}}
\nc{\Piar}[2][]{\TOMm{\Pi #1 \ar{#2}}}
\nc{\Siar}[2][]{\TOMm{\Sigma #1 \ar{#2}}}
\nc{\Omar}[2][]{\TOMm{\Omega #1 \ar{#2}}}
\nc{\Tear}[2][]{\TOMm{\Theta #1 \ar{#2}}}
\nc{\Upar}[2][]{\TOMm{\Upsilon #1 \ar{#2}}}
\nc{\Xiar}[2][]{\TOMm{\Xi #1 \ar{#2}}}

\nc{\txa}{\text{a}}
\nc{\txb}{\text{b}}
\nc{\txc}{\text{c}}
\nc{\txd}{\text{d}}
\nc{\txe}{\text{e}}
\nc{\txf}{\text{f}}
\nc{\txg}{\text{g}}
\nc{\txh}{\text{h}}
\nc{\txi}{\text{i}}
\nc{\txj}{\text{j}}
\nc{\txk}{\text{k}}
\nc{\txl}{\text{l}}
\nc{\txm}{\text{m}}
\nc{\txn}{\text{n}}
\nc{\txo}{\text{o}}
\nc{\txp}{\text{p}}
\nc{\txq}{\text{q}}
\nc{\txr}{\text{r}}
\nc{\txs}{\text{s}}
\nc{\txt}{\text{t}}
\nc{\txu}{\text{u}}
\nc{\txv}{\text{v}}
\nc{\txw}{\text{w}}
\nc{\txx}{\text{x}}
\nc{\txy}{\text{y}}
\nc{\txz}{\text{z}}

\nc{\txA}{\text{A}}
\nc{\txB}{\text{B}}
\nc{\txC}{\text{C}}
\nc{\txD}{\text{D}}
\nc{\txE}{\text{E}}
\nc{\txF}{\text{F}}
\nc{\txG}{\text{G}}
\nc{\txH}{\text{H}}
\nc{\txI}{\text{I}}
\nc{\txJ}{\text{J}}
\nc{\txK}{\text{K}}
\nc{\txL}{\text{L}}
\nc{\txM}{\text{M}}
\nc{\txN}{\text{N}}
\nc{\txO}{\text{O}}
\nc{\txP}{\text{P}}
\nc{\txQ}{\text{Q}}
\nc{\txR}{\text{R}}
\nc{\txS}{\text{S}}
\nc{\txT}{\text{T}}
\nc{\txU}{\text{U}}
\nc{\txV}{\text{V}}
\nc{\txW}{\text{W}}
\nc{\txX}{\text{X}}
\nc{\txY}{\text{Y}}
\nc{\txZ}{\text{Z}}

\nc{\vcxcutright}{\!\,\hs{\!}}
\nc{\vcxcutrightd}{\!\,\hs{\!\!}}
\nc{\vcxcutrightt}{\!\,\hs{\!\!\!}}
\nc{\vcxextleft}{\hspace{1pt}}
\nc{\vcxnormright}{\hs{\,}\,\!}
\nc{\vcxnormrightd}{\hs{\,\,}\,\!}
\nc{\vcxnormrightt}{\hs{\,\,\,}\,\!}

\nc{\vca}{\TOMm{\vc{a\vcxcutright}\vcxnormright}}
\nc{\vcb}{\TOMm{\vc{b\vcxcutright}\vcxnormright}}
\nc{\vcc}{\TOMm{\vc{c\vcxcutright}\vcxnormright}}
\nc{\vcd}{\TOMm{\vc{d\vcxcutright}\vcxnormright}}
\nc{\vce}{\TOMm{\vc{e\vcxcutright}\vcxnormright}}
\nc{\vcf}{\TOMm{\vc{f\vcxcutrightd}\vcxnormright}}
\nc{\vcg}{\TOMm{\vc{\vcxextleft g\vcxcutrightd}\vcxnormright}}
\nc{\vch}{\TOMm{\vc{h\vcxcutright}\vcxnormright}}
\nc{\vci}{\TOMm{\vc{i\vcxcutright}\vcxnormright}}
\nc{\vcj}{\TOMm{\vc{j\vcxcutrightd}\vcxnormright}}
\nc{\vck}{\TOMm{\vc{k\vcxcutright}\vcxnormright}}
\nc{\vcl}{\TOMm{\vc{l\vcxcutright}\vcxnormright}}
\nc{\vcm}{\TOMm{\vc{m\vcxcutright}\vcxnormright}}
\nc{\vcn}{\TOMm{\vc{n\vcxcutright}\vcxnormright}}
\nc{\vco}{\TOMm{\vc{o\vcxcutright}\vcxnormright}}
\nc{\vcp}{\TOMm{\vc{\vcxextleft p\vcxcutrightd}\vcxnormright}}
\nc{\vcq}{\TOMm{\vc{q\vcxcutright}\vcxnormright}}
\nc{\vcr}{\TOMm{\vc{r\vcxcutright}\vcxnormright}}
\nc{\vcs}{\TOMm{\vc{s\vcxcutright}\vcxnormright}}
\nc{\vct}{\TOMm{\vc{\vcxextleft t\vcxcutright}\vcxnormright}}
\nc{\vcu}{\TOMm{\vc{u\vcxcutright}\vcxnormright}}
\nc{\vcv}{\TOMm{\vc{v\vcxcutright}\vcxnormright}}
\nc{\vcw}{\TOMm{\vc{w\vcxcutright}\vcxnormright}}
\nc{\vcx}{\TOMm{\vc{x\vcxcutright}\vcxnormright}}
\nc{\vcy}{\TOMm{\vc{y\vcxcutrightd}\vcxnormright}}
\nc{\vcz}{\TOMm{\vc{z\vcxcutrightd}\vcxnormright}}

\nc{\vcA}{\TOMm{\vc{A\vcxcutright}\vcxnormright}}
\nc{\vcB}{\TOMm{\vc{B\vcxcutright}\vcxnormright}}
\nc{\vcC}{\TOMm{\vc{C\vcxcutrightd}\vcxnormrightd}}
\nc{\vcD}{\TOMm{\vc{D\vcxcutrightd}\vcxnormrightd}}
\nc{\vcE}{\TOMm{\vc{E\vcxcutrightd}\vcxnormrightd}}
\nc{\vcF}{\TOMm{\vc{F\vcxcutrightd}\vcxnormrightd}}
\nc{\vcG}{\TOMm{\vc{G\vcxcutrightd}\vcxnormrightd}}
\nc{\vcH}{\TOMm{\vc{H\vcxcutrightd}\vcxnormrightd}}
\nc{\vcI}{\TOMm{\vc{I\vcxcutrightd}\vcxnormright}}
\nc{\vcJ}{\TOMm{\vc{J\vcxcutrightd}\vcxnormright}}
\nc{\vcK}{\TOMm{\vc{K\vcxcutrightd}\vcxnormrightd}}
\nc{\vcL}{\TOMm{\vc{L\vcxcutright}\vcxnormright}}
\nc{\vcM}{\TOMm{\vc{M\vcxcutrightd}\vcxnormrightd}}
\nc{\vcN}{\TOMm{\vc{N\vcxcutrightd}\vcxnormrightd}}
\nc{\vcO}{\TOMm{\vc{O\vcxcutrightd}\vcxnormrightd}}
\nc{\vcP}{\TOMm{\vc{P\vcxcutrightd}\vcxnormright}}
\nc{\vcQ}{\TOMm{\vc{Q\vcxcutrightd}\vcxnormright}}
\nc{\vcR}{\TOMm{\vc{R\vcxcutright}\vcxnormright}}
\nc{\vcS}{\TOMm{\vc{S\vcxcutright}\vcxnormright}}
\nc{\vcT}{\TOMm{\vc{\vcxextleft T\vcxcutrightd}\vcxnormrightd}}
\nc{\vcU}{\TOMm{\vc{U\vcxcutrightd}\vcxnormrightd}}
\nc{\vcV}{\TOMm{\vc{V\vcxcutrightt}\vcxnormrightd}}
\nc{\vcW}{\TOMm{\vc{W\vcxcutrightd}\vcxnormrightd}}
\nc{\vcX}{\TOMm{\vc{X\vcxcutrightd}\vcxnormrightt}}
\nc{\vcY}{\TOMm{\vc{\vcxextleft Y\vcxcutrightt}\vcxnormrightt}}
\nc{\vcZ}{\TOMm{\vc{Z\vcxcutrightd}\vcxnormrightd}}

\nc{\vcal}{\TOMm{\vc\alpha}}
\nc{\vcbe}{\TOMm{\vc\beta}}
\nc{\vcga}{\TOMm{\vc\gamma}}
\nc{\vcde}{\TOMm{\vc\delta}}
\nc{\vcep}{\TOMm{\vc\epsilon}}
\nc{\vcvep}{\TOMm{\vc\varepsilon}}
\nc{\vcph}{\TOMm{\vc\phi}}
\nc{\vcvph}{\TOMm{\vc\varphi}}
\nc{\vcps}{\TOMm{\vc\psi}}
\nc{\vcet}{\TOMm{\vc\eta}}
\nc{\vcio}{\TOMm{\vc\iota}}
\nc{\vcka}{\TOMm{\vc\kappa}}
\nc{\vcla}{\TOMm{\vc\lambda}}
\nc{\vcmu}{\TOMm{\vc\mu}}
\nc{\vcnu}{\TOMm{\vc\nu}}
\nc{\vcpi}{\TOMm{\vc\pi}}
\nc{\vcro}{\TOMm{\vc\rho}}
\nc{\vcsi}{\TOMm{\vc\sigma}}
\nc{\vcta}{\TOMm{\vc\tau}}
\nc{\vcte}{\TOMm{\vc\theta}}
\nc{\vcvte}{\TOMm{\vc\vartheta}}
\nc{\vcom}{\TOMm{\vc\omega}}
\nc{\vcki}{\TOMm{\vc\chi}}
\nc{\vcxi}{\TOMm{\vc\xi}}
\nc{\vcze}{\TOMm{\vc\zeta}}

\nc{\vcGa}{\TOMm{\vc\Gamma}}
\nc{\vcDe}{\TOMm{\vc\Delta}}
\nc{\vcPh}{\TOMm{\vc\Phi}}
\nc{\vcPs}{\TOMm{\vc\Psi}}
\nc{\vcLa}{\TOMm{\vc\Lambda}}
\nc{\vcPi}{\TOMm{\vc\Pi}}
\nc{\vcSi}{\TOMm{\vc\Sigma}}
\nc{\vcOm}{\TOMm{\vc\Omega}}
\nc{\vcTe}{\TOMm{\vc\Theta}}
\nc{\vcUp}{\TOMm{\vc\Upsilon}}
\nc{\vcXi}{\TOMm{\vc\Xi}}

\nc{\opR}{\TOMm{\op{R}}}
\nc{\Tim}{\TOMm{\Ti{m}}}
\nc{\tij}{\TOMm{\ti{j}}}
\nc{\til}{\TOMm{\ti{l}}}
\nc{\tim}{\TOMm{\ti{m}}}
\nc{\tip}{\TOMm{\ti{p}}}
\nc{\tiz}{\TOMm{\ti{z}}}
\nc{\tiph}{\TOMm{\ti\phi}}
\nc{\tiet}{\TOMm{\ti\eta}}
\nc{\tiom}{\TOMm{\ti\omega}}
\nc{\tize}{\TOMm{\ti\zeta}}

\nc{\roz}{{\ro_0}}

\nc{\taz}{{\ta_0}}
\nc{\tao}{{\ta_1}}
\nc{\tat}{{\ta_2}}

\nc{\emptynumberone}[3]{#2}

\nc{\volp}[1][]{\TOMm{\emptynumberone{#1} {\vol{p}} {\vol[#1\!\!]{p}} } }
\nc{\volro}[1][]{\TOMm{\emptynumberone{#1} {\vol[\!]{\rho}} {\vol[#1\!\!\!]{\rho}} } }
\nc{\volOm}[1][]{\TOMm{\emptynumberone{#1} {\vol{\Omega}} {\vol[#1\!]{\Omega}} } }

\nc{\toflat}{\ltoartx[lim.]{flat}}
\nc{\toflatw}{\;\ltoartx[lim.]{flat}\;}
\nc{\toflatww}{\;\ltoartx[lim.]{flat}\;\;}

\nc{\regM}{\mathbb{M}}
\nc{\lregM}{_{\regM}}
\nc{\hregM}{^{\regM}}

\nc{\phR}[1][]{\TOMm{\ph^{ {#1} \mathbb{R} } }}
\nc{\phI}[1][]{\TOMm{\ph^{ {#1} \mathbb{I} } }}

\nc{\confman}[3]{\txM^{(#1,#2)}_{#3}}

\nc{\opAR}{\opA^{\reals}}
\nc{\opmcW}[5]{\op{\mc W}_{#1#2}\ar{#3_{#4},#3_{#5}}}
\nc{\mcWk}[6][]{\mc W^{#1}_{#2#3}\ar{#4_{#5},#4_{#6}}}

\nc{\sittzz}{\si^{\ta\ta}_{00}}

\nc{\cococab}[1][\vc k]{\coco{c^a_{#1}}c^b_{#1} \mn c^a_{#1}\coco{c^b_{#1}}}

\nc{\Recab}[1][\vc k]{\Repart (\coco{c^a_{#1}}c^b_{#1})}
\nc{\RecabC}[1][]{\Repart (\coco{c^{C,a}_{#1}}c^{C,b}_{#1})}

\nc{\Imcab}[1][\vc k]{\Impart (\coco{c^a_{#1}}c^b_{#1})}
\nc{\Imopcab}{\Impart (\coco{\opc^a}\opc^b)}
\nc{\ImcabC}[1][\vc k]{\Impart \bglrr{\coco{c^{C,a}_{#1}}c^{C,b}_{#1}} }
\nc{\ImopcabC}{\Impart \bglrr{\coco{\opc^{C,a}}\opc^{C,b}} }

\nc{\Sibar}{{\ovl\Si}}

\nc{\Sita}[1][]{{\Si_{\ta_{#1}}\,\!}}
\nc{\Sibarta}[1][]{{ \ovl{\Si_{\ta_{#1}}\! } }\,\!}
\nc{\hSita}[1][]{^{\Si_{\ta_{#1}}}}
\nc{\lSita}[1][]{_{\Si_{\ta_{#1}}}}
\nc{\lSibarta}[1][]{_{\ovl{\Si_{\ta_{#1}}}}}

\nc{\Siro}[1][]{{\Si_{\ro_{#1}}\,\!}}
\nc{\Sibarro}[1][]{{ \ovl{\Si_{\ro_{#1}}\! } }\,\!}
\nc{\hSiro}[1][]{^{\Si_{\ro_{#1}}}}
\nc{\lSiro}[1][]{_{\Si_{\ro_{#1}}}}

\nc{\Sir}[1][]{{\Si_{r_{#1}}\,\!}}
\nc{\Sibarr}[1][]{{ \ovl{\Si_{r_{#1}}\! } }\,\!}
\nc{\hSir}[1][]{^{\Si_{r_{#1}}}}
\nc{\lSir}[1][]{_{\Si_{r_{#1}}}}

\nc{\kgsol}[1]{\TOMt{S\smaa{OL}$(#1)$}}
\nc{\kgsolC}[1]{\TOMt{S\smaa{OL}$_{\complex}(#1)$}}

\nc{\solinpro}[3][]{\left\{#2,\,#3\right\}_{#1}}
\nc{\solinproi}[3][]{\bigl\{#2,\,#3\bigr\}_{#1}}
\nc{\solinproii}[3][]{\biigl\{#2,\,#3\biigr\}_{#1}}
\nc{\solinproiii}[3][]{\biiigl\{#2,\,#3\biiigr\}_{#1}}
\nc{\solinproiiii}[3][]{\biiiigl\{#2,\,#3\biiiigr\}_{#1}}

\nc{\lMink}{_{\smaaaa{\text{Mink}}}}
\nc{\hMink}{^{\smaaaa{\text{Mink}}}}

\nc{\AdS}[1]{\TOMt{AdS$_{1,#1}$}}
\nc{\RAdS}{\TOMm{R\lAdS}}
\nc{\lAdS}{_{\smaaaa{\text{AdS}}}}
\nc{\hAdS}{^{\smaaaa{\text{AdS}}}}
\nc{\delAdS}{\TOMt{$\del$AdS}}

\nc{\lmink}{_{\smaaaa{\text{Mink}}}}
\nc{\hmink}{^{\smaaaa{\text{Mink}}}}

\nc{\omag}[3]{\TOMm{\om\hs{#1}_{#2#3}}}
\nc{\omsubmag}[3]{\TOMm{\si^{#1}_{#2#3}}}

\nc{\Normlint}[4][]{\Norm{}^{#1}_{\scriptscriptstyle[#2_{#3},#2_{#4}]}}
\nc{\RNorm}[1][]{\biiglrrn{\RAdS^{d\mn1}\Norm[\pm]{nl}}{#1}}
\nc{\tRNorm}[1][]{\biglrr{\RAdS^{d\mn1}\Norm[\pm]{nl}}\,\!^{#1}}
\nc{\opRNorm}[1][]{\biiglrrn{\RAdS^{d\mn1}\opNorm[\pm]{\opn\opl}}{#1}}

\nc{\artrOm}{\ar{t,r,\Om}}
\nc{\artroOm}{\ar{t,\ro,\Om}}
\nc{\artrozOm}{\ar{t,\roz,\Om}}
\nc{\artOm}{\ar{t,\Om}}
\nc{\arroOm}{\ar{\ro,\Om}}

\nc{\artax}{\ar{\ta,\vc x}}
\nc{\artaz}{\ar{\ta_0}}
\nc{\artazx}{\ar{\ta_0,\vc x}}
\nc{\arrz}{\ar{r_0}}

\nc{\voltOm}{\volt\!\volOm}
\nc{\voltrOm}{\volt\!\volr\!\volOm}
\nc{\intvoltOm}[1][]{\int\!\!\volt\!\volOm[#1]}
\nc{\intvoltaOm}[1][]{\int\!\!\volta\!\volOm[#1]}
\nc{\intvolrOm}[1][]{\int\!\!\volr\!\volOm[#1]}
\nc{\intvolroOm}[1][]{\int\!\!\volro\!\volOm[#1]}
\nc{\intesumlm}{\int\!\!\dif\!E \sumliml{l,m_l}}
\nc{\intemsumlm}{\intlim{E>m}{}\,\volE\!\!\sumliml{l,m_l}\,}
\nc{\intesumvclm}{\intlim{}{}\volE\!\sumliml{\vc l,m_l}\,}
\nc{\intpsumlm}{\intlim{0}{\infty}\!\!\volp\!\!\sumliml{l,m_l}\,}
\nc{\intomsumvclm}{\int\!\!\dif\!\om\sumliml{\vc l,m_l}\,}
\nc{\intomsumlm}{\intlim{}{}\volom\!\!\sumliml{l,m_l}\,}
\nc{\inttiomsumlm}{\intlim{}{}\vol{\tiom}\!\!\sumliml{l,m_l}\,}
\nc{\inttiomsumvclm}{\intlim{}{}\!\!\vol{\tiom}\!\!
\sumliml{\vc l,m_l}\,}

\nc{\gaC}{\TOMm{\ga\htxs{C}}}
\nc{\gaS}{\TOMm{\ga\htxs{S}}}
\nc{\tigaS}{\TOMm{\tiga\htxs{S}}}

\nc{\alp}{\TOMm{\al_+}}
\nc{\alm}{\TOMm{\al_-}}
\nc{\alpm}{\TOMm{\al_\pm}}
\nc{\almp}{\TOMm{\al_\mp}}
\nc{\bep}{\TOMm{\be_+}}
\nc{\bem}{\TOMm{\be_-}}
\nc{\bepm}{\TOMm{\be_\pm}}
\nc{\bemp}{\TOMm{\be_\mp}}
\nc{\gaCp}{\TOMm{\gaC_+}}
\nc{\gaCm}{\TOMm{\gaC_-}}
\nc{\gaCpm}{\TOMm{\gaC_\pm}}
\nc{\gaCmp}{\TOMm{\gaC_\mp}}

\nc{\Timp}{\TOMm{\Tim_{\scriptscriptstyle +}}}
\nc{\Timm}{\TOMm{\Tim_{\scriptscriptstyle -}}}
\nc{\Timpm}{\TOMm{\Tim_{\scriptscriptstyle \pm}}}
\nc{\Timmp}{\TOMm{\Tim_{\scriptscriptstyle \mp}}}
\nc{\timp}{\TOMm{\tim_+}}
\nc{\timm}{\TOMm{\tim_-}}
\nc{\timpm}{\TOMm{\tim_\pm}}
\nc{\timmp}{\TOMm{\tim_\mp}}

\nc{\SCfun}[4]{\TOMm{\,\!\hs{#1\!}#2\htxs{#3}_{#4}}}
\nc{\oS}[1]{\SCfun 1 S {} {#1}}
\nc{\tSodd}[1]{\SCfun 2 S {odd} {#1}}
\nc{\tSeve}[1]{\SCfun 2 S {eve} {#1}}
\nc{\oC}[1]{\SCfun 1 C {} {#1}}
\nc{\tCnon}[1]{\SCfun 2 C {non} {#1}}
\nc{\tCint}[1]{\SCfun 2 C {int} {#1}}

\nc{\cSCfun}[5]{\TOMm{\,\!\ls{#2}\hs{\,#1\!}#3\htxs{#4}_{#5}}}
\nc{\tcSeve}[2]{\cSCfun 2 {#1} S {eve} {#2}}
\nc{\tcCint}[2]{\cSCfun 2 {#1} C {int} {#2}}

\nc{\Jac}[4][]{\TOMm{J#1\,\!\hbs{\!#2\!}_{#3#4}}}
\nc{\Jacar}[5][]{\TOMm{\Jac[#1]{#2}{#3}{#4}\ar{#5}}}

\nc{\msqBF}{m^2\ltx{BF}} 

\nc{\gind}[1][d]{g\hbs{#1}}
\nc{\ruutabsg}{\ruutabs g}
\nc{\ruutabsgt}[1]{\ruutabs{\gind	\ar{#1}}}
\nc{\ruutabsgtt}{\ruutabs{g^{tt}\!}}
\nc{\ruutabsggtt}[1]{\ruutabs{(\gind g^{tt}\!)\ar{#1}}}
\nc{\ruutabsgtata}{\ruutabs{g^{\ta\ta}\!}}
\nc{\ruutabsggtata}[1]{\ruutabs{(\gind g^{\ta\ta}\!)\ar{#1}}}
\nc{\ruutabsgroro}{\ruutabs{g^{\ro\ro}\!}}
\nc{\ruutabsggroro}[1]{\ruutabs{(\gind g^{\ro\ro}\!)\ar{#1}}}
\nc{\ruutabsgrr}{\ruutabs{g^{rr}\!}}
\nc{\ruutabsggrr}[1]{\ruutabs{(\gind g^{rr}\!)\ar{#1}}}

\nc{\sympot}[2]{\bgl[ #1 ,\, #2 \bgr] }
\nc{\sympott}[2]{\bigl[ #1 ,\, #2 \bigr]}
\nc{\sympottt}[2]{\biigl[ #1 ,\, #2 \biigr]}
\nc{\sympotttt}[2]{\biiigl[ #1 ,\, #2 \biiigr]}
\nc{\sympottttt}[2]{\biiiigl[ #1 ,\, #2 \biiiigr]}

\newcommand{\xd}{\mathrm{d}}
\newcommand{\cH}{\mathcal{H}}
\newcommand{\im}{\mathrm{i}}
\newcommand{\C}{\mathbb{C}}
\newcommand{\R}{\mathbb{R}}
\newcommand{\tens}{\otimes}
\newcommand{\one}{\mathbf{1}}
\nc{\change}[1]{{\bfseries\large #1}}

\begin{document}	
%
%
%
\title{Complex structures for an S-matrix of Klein-Gordon theory on AdS spacetimes}
\author{Max Dohse and Robert Oeckl}
\email{max/robert@matmor.unam.mx}
\affiliation{Centro de Ciencias Matem\'aticas,\\
	Universidad Nacional Aut\'onoma de M\'exico, Campus Morelia,\\
	C.P.~58190, Morelia, Michoac\'an, Mexico
	}
\date{2015-01-19}
%
%
\preprint{UNAM-CCM-2015-2}

\begin{abstract}
	%
	%

While the standard construction of the S-matrix fails on Anti-de~Sitter (AdS) spacetime, a generalized S-matrix makes sense, based on the hypercylinder geometry induced by the boundary of AdS. In contrast to quantum field theory in Minkowski spacetime, there is not yet a standard way to resolve the quantization ambiguities arising in its construction. These ambiguities are conveniently encoded in the choice of a complex structure. We explore in this paper the space of complex structures for real scalar Klein-Gordon theory based on a number of criteria. These are: invariance under AdS isometries, induction of a positive definite inner product, compatibility with the standard S-matrix picture and recovery of standard structures in Minkowski spacetime under a limit of vanishing curvature. While there is no complex structure that satisfies all demands, we emphasize two interesting candidates that satisfy most: In one case we have to give up part of the isometry invariance, in the other case the induced inner product is indefinite.

\end{abstract}

\maketitle

%
%
\noindent

\section{Introduction}
\label{introduction}

A key problem for quantum field theory in Anti-de-Sitter (AdS) spacetime arises from the failure of the standard notion of S-matrix. That is, due to the lack of a continuous spectrum of temporally asymptotic free states the usual construction of the S-matrix cannot capture interesting dynamics. In the context of the AdS/CFT conjecture an S-matrix type approach has been proposed using ``boundary states'' \cite{Gid:smatrixadscft}. Unfortunately, the quantum field theoretic meaning of this approach has remained unclear due to a lack of conceptual foundation for a notion of boundary state and due to other ad hoc ingredients. The boundary in this case has the geometry of a sphere in space extended over all of time, i.e., a hypercylinder. In particular, it is timelike.
A conceptual basis for a notion of states on timelike hypersurfaces \cite{Oe:timelike} has emerged in the context of the General Boundary Formulation (GBF) of quantum theory \cite{Oe:boundary,Oe:GBQFT}. A hypercylinder geometry in particular was considered in \cite{Oe:kgtl}. This lead to a first principles approach to generalizing the S-matrix with a first realization in Minkowski spacetime \cite{CoOe:spsmatrix,CoOe:smatrixgbf}. Application of this approach to de~Sitter spacetime \cite{Col:desitterletter,Col:desitterpaper} and discussion for a larger class of curved spacetimes followed \cite{CoDo:smatrixcsp}.
The application to AdS was outlined in \cite{CDO:adsproc}, based precisely on the observation that in AdS the hypercylinder geometry is a much more appropriate home for asymptotic free states than the early and late equal-time hypersurfaces of the conventional S-matrix. The present paper is dedicated to a realization of this application to AdS.

We limit our attention in this paper to a real scalar Klein-Gordon field. As usual, its quantum field theory is obtained by quantization from the classical Klein-Gordon field theory. In contrast to standard canonical quantization, however, a space of global solutions of the classical equations of motion is not a sufficient starting point for quantization in the GBF \cite{Oe:holomorphic}. Rather, spaces of solutions that cannot be extended over all of spacetime play an essential role \cite{Oe:kgtl}. The treatment of classical Klein-Gordon field theory on AdS in \cite{Doh:classads} contains precisely the required ingredients.

Following the strategy outlined in \cite{CDO:adsproc}, the key additional ingredient for quantizing the classical field theory and setting up the generalized S-matrix is a complex structure. More precisely, for any physical spacetime region of interest, we need to consider the hypersurfaces arising as its boundary (components). For any such hypersurface $\Sigma$ the classical field theory provides a symplectic vector space $(L_{\Sigma},\omega_{\Sigma})$ of solutions of the Euler-Lagrange equations in a neighborhood of $\Sigma$ \cite{KiTu:symplectic,Oe:holomorphic}. Quantization requires a complex structure $J_{\Sigma}$ on this vector space, compatible with the symplectic structure. This is intimately related to the Feynman propagator. In text book style canonical quantization this distinguishes between ``positive and negative energy'' solutions and is essentially fixed by requiring Poincaré symmetry.

The complex structure is the main technical focus of the present article. More precisely, it is the complex structure on the hypercylinder as this is the relevant geometry for the generalized S-matrix in AdS. As in standard canonical quantization, the main guiding principle in determining the complex structure is invariance with respect to isometries. This guarantees in turn the desired invariance of amplitudes and ultimately, the (generalized) S-matrix. An additional requirement in standard quantization is that the complex structure when combined with the symplectic structure yields a positive definite inner product on the respective space of solutions. This insures in turn a positive definite inner product on the Fock space of quantum states. However, this requirement, while convenient, turns out not to be necessary for a consistent theory \cite{Oe:freefermi}.

There are further guiding principle available in the present case. One arises from the fact that Minkowski spacetime is the flat limit of AdS spacetime. This can be translated into limiting relations for spaces of solutions, symplectic structures and also complex structures. Correspondence to well established structures in Minkowski spacetime can thus be used to constrain the structures on AdS. We exploit this to constrain the complex structure.

A further valuable consistency condition arises from the comparison of S-matrices associated to different asymptotic geometries. It is part of our point of course that there cannot be an equivalence between the standard S-matrix and the one associated to the hypercylinder geometry as exhibited in Minkowski spacetime in \cite{CoOe:spsmatrix,CoOe:smatrixgbf}. However, we show that a weaker version of such an equivalence does make sense here.

Our main results are the following: We survey the landscape of complex structures with a view to implementing as many of the desired features as possible. This leads us in particular to two candidates that satisfy most properties. (There is none that satisfies all.) In one case we sacrifice positive definiteness, in the other we sacrifice some (boost) isometry invariance.

We start in Section~\ref{sec:smatgbf} with a short review of the S-matrix and its relevant generalization. In Section~\ref{class_KG_AdS} we recall essentials of the geometry of AdS and review the relevant spaces of classical solutions of the Klein-Gordon equation. We proceed in Section~\ref{sec:smcomplex} to lay out the path to the generalized S-matrix in AdS with focus on the key role played by the complex structure. The core technical results on the complex structure and resulting amplitudes are presented in Section~\ref{complex_struct_AdS_KG}. A short summary is offered in Section~\ref{summary_outlook}. The appendices contain details concerning Klein-Gordon theory in Minkowski spacetime (Appendix~\ref{class_KG_AdS:_sols_mink}), flat limits of AdS Klein-Gordon solutions (Appendix~\ref{class_KG_AdS:_flatlim_sols}), isometries on AdS Klein-Gordon solutions (Appendix~\ref{class_KG_AdS:_iso_actions}), 
and imposing rotational and time-translation symmetry on the complex structure in AdS
(Appendix~\ref{zzz_complex_struct_AdS_KG:_essential_isometry}).

\section{S-matrix and GBF}
\label{sec:smatgbf}

\subsection{Conventional S-matrix}
\label{sec:stdsmat}

The main tool for extracting predictions from perturbative QFT is the S-matrix. It is usually constructed as the limit of the transition amplitude from a free initial state $\psi_{\text{in}}$ at time $t_{\text{in}}$ to a free final state $\psi_{\text{out}}$ at time $t_{\text{out}}$ when $t_{\text{in}}\to -\infty$ and $t_{\text{out}}\to \infty$. It will be useful to recall explicitly how the S-matrix is obtained using sources, the Feynman propagator and coherent states, see e.g., \cite{ItZu:qft}. Let $L$ be the phase space of the free field theory which we can identify here with the space of global solutions. For each element in $\eta\in L$ there is a normalized coherent state $\psi_\eta$ in the Hilbert space $\cH$ of the free theory, obtained by exponentiating the corresponding creation operator,
\begin{equation}
  \psi_\eta=\exp\left(\frac{1}{\sqrt{2}}a^\dagger_{\eta}\right)\psi_0 ,
 \label{eq:crcoh}
\end{equation}
where $\psi_0$ is the vacuum. Consider a source $\mu$ via adding to the free action a term
\begin{equation}
 D_\mu(\phi)=\int\xd x\, \mu(x)\phi(x) .
\label{eq:srcterm}
\end{equation}
We use notation suggesting a real scalar field.
The transition amplitude for the theory with source, between an initial coherent state $\psi_{\eta_\text{in}}$ and a final coherent state $\psi_{\eta_\text{out}}$ is then,
\begin{equation}
 \langle \psi_{\eta_\text{out}}, U^\mu_{[t_{\text{in}},t_{\text{out}}]} \psi_{\eta_{\text{in}}}\rangle
=\langle \psi_{\eta_{\mathrm{out}}}, \psi_{\eta_{\mathrm{in}}}\rangle
\exp\left(\im \int \xd x\, \mu(x) \hat{\eta}(x)\right)
 \exp\left(\frac{\im}{2}\int \xd x\xd x'\, \mu(x) G_F(x,x') \mu(x')\right) .
\label{eq:srctampl}
\end{equation}
$U^\mu_{[t_{\text{in}},t_{\text{out}}]}$ denotes here the unitary time evolution operator and $G_F$ the Feynman propagator. $\hat{\eta}$ is a en element of the complexified phase space $L^\C$. It is sometimes called the ``classical asymptotic field''. Concretely,
\begin{equation}
 \hat{\eta}=\frac{1}{2} (\eta_{\text{in}}+\eta_{\text{out}})+\frac{\im}{2} (J\eta_{\text{in}}-J\eta_{\text{out}}).
\label{eq:asymfield}
\end{equation}
Here, $J:L\to L$ is the complex structure that multiplies positive energy solutions with $\im$ and negative energy solutions with $-\im$. In other words, $\hat{\eta}$ coincides with $\eta_{\text{in}}$ in its negative energy component and with $\eta_{\text{out}}$ in its positive energy component.

Due to the parametrization of coherent states in terms of global solutions (interaction picture), the expression (\ref{eq:srctampl}) is independent of the choice of $t_{\text{in}}$ and $t_{\text{out}}$, as long as the source is contained completely in the interval $[t_{\text{in}},t_{\text{out}}]$. Consider an interaction that contributes to the action through a potential term,
\begin{equation}
 \int \xd x\, V(\phi(x)) .
\end{equation}
The corresponding S-matrix, denoted $U^V$ here, is then given by the formal expression,
\begin{equation}
\langle \psi_{\eta_\text{out}}, U^V \psi_{\eta_{\text{in}}}\rangle
 =  \exp\left(\im\int\xd x\,
V\left(-\im\frac{\delta}{\delta \mu(x)}\right)\right)
\langle \psi_{\eta_\text{out}}, U^\mu \psi_{\eta_{\text{in}}}\rangle
\bigg|_{\mu=0} .
\label{eq:intsmat}
\end{equation}

Before proceeding to a more general perspective on the S-matrix, we motivate it by some simple observations and changes of notation. Instead of representing transition amplitudes through evolution operators $U:\cH\to\cH$ we represent them as maps $\rho:\cH\tens\cH^*\to\C$, where $\cH^*$ is the dual Hilbert space of $\cH$,
\begin{equation}
 \rho(\psi\tens\psi')=\langle \psi', U \psi\rangle .
\end{equation}
We call them \emph{amplitude maps}.
More specifically we shall write here $\rho$, $\rho^\mu$ and $\rho^V$ for the amplitude maps of the free theory, the theory with source $\mu$ and the theory with interaction potential $V$, respectively. The Hilbert space $\cH$ is the Fock space over $L$, where $L$ is equipped with the Hilbert space structure of the 1-particle space. Correspondingly, the tensor product Hilbert space $\cH_{\partial}\defeq\cH\tens\cH^*$ is the Fock space over the complex Hilbert space $L_{\partial}\defeq L\oplus \overline{L}$. Here $\overline{L}$ denotes $L$ with reversed complex structure and complex conjugated inner product. What is more, the product of a coherent state in $\cH$ and a coherent state in $\cH^*$ is a coherent state in $\cH_\partial$, parametrized by an element $\eta=(\eta_{\text{in}},\eta_{\text{out}})$ in $L_{\partial}=L\oplus \overline{L}$ and generated by an operator the form (\ref{eq:crcoh}) on $\cH_{\partial}$. We write,
\begin{equation}
 \psi_{\eta}=\psi_{(\eta_{\text{in}},\eta_{\text{out}})}=\psi_{\eta_{\text{in}}}\tens \psi_{\eta_{\text{out}}} .
\end{equation}
With this notation, formula (\ref{eq:srctampl}) for the amplitude with source takes the form,
\begin{equation}
\rho^{\mu}(\psi_{\eta})
=\rho(\psi_{\eta})
\exp\left(\im \int \xd x\, \mu(x) \hat{\eta}(x)\right)
 \exp\left(\frac{\im}{2}\int \xd x\xd x'\, \mu(x) G_F(x,x') \mu(x')\right) .
\label{eq:srcampl}
\end{equation}
Similarly, the S-matrix formula (\ref{eq:intsmat}) rewrites as,
\begin{equation}
\rho^V(\psi) =  \exp\left(\im\int\xd x\,
V\left(-\im\frac{\delta}{\delta \mu(x)}\right)\right)
\rho^\mu(\psi)\bigg|_{\mu=0} .
\label{eq:intampl}
\end{equation}

\subsection{Generalized amplitudes}
\label{sec:genampl}

As the notation suggests, the formulas (\ref{eq:srcampl}) and (\ref{eq:intampl}) apply much beyond the context given in the previous section. Underlying this are generalized notions of amplitude, observable and S-matrix. These can be made sense of in the General Boundary Formulation of quantum theory (GBF) \cite{Oe:boundary,Oe:GBQFT}. In particular, amplitudes can be associated to spacetime regions that are not of the special form $[t_{\text{in}},t_{\text{out}}]\times \R^3$ (or $[-\infty,\infty]\times \R^3$). To do this we need a state space $\cH_{\partial M}$ assigned to the boundary $\partial M$ of the region $M$. The amplitude map on $M$ is then a linear map $\cH_{\partial M}\to\C$. In the special case of a region $[t_{\text{in}},t_{\text{out}}]\times \R^3$, the boundary decomposes into two connected components, one at $t_{\text{in}}$ and one at $t_{\text{out}}$. The boundary state space correspondingly decomposes into a tensor product of two Hilbert spaces $\cH_{\text{in}}\tens\cH_{\text{out}}^*$, the second one being here dual to the first. The amplitude map is then a conventional transition amplitude and corresponds to an operator. In general, however, a single Hilbert (or Krein) space accommodates both, incoming and outgoing particles \cite{Oe:timelike}.

It turns out that formula (\ref{eq:srcampl}) provides the amplitude for a free bosonic field theory with a source $\mu$ in an arbitrary spacetime region $M$ in an arbitrary spacetime, given that its ingredients are defined \cite{Oe:affine,Oe:feynobs}. We proceed to explain this starting with a free classical field theory on an unspecified spacetime. Given a hypersurface $\Sigma$ denote by $L_{\Sigma}$ the real vector space of solutions of the equations of motion in a neighborhood of $\Sigma$. (Strictly speaking we should consider ``germs'' of solutions.) The second variation of the action yields an anti-symmetric bilinear form $\omega_{\Sigma}:L\times L\to\R$, the \emph{symplectic structure}. We assume this to be non-degenerate. (Otherwise one has to resort to symplectic reduction.) This makes $L_{\Sigma}$ into a symplectic vector space. It is the \emph{phase space} on $\Sigma$.
For a spacetime region $M$ we denote by $L_M$ the real vector space of solutions of the equations of motion in $M$. Let $r_M:L_M\to L_{\partial M}$ be the map that restricts a solution in $M$ to a neighborhood of the boundary $\partial M$. Remarkably, the subspace $L_{\tilde{M}}\defeq r(L_M)\subseteq L_{\partial M}$ is generically a \emph{Lagrangian subspace} \cite{KiTu:symplectic}. That is, the symplectic form $\omega_{\partial M}$ vanishes on $L_{\tilde{M}}$, and $L_{\tilde{M}}$ is a maximal subspace with this property.

To quantize the theory we need a compatible complex structure $J_{\Sigma}$ on $L_{\Sigma}$ for each hypersurface $\Sigma$. That is $J_{\Sigma}$ must satisfy $J_{\Sigma}^2=-\one_{\Sigma}$ and $\omega_{\Sigma}(J_{\Sigma}\phi_1,J_{\Sigma}\phi_2)=\omega_{\Sigma}(\phi_1,\phi_2)$. Then,
\begin{equation}
  g_{\Sigma}(\phi_1,\phi_2)\defeq 2\omega_{\Sigma}(\phi_1,J_{\Sigma}\phi_2)\qquad\text{and}\qquad
  \{\phi_1,\phi_2\}_{\Sigma}\defeq g_{\Sigma}(\phi_1,\phi_2)+2\im \omega_{\Sigma}(\phi_1,\phi_2)
\label{eq:ip}
\end{equation}
define a real and a complex inner product on $L_{\Sigma}$, respectively. In standard quantization the complex structure is such that $g_{\Sigma}$ (and thus also $\{\cdot,\cdot\}_{\Sigma}$) is positive definite. However, a quantization where these structures are indefinite is perfectly consistent, provided we respect associated superselection rules when extracting probabilities and expectation values \cite{Oe:freefermi}. (In fact, in fermionic field theories indefinite inner products appear necessarily when giving up the restriction to spacelike hypersurfaces \cite{Oe:freefermi}.) With the complex inner product (and upon completion), $L_{\Sigma}$ is a complex Hilbert space (or Krein space in the indefinite case). This is the ``one-particle'' space. The state space $\cH_{\Sigma}$ is the bosonic Fock space over $L_{\Sigma}$. (In the indefinite case, the Fock space is also indefinite and a Krein space.)

Given a spacetime region $M$, the fact that $L_{\tilde{M}}\subseteq L_{\partial M}$ is Lagrangian has an important consequence: $L_{\partial M}$ decomposes as a direct sum $L_{\partial M}=L_{\tilde{M}}\oplus J_{\partial M} L_{\tilde{M}}$ over $\R$ \cite{Oe:holomorphic}. For $\eta\in L_{\partial M}$ we write this as $\eta=\eta^{\mathrm{R}}+J_{\partial M}\eta^{\mathrm{I}}$ with $\eta^{\mathrm{R}},\eta^{\mathrm{I}}\in L_{\tilde{M}}$. Define the element $\hat{\eta}\defeq\eta^{\mathrm{R}}-\im \eta^{\mathrm{I}}$ in the complexified subspace $L_{\tilde{M}}^\C\subseteq L_{\partial M}^\C$. Remarkably, the amplitude map for the region $M$ can be expressed in closed form, using these ingredients. Given $\eta\in L_{\partial M}$, the amplitude of the associated coherent state $\psi_{\eta}\in\cH_{\partial M}$ is \cite{Oe:holomorphic},
\begin{equation}
 \rho_M(\psi_{\eta})=\exp\left(\frac{1}{4}g_{\partial M}(\hat{\eta},\hat{\eta})\right) .
 \label{eq:freeampl}
\end{equation}
Even though we consider a free theory where everything is supposed to be simple, this result is still striking: It applies irrespective of the shape of the spacetime region $M$ and without even specifying what kind of spacetime we are actually in. Of course, restrictions on both are hidden in the assumptions we have made on the various ingredients, in particular the complex structure. We are still very far from a general understanding of these restrictions in terms of spacetime structure and field theory.

If we add to the action a linear functional $D$ on field configurations $K_M$ in $M$, i.e., $D:K_M\to\R$ linear, then the equations of motion in $M$ are modified. There is a special solution $\xi_D$ of these modified equations with the property that its restriction to the boundary lies in $J_{\partial M} L_{\tilde{M}}\subseteq L_{\partial M}$. The amplitude for the thus modified theory can be shown to take the form \cite{Oe:affine,Oe:feynobs},
\begin{equation}
  \rho_M^D(\psi_\eta)=\rho_M(\psi_\eta) \exp\left(\im D(\hat{\eta})\right)
   \exp\left(\frac{\im}{2} D(\xi_D)-\frac{1}{2}g_{\partial M}(r(\xi_D),r(\xi_D))\right) .
\label{eq:dampl}
\end{equation}
Introducing a source term $\mu$ via (\ref{eq:srcterm}) is a special case of this. The expression (\ref{eq:dampl}) turns then into the expression (\ref{eq:srcampl}) with the three factors on the right hand side in exact correspondence. For the third factor, this can be seen \cite{Oe:feynobs} through the relation between complex structure and Feynman propagator \cite{Lic:propquantgr}.

We return to the setting of Section~\ref{sec:stdsmat} to see how it fits into the framework just presented. The vector space $L_{\partial}$ may be viewed as the space of solutions of the classical equations of motion in a neighborhood of the boundary of the spacetime region $[t_{\text{in}},t_{\text{out}}]\times \R^3$ (or $[-\infty,\infty]\times \R^3$). Because of the Cauchy property such a solution is equivalent to a pair of two (generically distinct) global solutions. Hence, $L_{\partial}=L_{\text{in}}\oplus L_{\text{out}}$ with each summand equivalent to $L$. $L_{\text{in}}$ inherits the complex structure of $L$ while $L_{\text{out}}$ has the opposite complex structure, due to its opposite orientation as a boundary component of the region. Combining the two yields the complex structure $J_{\partial}:L_{\partial}\to L_{\partial}$ given by $J_{\partial} (\eta_{\text{in}},\eta_{\text{out}})=(J\eta_{\text{in}},-J \eta_{\text{out}})$. The space of solutions inside the region is again equivalent to $L$. Thus, given $\eta\in L_{\partial}$, $\hat{\eta}=\eta^{\mathrm{R}}-\im \eta^{\mathrm{I}}$ is an element of $L^\C$. As is straightforward to verify now, it is precisely given by formula (\ref{eq:asymfield}).

Given a quantum field theory and a spacetime region $M$, the availability of a compatible notion of source amplitude in $M$ via formula (\ref{eq:dampl}) or (\ref{eq:srcampl}) depends crucially on the availability of a suitable complex structure on $\partial M$. This is a given for standard QFTs and spacelike hypersurfaces in Minkowski spacetime. There, invariance under Poincaré transformations determines a complex structure essentially uniquely. For some relevant results for time-like hypersurfaces (not necessarily explicitly using the language of a complex structure), see \cite{Oe:timelike,Oe:kgtl,Oe:holomorphic,CoRa:qftrindler}.

\subsection{Generalized S-matrix}
\label{sec:gensm}

We are interested here in the particular case of spacetime regions extended to infinity to cover all of spacetime. In this case the interacting theory can be described perturbatively through formula (\ref{eq:intampl}). As recalled above, the usual S-matrix in Minkowski space is obtained by taking a time-interval region $[t_{\text{in}},t_{\text{out}}]$ and sending the boundaries to infinity, $t_{\text{in}}\to -\infty$ and $t_{\text{out}}\to \infty$. However, this is not the only possibility. A particularly compelling setup is the following: Consider the sphere $\sphere[2]_r$ of radius $r$ centered at the origin of space, extend this over all of time in Minkowski spacetime. This yields a hypercylinder $\R\times\sphere[2]_r$. The interior is the region spanned by a ball of radius $r$ extended over all of time, $\R\times\ball[3]_r$. We call this type of region a rod region.
Physically, we are considering an experiment that is spatially confined, but may run continuously. We are injecting and detecting particles from a distance $r$ from the center, but at any time. The asymptotic idealization is achieved by letting $r$ go to infinity, moving out from the interaction region, where it is well justified to consider particles as free. One might even argue that this setup is more physically compelling than the usual asymptotics in time. It was shown in \cite{CoOe:smatrixgbf} that the resulting asymptotic amplitude is in fact precisely equivalent to the usual S-matrix.

Physically, the equivalence is based on a correspondence between asymptotic classical solutions. In the standard S-matrix setting these solutions are pairs of global solutions, one at early times and one at late times. We have already identified this space of solutions as $L_{\partial T}=L_{\text{in}}\oplus L_{\text{out}}$, (but note the slight change of notation). In the hypercylinder setting the solutions live in the space $L_{\partial R}$, arising as the limit of the space of solutions $L_r$ in a neighborhood of the hypercylinder
$\R\times\sphere[2]_r$ of radius $r$, when $r$ goes to infinity. In this case there is a subtlety. In addition to the usual propagating solutions,
$L_r$ contains evanescent solutions (that is, solutions showing exponential behavior in space, and thus well-defined in a
neighborhood of the hypercylinder with finite radius $r$,
while diverging for $r\to\infty$).
The evanescent solutions are absent, however, in $L_{\partial R}$. 
Taking this into account, there is a one-to-one correspondence between the elements of $L_{\partial R}$ and those of $L_{\partial T}$. More precisely, there is an equivalence between $L_{\partial R}$ and $L_{\partial T}$ \emph{as symplectic vector spaces}. For this equivalence to survive quantization we need to choose corresponding complex structures on $L_{\partial R}$ and $L_{\partial T}$. Then $L_{\partial R}$ and $L_{\partial T}$ are equivalent \emph{as complex Hilbert (or Krein) spaces}. Consequently, the state spaces of the quantum theory, i.e., the Fock spaces over $L_{\partial R}$ and $L_{\partial T}$ will also be equivalent as complex Hilbert (or Krein) spaces. What is more, as a consequence of the classical correspondence, the amplitudes will be the same, without and with sources. For later use, we refer to this equivalence as \emph{amplitude equivalence}. For examples of amplitude equivalence in curved space times, see \cite{Col:desitterletter,Col:desitterpaper,CoDo:smatrixcsp,CoRa:qftrindler}.

The availability of different asymptotic regimes in a given spacetime becomes particularly interesting when they are inequivalent. This is the case of AdS spacetime. As is well known the conventional S-matrix approach fails due to the lack of temporally asymptotically free states. From the present perspective this manifests itself as follows. The space $L_t$ of admissible solutions in a neighborhood of the equal-time hypersurface at time $t$ is rather small and admits only discrete energy levels. There is a larger continuum of solutions, but these do not decay sufficiently fast at spatial infinity to be ``normalizable''. The negative curvature of AdS makes the solutions behave
akin to being in a box potential: only those solutions
that vanish at radial infinity (the ``wall of the box'')
are normalizable.

On the other hand, it was shown in \cite{Doh:classads} that a hypercylinder geometry leads to a very different picture. The space of admissible solutions $L_r$ in a neighborhood of the hypercylinder of radius $r$ contains a full continuum of solutions. This suggests to build the physical S-matrix in AdS on the asymptotic hypercylinder geometry rather than the conventional asymptotic time-interval geometry \cite{CDO:adsproc}. While the requisite classical theory was developed in \cite{Doh:classads}, the main ingredient for quantization is the complex structure. This is the main focus of the present article.

\section{Classical Klein-Gordon theory on AdS}%
\label{class_KG_AdS}%
%
%
\subsection{Essential geometry of AdS and Minkowski}%
\label{class_KG_AdS:_geom}%

As a preparation for the classical field theory,
we summarize here the presentation in \cite{Doh:classads}.
We denote by AdS what is more precisely denoted as C\AdS d,
that is: $(1\pn d)$-dimensional Anti-de~Sitter spacetime
with Lorentzian signature in the universal covering version.
Just as Minkowski spacetime, AdS then has the topology of 
$\reals[1\pn d]$ and no closed timelike curves. We only consider AdS 
with \emph{odd} spatial dimension $d\geq3$. We use global coordinates
with the time coordinate $t \elof (-\infty,+\infty)$,
a radial coordinate $\rho \elof [0,\fracwss{\piu}{2})$,
and denote the $(d\mn1)$ angular coordinates on \sphere[d\mn1]
collectively by $\Omega$.
In contrast to Minkowski spacetime, AdS at $\ro \eq \piu/2$
has a timelike boundary, which we denote by \delAdS.
Its topology is that of a hypercylinder:
$\delAdS \eq \reals_t \cartprod \sphere[d\mn1]$.
As already indicated, we are interested in two distinct ways
that the ''region'' of ''all of AdS'' can be obtained:
The first (corresponding to the usual S-matrix approach) is the limit $t_0\ton\infty$
of the region $\regM\hAdS_{[\mn t_0,\pn t_0]}$ consisting of the time-interval $[\mn t_0,\pn t_0]$, extended over all of space. The second is the limit $\ro_0\ton\tfrac\piu2$
of the rod region $\regM\hAdS_{\ro_0}$, given by the interior of the hypercylinder of radius $\ro_0$.
With $R\lAdS$ denoting the curvature radius of AdS and 
$\dif\!s^2_{\sphere[d\mn1]}$ the metric on the $(d\mn1)$-dimensional
unit sphere, the AdS metric writes $\dif\!s^2\lAdS
\eq \tfrac{R^2\lAdS}{\cosn2 \rho} \, \bigl(-\dif\!t^2 
+ \dif\!\rho^2 + \sinn2 \rho \; \dif\!s^2_{\sphere[d\mn1]} \bigr)$.

In order to consider the limit of large curvature radius
$R\lAdS$, we introduce the rescaled global coordinates
$r \defeq R\lAdS \,\ro$ with $r \elofs [0,\tfrac\piu2\,R\lAdS)$
and $	\ta \defeq R\lAdS \,t$ with $\ta \elofs (-\infty,+\infty)$.
Then, for large $\RAdS$ the AdS metric
approximates the Minkowski metric in the rescaled coordinates
$\ta$ and $r$, that is:
$\dif\! s^2\lAdS \toflatw \dif\!s^2\lmink \eq -\dif\!\ta^2 
+ \dif\!r^2 + r^2\, \dif\!s^2_{\sphere[d\mn1]}$.
Therefore the large-$\RAdS$ limit is also called flat limit.
The Laplace-Beltrami operator on AdS is
$\beltrami\lAdS \eq R\lAdS^{-2}\,\biglrr{ -\cossq\ro\,\del_t^2 + \cossq\ro\, \del_\ro^2 +\tfracw{(d\mn1)}{\tan\ro} \del_\ro 
+ \tann{-2}\ro \:\beltrami_{\sphere[d\mn1]} }$,
and as its flat limit we obtain the Laplace-Beltrami 
on Minkowski spacetime
$\beltrami\lAdS \toflatw  \beltrami\lmink \eq -\del_\ta^2 +\del_r^2 
+ \tfracw{(d\mn1)}r \del_r	+ r\inv[2] \:\beltrami_{\sphere[d\mn1]}$.

On AdS we have the following Killing vector fields
with $j,k \elof \{1,\ldots,d\}$:
the time translation $K_{d\pn1,0} \eq \del_t$,
the rotations $K_{jk} \eq \xi_j\del_{\xi_k} - \xi_k\del_{\xi_j}$
(with $\xi_j$ the constrained coordinates
$\vc\xi^2=1$ on $\sphere[d\mn1]$),
and two kinds of boosts: the ''0-boosts''
$K_{0j} \eq \mn\xi_j \cos t\,\sin\ro\;\del_t 
- \xi_j \sin t \,\cos \ro \;\del_\ro -
\fracw{\sin t}{\sin\ro}(\del_{\xi_j}-\xi_j\xi_i\,\del_{\xi_i})$
and the ''$(d\pn1)$-boosts'' $
K_{d\pn1,j} \eq \mn\xi_j\sin t\,\sin\ro\;\del_t 
+ \xi_j \cos t \,\cos \ro \;\del_\ro +
\fracw{\cos t}{\sin\ro}(\del_{\xi_j}-\xi_j\xi_i\,\del_{\xi_i})$.
The AdS Killing vectors are the generators of the
isometry group $\SOgroup{2,d}$ of AdS$_{1,d}$.
In the flat limit the AdS Killing vectors become the Minkowski Killing vectors: the AdS time-translation becomes the Minkowski one:
$K_{d\pn1,0} \toflatw \RAdS\,\del_\ta$,
ditto the rotations: $K_{jk} \toflatw 
\xi_j\del_{\xi_k} - \xi_k\del_{\xi_j}$,
while the AdS $(d\pn1)$-boosts become Minkowski $x_j$-translations:
$K_{d\pn1,j} \toflatw \RAdS \biglrr{\xi_j \,\del_r
+ \tfracw1r (\del_{\xi_j}-\xi_j\xi_i\, \del_{\xi_i}) }$,
and the AdS 0-boosts become Minkowski boosts
in the $(t,x_j)$-plane:
$K_{0j} \toflatw -\xi_j\,r\;\del_\ta - \xi_j \,\ta \;\del_r
- \tfracw\ta r (\del_{\xi_j}-\xi_j\xi_i\,\del_{\xi_i})$.

%
\subsection{Classical Klein-Gordon solutions on AdS}%
\label{class_KG_AdS:_sols_AdS}%

The action for a free, real, scalar field $\phi\ar{x}$ on AdS is $S(\phi) = \int \vol[d\pn1]{\!\!x} \ruutabs g\;
\tfrac12\,[ -g^{\mu\nu} (\del_\mu\phi)(\del_\nu\phi) - m^2 \phi^2]$, with $m$ the field's mass, and its Euler-Lagrange equation
is the free Klein-Gordon equation
$0 \eq (-\dalembertian\lAdS + m^2  ) \, \phi$.
We are only interested in solutions which are well-defined
and bounded on our respective AdS regions or their boundaries.
We need two types of modes
which we call hypergeometric $S^a$ and $S^b$-modes:
\bal{\label{AdS_KG_solutions_501}
	\mu\hbs{S,a}_{\om\vc l m_l}\artroOm
	& \eq \eu^{-\iu \om t}\; \spherharmonicar{m_l}{\vc l}\Om\;
		S^a_{\om l}\ar \ro
		&
	\mu\hbs{S,b}_{\om\vc l m_l}\artroOm
	& \eq \eu^{-\iu \om t}\; \spherharmonicar{m_l}{\vc l}\Om\;
		S^b_{\om l}\ar \ro.
	}
Therein, $\spherharmonicar{m_l}{\vc l}\Om$ denote the hyperspherical harmonics, and with $\hypergeo abcx$ denoting the
hypergeometric function, we use the 
linear independent radial functions
\bal{\label{kg_ads_solutions_511}
	S^a_{\om l}\ar\ro & =
		\sinn l\ro\; \cosn{\timp}\!\ro\;\;
		\hypergeo{\al^{S,a}}{\be^{S,a}}{\ga^{S,a}}{\sinn2\ro}
		&
	S^b_{\om l}\ar\ro & =
		-(\sin\ro)^{2\mn l\mn d\!} \cosn{\timp\!}\!\ro\;
		\hypergeo{\al^{S,b}}{\be^{S,b}}{\ga^{S,b}}{\sinn2\ro}.
	}
The hypergeometric parameters therein are given by
\bal{\label{kg_ads_solutions_74_parameters_SC_ab}
	\al^{S,a} & \eq \fracsss12(l\pn\Timp\mn\om)
		&
	\be^{S,a} & \eq \fracsss12(l\pn\Timp\pn\om)
		&
	\ga^{S,a} &\eq l\pn\tfrac d2
		&
	\Timp &\eq \tfrac d2\pn\nu
		\\
	\al^{S,b} & \eq \al^{S,a}\mn\ga^{S,a}\pn1
		&
	\be^{S,b} & \eq \be^{S,a}\mn\ga^{S,a}\pn1
		&
	\ga^{S,b} & \eq 2-\ga^{S,a}
		&
	\nu & \eq \ruut{d^2/4+m^2\RAdS^2}.
		\notag
	}
Whenever the frequency $\om$ is one of the discrete values
dubbed magic frequencies in \cite{bagidlaw:_what_cft_tell_about_ads}:
$\omag+nl \eq 2n+l+\Timp$,
then the $S^a$-modes take on a special form called (ordinary)
Jacobi modes, because the hypergeometric function then writes
as a Jacobi polynomial:
\bal{\label{AdS_KG_solutions_101}
	\mu\hbs+_{n\vc l m_l}\artroOm
	\eq \mu\hbs{S,a}_{\omag+ nl \vc l m_l}\artroOm
	\eq \eu^{-\iu \omag+nl t}\; \spherharmonicar{m_l}{\vc l}\Om\;
	\Jacar+nl\ro\;.
	}
The Jacobi modes are well-defined and bounded both on time axis and boundary.
We call $\mu\hbs+_{n\vc l m_l}\artroOm$ positive frequency modes 
and $\coco{\mu\hbs+_{n\vc l m_l}\artroOm}$ negative frequency modes.
By $\jacobipolyar{a}{b}{n}{x}$ we denote the Jacobi polynomials,
by $\pochhammer an$ the Pochhammer symbols, and then
\begin{equation}
\Jacar+nl\ro \eq \tfracw{n!}{\pochhammer{l\pn d/2}{n}}
\sinn l\ro\; \cosn{\timp\!}\ro\;
\jacobipolyar{l\pn d/2\mn1}{\nu}{n}{\cos 2\ro} .
\end{equation}
The hypergeometric $S^a$ and $S^b$-modes are evanescent modes
(except for the magic frequencies):
when approaching the boundary $\ro \eq \tfrac\piu2$
they grow like $\exp(\tilde\ro(\nu-\tfrac d2))$,
wherein $\tilde\ro$ is a radial coordinate of noncompact range
$\tilde\ro\in[0,\infty)$.
On the time axis $\ro\equiv 0$ the $S^a$-modes are regular 
(like Bessel modes on Minkowski spacetime)
while the $S^b$-modes are singular there
(like Neumann modes on Minkowski spacetime).
Since we wish to consider asymptotic solutions, the latter are also important.
More explicitly, when we consider the case of KG theory 
with some source field of compact support
(say within some hypercylinder surface $\Si_\roz$),
then any solution of this inhomogeneous KG equation
coincides with some free solution outside of the source region.
This \emph{free} solution generically contains
regular and singular modes.

We proceed to compile which solutions are admissible on
time-interval and rod regions, and near their boundaries. For the rod region
$\regM\hAdS_{\ro_0} \eq \reals_t\cartprod \ball[d\mn1]_{\ro_0}$,
we need solutions that are bounded for all of time
while in space we only need them bounded on $[0,\ro_0]$.
The solutions with these properties are precisely the hypergeometric $S^a$-modes.
We expand an arbitrary complexified KG solution $\ph$
(see \exgra Section 2.3 in \cite{Oe:affine})
on the rod region as an integral over these modes,
which we call rod expansion ($\ph\artroOm$ becomes real 
if and only if $\ph^{S,a}_{\om \vc l m_l} 
\eq \coco{\ph^{S,a}_{\mn\om, \vc l,\mn m_l}}$):
\bal{\label{kg_ads_solutions_rod_520}
	\ph \artroOm & \eq \sma{\intomsumvclm} 
		\ph^{S,a}_{\om \vc l m_l}\;
		\mu\hbs{S,a}_{\om \vc l m_l}\artroOm \;.
	}
This determines the space $L\hAdS_{\ro_0}$
of KG solutions for the rod region.
If we consider solutions near the boundary hypercylinder of the rod,
then again we need them bounded for all of time
but in space only for an interval like $(\ro_0\mn\ep,\ro_0\pn\ep)$.
Thus we can use $S^a$ and $S^b$-modes here,
with the frequency $\om$ being real.
We expand an arbitrary complexified KG solution $\ph$
near the rod's boundary as an integral over these modes,
which we call $S$-expansion
(iff $\coco{\ph^{S,a}_{\om \vc l m_l}} \eq
\ph^{S,a}_{\mn\om, \vc l,\mn m_l}$ and the same for
${\ph^{S,b}_{\om \vc l m_l}}$, then $\ph\artroOm$ is real):
\bal{\label{kg_ads_solutions_520}
	\ph \artroOm  & \eq \sma{\intomsumvclm}
		\biiglrr{\ph^{S,a}_{\om \vc l m_l}\;
			\mu\hbs{S,a}_{\om \vc l m_l}\artroOm
			+\ph^{S,b}_{\om \vc l m_l}\;
			\mu\hbs{S,b}_{\om \vc l m_l}\artroOm
			}.
	}
This determines the space $L\hAdS_{\del\ro_0}$ of KG solutions on the boundary of the rod region, the hypercylinder.
For the time-interval region $\regM\hAdS\lintval t12 
\eq \intval t12\cartprod \ball[d\mn1]_{\piu/2}$,
that is: time interval times all of space,
we need KG solutions that are bounded on all of space.
Thus we can only use Jacobi modes here,
and expand any complexified KG solution $\ph$ 
as a sum of Jacobi modes, which we call ordinary Jacobi expansion
(iff $\ph^+_{n \vc l m_l} \eq \ph^-_{n \vc l m_l}$,
then $\ph\artroOm$ becomes real):
\bal{\label{AdS_KG_solutions_202}
	\ph \artroOm & \eq \sumliml{n \vc l m_l}
 		\biiglrr{\ph^+_{n \vc l m_l}\;\mu\hbs+_{n\vc l m_l}\artroOm
			+ \coco{\ph^-_{n \vc l m_l}}\;\coco{\mu\hbs+_{n\vc l m_l}\artroOm}\,
			} \;.
	}
This determines the space $L\hAdS\lintval t12$ of KG solutions for the time-interval region $\regM\hAdS\lintval t12$. At the same time, this is also the space $L\hAdS_t$ of KG solutions near an equal-time hypersurface.
The space of solutions near the boundary of the time-interval region
then consists of two copies of $L\hAdS_t$ as in
$L\hAdS_{\partial\intval t12}=L\hAdS_{t_1}\oplus L\hAdS_{t_2}$.
The ordinary Jacobi modes are propagating modes, 
well-defined on the whole spacetime.
Since the Jacobi modes are special cases of the $S^a$-modes,
the space of KG solutions on time-interval regions
is contained in the space of solutions on a rod regions as a subspace.

We recall the symplectic structures
obtained in \cite{Doh:classads}
for the spaces $L\hAdS_t$ and $L\hAdS_{\rho}$:
\bal{\om_{t}(\et,\ze) 
	& \eq -\tfrac12 \int_{\Si_t}\!\!\volro \volOm[d\mn1]\,
		 R\lAdS^{d\mn1} \tan^{d\mn1}\!\ro \; 
		\biglrr{\et\,\del_t\ze - \ze\,\del_t\et}\ar{{t},\ro,\Om}
		\\
	\label{zzz_AdS_structures_100}
	& \eq \iu\RAdS^{d\mn1} \sumliml{n \vc l m_l}
			\omag+nl \Norm[+]{n l}\,
		 \biiiglrr{\coco{\et^-_{n \vc l m_l}}\,\ze^+_{n \vc l m_l}
			-\et^+_{n \vc l m_l}\, \coco{\ze^-_{n \vc l m_l}}}\; ,
		\\
	\label{zzz_AdS_structures_421_0}
	\om_{\ro}(\et,\ze) 
	& \eq \tfrac12 \int_{\Si_\ro}\!\! \dif\!t\, \volOm[d\mn1]\,
		R\lAdS^{d\mn1} \tan^{d\mn1}\!\ro\;\,
		\biglrr{\et\,\del_\ro\ze - \ze\,\del_\ro\et}\ar{t,\ro,\Om}
		\\
	\label{zzz_AdS_structures_421}
	& \eq \piu \RAdS^{d\mn1} 
		\int\!\!\dif\!\om \sumliml{\vc l, m_l} (2l\pn d\mn2)\,
		\biiglrr{\et^{S,a}_{\om \vc l m_l}\,\ze^{S,b}_{-\om,\vc l,-m_l}
				-\et^{S,b}_{\om \vc l m_l}\,\ze^{S,a}_{-\om,\vc l,-m_l}
				}.
	}
Therein, $\Norm[+]{nl}$ is a factor arising from the radial integration
of the Jacobi modes \eqref{AdS_KG_solutions_101} in $\om_t$:
\bal{\label{zzz_AdS_normalize_equal_time_20}
	\Norm[+]{nl} \defeq 
	\sma{\intlim 0{\piu/2}} \volro \tann{d\mn1}\ro\;
	\biglrrn{\Jacar+nl\ro}{2}
	\eq \fracws{n!\;\Gaar{\ga^{S,a}}^2\,\Gaar{n\pn\nu\pn1}}
	{2\omag+nl\; \Gaar{n\pn\ga^{S,a}}\; \Gaar{n\pn\nu\pn\ga^{S,a}}} \;.
	}
Note that despite the labels $t$ and $\ro$, both $\om_t$ and $\om_\ro$
are independent of the values of the time $t$ respectively the radius $\ro$.
The invariance under all isometry actions of AdS has been shown
for both symplectic structures in \cite{Doh:classads}.
The corresponding expressions for the field theory on
Minkowski spacetime are listed in Appendix~\ref{class_KG_AdS:_sols_mink}.

\section{Towards S-matrices on AdS via complex structures}
\label{sec:smcomplex}

\subsection{Isometries and complex structure}

As recalled in Section~\ref{sec:stdsmat}, a key ingredient in the quantization of a classical field theory is the complex structure \cite{ash_mag:_qf_curved_spacetimes}. It has to be compatible with the symplectic structure encoding classical dynamics, yielding the inner product (\ref{eq:ip}) on $L$, which in turn determines the Fock space of states. To make spacetime symmetries of the classical theory also symmetries of the quantum theory, the complex structure should be invariant under these. In Minkowski spacetime with its isometry group of Poincaré transformations, the complex structure on an equal-time hypersurface is essentially uniquely determined by these requirements. For spacelike hypersurfaces in more general curved spacetimes the situation is more complicated \cite{ash_mag:_qf_curved_spacetimes}.

In the standard approach, the space $L_t$ where the complex structure lives is thought of as a space of global solutions. It is then clear how isometries act and what isometry invariance of the complex structure means. In contrast, in the GBF that we make use of here, the complex structure is seen as intrinsically associated to the hypersurface and to solutions in its neighborhood. There is a straightforward action then only for isometries that map the hypersurface to itself. If we restrict to infinitesimal isometries, however, the fact that solutions are defined not only on the hypersurface itself, but on a neighborhood, is enough to make their actions well defined. (In this way the isometry invariance of symplectic structures on hypersurfaces on AdS was understood in \cite{Doh:classads}.)

The complex structure for KG theory on the hypercylinder in Minkowski spacetime has this same essential uniqueness property,
as exhibited implicitly in \cite{Oe:kgtl} and explicitly in \cite{Oe:holomorphic} for propagating solutions.
This gives additional motivation for pursuing the same isometry invariance criterion for selecting reasonable complex structures on AdS spacetime.

\subsection{Minkowski limit and amplitude equivalence}
\label{sec:mlampleq}

Since QFT in Minkowski spacetime is much better understood than in curved spacetime, we shall make use of the fact that Minkowski spacetime arises from AdS in a flat limit, in the sense of Section~\ref{class_KG_AdS:_geom}. Concretely, we shall require that 
the flat limit of the AdS amplitudes reproduces 
the respective Minkowski amplitudes.
From (\ref{eq:freeampl}) and (\ref{eq:dampl})
it is easy to see that this holds,
if for Klein-Gordon solutions on AdS the flat limit of the inner product $g$ is the same as the Minkowski inner product
of the solutions' flat limits. 
In turn, this holds if the limit of the AdS complex structure is the Minkowski complex structure for the relevant class of hypersurfaces,
as sketched in diagrams \eqref{diag_J_t_AdS_Mink} and \eqref{diag_J_ro_AdS_Mink}.
This turns out to make sense for both equal-time hypersurfaces and hypercylinders, yielding a limit both on rod regions and on time-interval regions, which we work out in Sections \ref{complex_struct_AdS_KG:_flatlim_rod} and \ref{complex_struct_AdS_KG:_flatlim_eqtime}
respectively.

\begin{figure}[H]
	\centering
	\igx[width=0.61\linewidth]{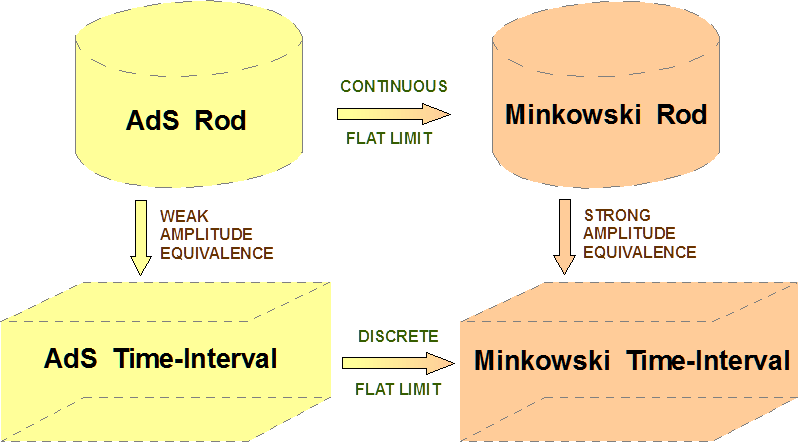}
	\caption{Relations of amplitudes for time-interval and rod regions on AdS and Minkowski spacetime.}
\label{AdS_Mink_Slice_Rod}
\end{figure}

As mentioned in Section~\ref{sec:gensm} the standard asymptotic time-interval geometry and the asymptotic rod geometry lead to equivalent amplitudes in Minkowski spacetime \cite{CoOe:smatrixgbf}. On the other hand, the corresponding geometries are inequivalent in AdS. In particular, the relevant asymptotic solution spaces are rather different. The space $L\hAdS_{\rho}$ of solutions on the hypercylinder is continuous, while the space $L\hAdS_{t_1}\oplus L\hAdS_{t_2}$ of solutions on the boundary of a time-interval is much smaller and discrete. In particular, there can be no amplitude equivalence. However, it turns out that there is a suitable embedding of the discrete space $L\hAdS_{t_1}\oplus L\hAdS_{t_2}$ into the continuous one $L\hAdS_{\rho}$ allowing for a weak form of amplitude equivalence, which we work out in Section \ref{complex_struct_AdS_KG:_ampli_equi}. 
What is more, this \emph{weak amplitude equivalence} can in the flat limit be brought into congruence with the (strong) amplitude equivalence in Minkowski spacetime.
The ensuing relations between amplitudes (and thus complex structures) are illustrated in Figure~\ref{AdS_Mink_Slice_Rod}.

\begin{figure}[H]
	\centering
	\igx[width=0.5\linewidth]{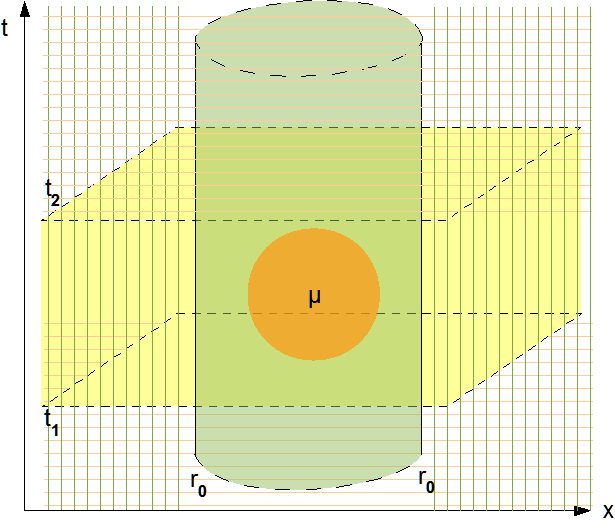}
	\caption{Time-interval and rod regions with source $\mu$.}
	\label{slice_rod_source}
\end{figure}

We proceed to explain in more detail how amplitude equivalence relates complex structures. Suppose we have two spacetime regions $M$, $N$ to describe the same physics in the interior (of both). (This is illustrated for a time-interval and a rod region in Figure~\ref{slice_rod_source}.) Comparing solutions outside the regions (or asymptotically) should yield an equivalence between the boundary solution spaces $L_{\partial M}$ and $L_{\partial N}$. The symplectic structures on the boundaries should be the same under this equivalence,
\begin{equation}
 \omega_{\partial M}(\phi_1,\phi_2)=\omega_{\partial N}(\phi_1,\phi_2) ,
\end{equation}
for ``outside'' solutions $\phi_1,\phi_2$. While this equality follows straightforwardly from Lagrangian field theory for compact regions, the case of non-compact regions is less trivial. To obtain equality of the inner products on $\partial M$ and $\partial N$ under the equivalence we require in addition,
\begin{equation}
 \omega_{\partial M}(\phi_1,J_{\partial M}\phi_2)=\omega_{\partial N}(\phi_1,J_{\partial N}\phi_2) .
 \label{eq:eqgip}
\end{equation}
This means that the complex structures on $L_{\partial M}$ and $L_{\partial N}$ need to be related by the same map between $L_{\partial M}$ and $L_{\partial N}$ that establishes the equivalence. A trouble with this setting is that the nature of the spaces $L_{\partial M}$ and $L_{\partial N}$ depends itself on the choice of complex structure. This is so because $L_{\partial M}$ and $L_{\partial N}$ are supposed to be Hilbert (or Krein) spaces with their inner products. Thus, the complex structure itself determines to some extent what is the nature of the elements in $L_{\partial M}$ and $L_{\partial N}$. Usually, these are some kind of $L^2$ spaces, i.e., the elements are equivalence classes of square integrable functions.

On the other hand, the spaces of free solutions in the interior of $M$ and of $N$, $L_{\tilde{M}}$ and $L_{\tilde{N}}$ should coincide by assumption. Moreover, these must give rise to Lagrangian subspaces of $L_{\partial M}$ and $L_{\partial N}$. We also recall the decomposition $L_{\partial M}=r(L_{\tilde{M}})\oplus J_{\partial M} r(L_{\tilde{M}})$ and the corresponding one for $N$. We thus see that it is sufficient to require the equality (\ref{eq:eqgip}) for elements of $L_{\tilde{M}}=L_{\tilde{N}}$.
Suppose in particular that the complex structure on $L_{\partial M}$ is given and we wish to construct an equivalent one on $L_{\partial N}$. Once we have chosen a complement of $r(L_{\tilde{N}})$ in $L_{\partial N}$ this equivalent complex structure on $L_{\partial N}$ is completely determined by equation (\ref{eq:eqgip}) on $L_{\tilde{M}}=L_{\tilde{N}}$.

\section{Complex structures for Klein-Gordon solutions on AdS}
\label{complex_struct_AdS_KG}

For the time-interval region, the amplitudes $\ro_t$
are determined by the real inner product $g_t$.
The boundary of this region are equal-time hypersurfaces,
and $g_t$ is determined by the complex structure $J_t$
associated to these hypersurfaces. 
Here, there is a long-known standard choice for $J_t$
and therefore we only need to check that it commutes with the
isometries and has the correct flat limit.

For the rod region, the amplitudes $\ro_\ro$
are determined by the real inner product $g_\ro$.
The boundary of this region is a hypercylinder,
and $g_\ro$ is determined by the complex structure $J_\ro$
associated to this hypersurface.
Since here there is no standard choice,
we need to construct it.
With this goal we implement the requirements of 
Section \ref{sec:smcomplex} in the following sequence.
First, we impose commutation of $J_\ro$ with the isometries' actions,
because this already fixes the form of $J_\ro$ to a great degree.
Using this preliminary form we implement 
a weak version of amplitude equivalence,
because this completely fixes the action of $J_\ro$
on the Jacobi modes (the modes with magic frequencies $\omag+nl$).
Then, we present two choices how to extend the action of $J_\ro$
to all real frequencies $\om$ (that is, two ways of
completely fixing $J_\ro$) and study their properties.

\subsection{Complex structure for time-interval regions}
\label{complex_struct_AdS_KG:_time_interval}

We start by considering the complex structure $J_t$ for equal-time hyperplanes. In terms of the parametrization (\ref{AdS_KG_solutions_202}) of solutions this takes the form:
\bal{\label{complex_struct_AdS_KG:_ess_iso_J_t}
	\biglrr{J_t\, \ph}^\pm_{n\vcl m_l}
	\eq -\iu\,\ph^\pm_{n\vcl m_l} \;.
	}
This is in complete analogy to the standard case of equal-time hyperplanes in Minkowski spacetime.
It can be seen as an instance of the standard
complex structure $J_t \eq \mc L_t/\ruut{-\mc L^2_t}$
of Ashtekar and Magnon \cite{ash_mag:_qf_curved_spacetimes}
for stationary spacetimes
($\mc L_t$ is the Lie derivative in direction
 of the unit normal to $\Si_t$).
This operator is defined by its eigenvalues $\mp\iu$
when acting on a mode with frequency $\om \gtrless 0$,
inducing thereby the above action in the momentum representation.
Using the Jacobi expansion \eqref{AdS_KG_solutions_202}
and symplectic structure $\om_t$ of \eqref{zzz_AdS_structures_100},
it is straightforward to verify that this $J_t$ fulfills 
the essential properties of $J^2=-\One$, compatibility
with $\om_t$, and maps real solutions to real solutions.
Using \eqref{zzz_AdS_isometries_time_81},
\eqref{zzz_AdS_isometries_symplectic_rot_100}
and \eqref{zzz_AdS_isometries_boosts_171},
it is also straightforward to check that this $J_t$ indeed
commutes with with all isometries of AdS.
The real inner product $g_t$ for the equal-time hypersurface 
induced by this $J_t$ through 
$g_{t}(\et,\ze) \eq 2\om_{t}(\et,J_t\ze)$ is given by
\bal{\label{AdS_structures_180_n_nu}
	g_{t}(\et,\ze)
	& \eq \RAdS^{d\mn1}\,\sumliml{n \vc l m_l}
		\biiiglrc{\coco{\et^-_{n \vc l m_l}}\,\ze^+_{n \vc l m_l}
					+\et^+_{n \vc l m_l}\, \coco{\ze^-_{n \vc l m_l}}
					}		
		\fracw{n!\;\Gaar{\ga^{S,a}}^2\,\Gaar{n\pn\nu\pn1}}
				{\Gaar{n\pn\ga^{S,a}}\; \Gaar{n\pn\nu\pn\ga^{S,a}}}
		\\
	\label{AdS_structures_180_al_be}
	& \eq \RAdS^{d\mn1}\,\sumliml{n \vc l m_l}
		\biiiglrc{\coco{\et^-_{n \vc l m_l}}\,\ze^+_{n \vc l m_l}
					+\et^+_{n \vc l m_l}\, \coco{\ze^-_{n \vc l m_l}}
					}\,
		\Gaar{\ga^{S,a}}^2
		\fracw{\Gaar{1\mn\al^{S,b}}\;\Gaar{\be^{S,b}}}
				{\Gaar{1\mn\al^{S,a}}\; \Gaar{\be^{S,a}}}.
	}
In the last line, the parameters $\al^{S,\cdot},\be^{S,\cdot}$
and $\ga^{S,a}$ are understood as evaluated at the respective values
of $l$ and at positive magic frequencies $\om \eq +\omag+nl$.
The last line results from the first by plugging in relations
\eqref{kg_ads_solutions_74_parameters_SC_ab}.
For a real solution $\ph$
with $\ph^-_{n \vc l m_l} \eq \ph^+_{n \vc l m_l}$
we obtain from \eqref{AdS_structures_180_n_nu}
using $\Norm[+]{n l}$ from \eqref{zzz_AdS_normalize_equal_time_20}:
\bal{g_{t}(\ph,\ph) 
	& \eq \sumliml{n \vc l m_l}
			4 \omag+nl \RAdS^{d\mn1} \Norm[+]{n l}
		 \absq{\ph^+_{n \vc l m_l}},
	}
which is positive for all Jacobi modes
(since $\Norm[+]{n l}$ is always positive).
The same happens of course for the real inner product $g_t$ on a Minkowski
equal-time hyperplane. $g_t$ is positive definite for AdS as for Minkowski spacetime.

\subsubsection{Flat limits of time-interval amplitudes}
\label{complex_struct_AdS_KG:_flatlim_eqtime}

A good choice of the complex structure
$J_t$ for AdS should induce a real inner product $g_t$
whose flat limit recovers the Minkowskian $g_t$,
thereby letting the amplitude \eqref{eq:freeampl}
for the AdS time-interval
reproduce the Minkowskian time-interval amplitude in the flat limit.
Our goal is thus to recover in the flat limit 
the $g_t$ for two solutions near an equal-time plane
in Minkowski spacetime given in
\eqref{zzz_AdS_invar_complex_isometries_2642_Mink_t_txg}.
Since near an equal-time plane on AdS
only the Jacobi modes are well-defined
(the modes with magic frequencies $\omag+nl$),
our limit is discrete: it maps the discrete magic frequencies
of the AdS theory to discrete frequencies of the Minkowski theory.
For $\RAdS\to\infty$ the Minkowski frequencies become dense.
We start with the real inner product 
\eqref{AdS_structures_180_n_nu} for AdS,
and inserting the flat Jacobi representation
\eqref{class_flat_eqtim_4734},
with \eqref{zzz_AdS_normalize_equal_time_20} we obtain
\bals{g_{t}(\et,\ze) 
	& \eq \sumliml{n \vc l m_l}
			\biiiglrr{\coco{\et^{\txF,-}_{\tip \vc l m_l}}\,
									\ze^{\txF,+}_{\tip \vc l m_l}
							+\et^{\txF,+}_{\tip \vc l m_l}\,
							\coco{\ze^{\txF,-}_{\tip \vc l m_l}}
							}
			\RAdS^{d\mn3}
			\fracw{n!\;\Gaar{\ga^{S,a}}^2\,\Gaar{n\pn\nu\pn1}}
					{\Gaar{n\pn\ga^{S,a}}\; \Gaar{n\pn\nu\pn\ga^{S,a}}}
			\fracw{(p\hreals_{nl})^{2l}\;8(\omag+nl)^2}
			{\piu\,((2l\pn d\mn2)!!)^2}\,.
	}
Setting $d=3$, and using that for odd $k$ we have
$k!! \eq \Gaar{\tfrac k2\pn1}2^{\frac{k\pn1}{2}}/\ruut\piu$, we get
\bals{g_{t}(\et,\ze) 
	& \eq \sumliml{n l m_l}
			\biiiglrr{\coco{\et^{\txF,-}_{\tip l m_l}}\,
									\ze^{\txF,+}_{\tip l m_l}
							+\et^{\txF,+}_{\tip l m_l}\,
							\coco{\ze^{\txF,-}_{\tip l m_l}}
							}
			\fracw{\Gaar{n\pn1}\;\Gaar{n\pn\nu\pn1}}
					{\Gaar{n\pn l\pn\tfrac32}\;
					 \Gaar{n\pn\nu\pn l\pn\tfrac32}
					}
			\fracw{(p\hreals_{nl})^{2l}\;8(\omag+nl)^2}
			{2^{2l\pn2}}\,.
	}
In the next lines we perform the flat limit, which involves
$\nu\to mR$ (wherein, $R$ is short for $\RAdS$)
and making the sum into an integral.
To this end, from $\omag+nl\defeq \timp+2n+l$
we also substitute $n\to\tfrac12(\omag+nl-\timp-l)$.
Since $\omag+nl\geq\timp$ and thus $\tiom>m$, for large $R$ 
here we have no negative signs in the Gamma's arguments.
\bals{g_{\Si_t}(\et,\ze) 
	& \eq \sumliml{l m_l}\sumlim{\omag+nl}{\De\omag+nl=2}
			\biiiglrr{\coco{\tiet^{\txF,-}_{\tip l m_l}}\,
									\tize^{\txF,+}_{\tip l m_l}
							+\tiet^{\txF,+}_{\tip l m_l}\,
							\coco{\tize^{\txF,-}_{\tip l m_l}}
							}
			\fracw{(p\hreals_{nl})^{2l}\;8(\omag+nl)^2}{R^2\,2^{2l\pn2}}
			\fracw{\Gaar{\tfrac12(\omag+nl\mn\nu\mn l\pn\frac12)}}
					{\Gaar{\tfrac12(\omag+nl\mn\nu\pn l\pn\frac32)}}
			\frac{\Gaar{\tfrac12(\omag+nl\pn\nu\mn l\pn\frac12)}}
					{\Gaar{\tfrac12(\omag+nl\pn\nu\pn l\pn\frac32)}}
		\\
	\toflatw & \sumliml{l m_l}\intlim0\infty\!\! \dif\!\tip\;
			\fracw R2 \frac{\tip}{\tiom_\tip}
			\biiiglrr{\coco{\tiet^{\txF,-}_{\tip l m_l}}\,
									\tize^{\txF,+}_{\tip l m_l}
							+\tiet^{\txF,+}_{\tip l m_l}\,
							\coco{\tize^{\txF,-}_{\tip l m_l}}
							}
			\fracw{p^{2l}\;8(\om_p)^2}{R^2\,2^{2l\pn2}}
			\fracw{\Gaar{\tfrac12(R(\tiom\mn m)\mn l\pn\frac12)}}
					{\Gaar{\tfrac12(R(\tiom\mn m)\pn l\pn\frac32)}}
			\frac{\Gaar{\tfrac12(R(\tiom\pn m)\mn l\pn\frac12)}}
					{\Gaar{\tfrac12(R(\tiom\pn m)\pn l\pn\frac32)}}
		\\
	& \approx \sumliml{l m_l}\intlim0\infty\!\! \dif\!\tip\;
			\biiiglrr{\coco{\tiet^{\txF,-}_{\tip l m_l}}\,
									\tize^{\txF,+}_{\tip l m_l}
							+\tiet^{\txF,+}_{\tip l m_l}\,
							\coco{\tize^{\txF,-}_{\tip l m_l}}
							}
			\frac{\tip}{\tiom_\tip}
			\fracw{p^{2l}\;4(\om_p)^2}{R\,2^{2l\pn2}}
			\biiglrr{\tfrac12 R(\tiom\mn m)}^{-l-\frac12}
			\biiglrr{\tfrac12 R(\tiom\pn m)}^{-l-\frac12}
	}
Now we can simplify
$\biglrr{\tfrac12 R(\tiom\mn m)}^{-l-\frac12}
\biglrr{\tfrac12 R(\tiom\pn m)}^{-l-\frac12}
\eq \biglrr{\tfrac14 R^2(\tiom^2\mn m^2)}^{-l-\frac12}
\eq 2^{2l\pn1}\, p^{\mn2l\mn1}$,
giving us 
\bals{g_{\Si_t}(\et,\ze) \toflatw 
	\sumliml{l m_l}\intlim0\infty\!\! \dif\!\tip\;2\tiom_\tip\;
			\biiiglrr{\coco{\tiet^{\txF,-}_{\tip l m_l}}\,
									\tize^{\txF,+}_{\tip l m_l}
							+\tiet^{\txF,+}_{\tip l m_l}\,
							\coco{\tize^{\txF,-}_{\tip l m_l}}
							}\;.
	}
This is precisely the real inner product
\eqref{zzz_AdS_invar_complex_isometries_2642_Mink_t_txg}
for two solutions near a Minkowski equal-time hyperplane.
Further, using the flat Jacobi expansion
it is easy to verify that for equal-time hypersurfaces
the complex structure and the flat limit ''commute''
because for both Minkowski and AdS we use a parametrization making the
complex structure take the form $(J_t\ph)^\pm \eq -\iu\ph^\pm$. That is, 
the standard complex structures make the below diagram commutative.
\bal{\label{diag_J_t_AdS_Mink}
	\bdia{\node{\ph\htx{AdS}}
				\arrow[2]{e,t}{J\htx{AdS}_t}
				\arrow[1]{s,l}{\text{disc. flat lim.}}
				\node{}
				\node{J\htx{AdS}_t \ph\htx{AdS}}
				\arrow[1]{s,r}{\text{disc. flat lim.}}
					\\
				\node{\ph\htx{Mink}}
				\arrow[2]{e,t}{J\htx{Mink}_t}
				\node{}
				\node{J\htx{Mink}_t \ph\htx{Mink}}
				}
		}

\subsection{Constructing $J_\ro$: essential properties and isometry-invariance}
\label{complex_struct_AdS_KG:_essential_isometry}

Since the most general form of the complex structure $J_\ro$
is rather unwieldy, we accommodate it in Appendix~\ref{zzz_complex_struct_AdS_KG:_essential_isometry}
and constrain it considerably therein through the following steps.
First, we require $J_\ro$ to fulfill the essential properties:
$J_\ro^2 = -\One$, compatibility 
$\om_\ro(J_\ro\,\cdot,\, J_\ro\,\cdot) = \om_\ro(\cdot,\cdot)$
with the symplectic structure
$\om_\ro$ of \eqref{zzz_AdS_structures_421},
and mapping real solutions $\ph$ to real solutions $J_\ro\ph$.
Second, we require $\txJ_\ro$ to commute with those AdS isometries,
that map the hypercylinder to itself, that is:
time translations and spatial rotations.
In \eqref{zzz_AdS_invar_complex_isometries_Jgen_X13}
these steps result in the following simpler form
for the complex structure of an AdS hypercylinder:
\bal{\label{AdS_invar_complex_isometries_Jgen_X13}
	\biglrr{J_\ro\ph}^{S,a}_{\om \vc l m_l}
	& \eqn j^{S,aa}_{\om l}\,\ph^{S,a}_{\om \vc l m_l}
			\pn j^{S,ab}_{\om l}\,\ph^{S,b}_{\om \vc l m_l}
		&	
	\biglrr{J_\ro\ph}^{S,b}_{\om \vc l m_l}
	& \eqn j^{S,ba}_{\om l}\,\ph^{S,a}_{\om \vc l m_l}
			\mn j^{S,aa}_{\om l}\,\ph^{S,b}_{\om \vc l m_l}
		&
	\biglrr{j^{S,aa}_{\om l}}^2
	& \eqn - j^{S,ab}_{\om l} j^{S,ba}_{\om l} \mn 1.
	}
That is, the action of $J_\ro$ is determined
by the choice of factors $j^{S,\cdot\cdot}_{\om l}$.
By plugging this into the symplectic structure
\eqref{zzz_AdS_structures_421}, we can read off
that the induced real inner product 
$g_\ro(\cdot,\cdot) = 2\om_\ro(\cdot,\,J_\ro\,\cdot)$
becomes positive definite for modes $\mu^{S,a/b}_{\om \vc l m_l}$
whenever we have $j^{S,ab}_{\om l} < 0$
with $j^{S,ba}_{\om l} > 0$. So far, 
nothing forces the diagonal element $j^{S,aa}_{\om l}$ to vanish.
The form \eqref{AdS_invar_complex_isometries_Jgen_X13}
already lets $J_\ro$ commute with time-translations and spatial
rotations. Requiring in addition commutation
with the boost actions \eqref{zzz_AdS_isometries_boosts_5201},
amounts to imposing the additional conditions
\eqref{zzz_AdS_invar_complex_isometries_681_m}
\bal{\label{AdS_invar_complex_isometries_681_m}
	j^{S,ba}_{\om\mn1,l\pn1}
	& \eqn -j^{S,ba}_{\om l}\,
		\fracwss{(2l\pn d)\,(2l\pn d\mn2)}
				{(\Timp \pn\om \mn l \mn d)\,(\Timp \mn\om \pn l)}
		&
	j^{S,ba}_{\om\pn1,l\pn1}
	& \eqn - j^{S,ba}_{\om l}\,
		\fracwss{(2l\pn d)\,(2l\pn d\mn2)}
				{(\Timp \mn\om \mn l \mn d)\,(\Timp \pn\om \pn l)},
	}
and the similar conditions \eqref{zzz_AdS_invar_complex_isometries_443}
for $j^{S,ab}_{\om l}$.

Thus, after imposing the essential properties
and commuting with time-translations and rotations,
for the space of solutions near an AdS hypercylinder
we have a complex structure $J_\ro$
of the form \eqref{AdS_invar_complex_isometries_Jgen_X13}.
Therein, for commuting also with the boosts
the factor $j^{S,ba}$ must obey
\eqref{AdS_invar_complex_isometries_681_m} and $j^{S,ab}$ obey
similar conditions obtained therefrom by setting 
$j^{S,ab} = -1/j^{S,ba}$.

\subsection{Implementing weak amplitude equivalence}
\label{complex_struct_AdS_KG:_ampli_equi}

We proceed to implement the weak form of the amplitude equivalence
between the complex structures $J_\ro$ and  $J_t$
discussed in Section \ref{sec:mlampleq}. To this end we have to restrict the hypercylinder solutions to the discrete set of solutions with magic frequencies as only for those a comparison makes sense. (Recall that this is where the adjective ``weak'' comes from.) To implement this in the continuous integral expansion \eqref{kg_ads_solutions_520} we use delta-functions,
\bal{\label{zzz_AdS_invar_complex_isometries_1538_disc_S_rep_a}
	\xi^{S,a}_{\om \vc l m_l}
	& \eq \sumlim{n=0}{\infty}
		\biiglrr{\xi^{a}_{+n, \vc l, m_l}\de(\om\mn\omag+nl)
					+ \xi^{a}_{-n, \vc l, m_l}\de(\om\pn\omag+nl)
				}
		\\
	\label{zzz_AdS_invar_complex_isometries_1538_disc_S_rep_b}
	\xi^{S,b}_{\om \vc l m_l}
	& \eq \sumlim{n=0}{\infty}
		\biiglrr{\xi^{b}_{+n, \vc l, m_l}\de(\om\mn\omag+nl)
					+ \xi^{b}_{-n, \vc l, m_l}\de(\om\pn\omag+nl)
				}.
	}
With this the $S$-expansion
\eqref{kg_ads_solutions_520}
of a solution consisting of $a$ and $b$-modes
of only magic frequencies becomes
\bal{\label{zzz_AdS_invar_complex_isometries_1558_disc_S-exp}
		\xi \artrOm  & \eq \sumliml{\vc l,m_l}\sumlim{n=0}{\infty}
		\biiglrc{\xi^{a}_{+n, \vc l, m_l}\;
			\mu\hb{S,a}_{\omag+nl \vc l m_l}\artroOm
			+\xi^{a}_{-n, \vc l, m_l}\;
			\mu\hb{S,a}_{-\omag+nl, \vc l, m_l}\artroOm
			\\
	& \qquad\qquad\qquad
			+ \xi^{b}_{+n, \vc l, m_l}\;
			\mu\hb{S,b}_{\omag+nl \vc l m_l}\artroOm
			+\xi^{b}_{-n, \vc l, m_l}\;
			\mu\hb{S,b}_{-\omag+nl, \vc l, m_l}\artroOm
			}.
		\notag
	}
Let us call this the discrete $S$-expansion.
For global solutions we have $\xi^{b}_{\pm n, \vc l, m_l} \equiv 0$,
then the Jacobi and the discrete $S$-expansion are equivalent
and we can translate them into each other:
\bal{\label{zzz_AdS_invar_complex_isometries_1527}
	\xi^a_{+n, \vc l, m_l} 
		& \eq \xi^+_{n \vc l m_l}
		&
	\xi^a_{-n, \vc l, \mn m_l} 
		& \eq \coco{\xi^-_{n \vc l m_l}}.
	}

Applying the equivalence equation \eqref{eq:eqgip} 
for a time-interval and a rod region, we find
that amplitude equivalence holds iff $g_t(\et,\ze) 
= \tfrac12\, g_\ro(\et,\ze)$,
for all \emph{global solutions} $\et$ and $\ze$ in AdS.
To compute
$g_\ro(\et,\ze) = 2 \om_\ro(\et,\,J_\ro \ze)$ we let in the symplectic structure
\eqref{zzz_AdS_structures_421} act on $\ze$ a complex structure
as in \eqref{zzz_AdS_invar_complex_isometries_Jgen_X13}
wherein $(j^{S,aa}_{\om l})^2 \eq 
- j^{S,ab}_{\om l}\,j^{S,ba}_{\om l} -1 \geq 0$:
\bals{\biglrr{J_\ro\ph}^{S,a}_{\om \vc l m_l}
	& \eq j^{S,aa}_{\om l}\,\ph^{S,a}_{\om \vc l m_l}
			+ j^{S,ab}_{\om l}\,\ph^{S,b}_{\om \vc l m_l}
		&
	j^{S,ab}_{\mn\om,l} 
	& \eq j^{S,ab}_{\om l}
		\\
	\biglrr{J_\ro\ph}^{S,b}_{\om \vc l m_l}
	& \eq j^{S,ba}_{\om l}\,\ph^{S,a}_{\om \vc l m_l}
			- j^{S,aa}_{\om l}\,\ph^{S,b}_{\om \vc l m_l}
		&
	j^{S,ba}_{\mn\om,l} 
	& \eq j^{S,ba}_{\om l}.
	}
Using the discrete $S$-expansion
\eqref{zzz_AdS_invar_complex_isometries_1558_disc_S-exp},
for global solutions $\et,\ze$ we then obtain
(going from second to third line we use the frequency symmetry
$j^{S,ba}_{\omag+nl l} \eq j^{S,ba}_{-\omag+nl l}$, and in the last
line the factor $\tfrac12$ comes from the scaling property
of the Dirac delta:
$\de(\omag+nl-\omag+{n',}l) = \de(2n-2n')=\tfrac12\,\de(n-n')$):
\bals{\tfracw12 g_{\ro}(\et,\ze) 
	& \eq \RAdS^{d\mn1} \sumliml{n,\vc l,m_l}\!\piu(2l\pn d\mn2)\;
		\biiiglrc{\et^{a}_{+n, \vc l, m_l}
					\biglrr{J_\ro\ze}^{S,b}_{-\omag+nl,\vc l,-m_l}
				+ \et^{a}_{-n, \vc l,-m_l}
					\biglrr{J_\ro\ze}^{S,b}_{\omag+nl,\vc l,m_l}
		\\
	& \qquad\qquad\qquad\qquad\qquad\quad
				-\unb{\et^{S,b}_{\omag+nl, \vc l, m_l}}_0
					\biglrr{J_\ro\ze}^{a}_{-n,\vc l,-m_l}
				-\unb{\et^{S,b}_{-\omag+nl, \vc l,-m_l}}_0
					\biglrr{J_\ro\ze}^{a}_{+n,\vc l,m_l}
				}
		\\ 
	& \eq \RAdS^{d\mn1} \sumliml{n,\vc l,m_l}\!\piu(2l\pn d\mn2)\;
		\biiiglrc{\et^{a}_{+n, \vc l, m_l}
					\biiglrr{j^{S,ba}_{\omag+nl l}
						\ze^{S,a}_{-\omag+nl,\vc l,-m_l}
						- j^{S,aa}_{\omag+nl l}
						\unb{\ze^{S,b}_{-\omag+nl,\vc l,-m_l}}_0
						}
		\\
	& \qquad\qquad\qquad\qquad\qquad\quad
					+ \et^{a}_{-n, \vc l, -m_l}
					\biiglrr{j^{S,ba}_{\omag+nl l}
						\ze^{S,a}_{\omag+nl,\vc l,m_l}
						- j^{S,aa}_{\omag+nl l}
						\unb{\ze^{S,b}_{\omag+nl,\vc l,m_l}}_0
						}
				}
		\\ 
	& \eq \RAdS^{d\mn1} \sumliml{n,\vc l,m_l}\!
		\piu(2l\pn d\mn2) j^{S,ba}_{\omag+nl l}\;
		\biiiglrc{\et^{a}_{+n, \vc l, m_l}
						\ze^{S,a}_{-\omag+nl,\vc l,-m_l}
						+ \et^{a}_{-n, \vc l, -m_l}
						\ze^{S,a}_{\omag+nl,\vc l,m_l}
				}
		\\ 
	& \eq \RAdS^{d\mn1} \sumliml{n,\vc l,m_l}\!
		\tfracw\piu2 (2l\pn d\mn2)\, \de(0)\, j^{S,ba}_{\omag+nl l}\;
		\biiiglrc{\et^{a}_{+n, \vc l, m_l}
						\ze^{a}_{-n,\vc l,-m_l}
						+ \et^{a}_{-n, \vc l, -m_l}
						\ze^{a}_{+n,\vc l,m_l}
				}
		.
	}
That is, if our solutions consist only of magic frequency modes,
then the inner product $g_\ro$ on the hypercylinder 
has a $\de$-divergence.
The reason for this is the following: in the definition 
\eqref{zzz_AdS_structures_421_0} of the symplectic structure $\om_\ro$,
and thus also for the inner product $g_\ro$,
we integrate $\et\,\del_\ro\ze - \ze\,\del_\ro\et$
over the hypercylinder, which is infinite in $t$-direction.
However, using only modes with the discrete magic frequencies
we cannot form wave packets that are compactly supported
(resp.~decay sufficiently fast) on the hypercylinder
(analogous to discrete Fourier transformation).
Hence the integral diverges.
On the other hand, we could have easily avoided this factor by not using the
Dirac delta (distribution) but rather its square root (distribution)
in \eqref{zzz_AdS_invar_complex_isometries_1538_disc_S_rep_a}. That is, the issue has to do with how exactly we bring into correspondence the modes on the hypercylinder with the modes on the equal-time hyperplane. The freedom is justified since we are only after a \emph{weak} amplitude equivalence and we shall thus remove the $\de(0)$-factor from here onwards.
We proceed by plugging \eqref{zzz_AdS_invar_complex_isometries_1527}
into the last line and obtain for global solutions $\et,\ze$
\bals{\tfracw12 g_{\ro}(\et,\ze) 
	& \eq \RAdS^{d\mn1} \sumliml{n,\vc l,m_l}\!
		\piu(\ga^{S,a}\mn1)\,  j^{S,ba}_{\omag+nl l}\;
		\biiiglrc{\et^+_{n \vc l m_l}
						\coco{\ze^-_{n \vc l m_l}}
						+ \coco{\et^-_{n \vc l m_l}}
						\ze^+_{n \vc l m_l}
				}.
	}
We can read off that this agrees with
\eqref{AdS_structures_180_n_nu} precisely if
\bal{\label{zzz_AdS_invar_complex_isometries_1691_j_cond}
	j^{S,ba}_{\omag+nl l} \eq j^{S,ba}_{-\omag+nl l}
	& \eq \fracw1{\piu(\ga^{S,a}\mn1) }
			\fracw{n!\;\Gaar{\ga^{S,a}}^2\,\Gaar{n\pn\nu\pn1}}
					{\Gaar{n\pn\ga^{S,a}}\; \Gaar{n\pn\nu\pn\ga^{S,a}}}
		\\
	& \eq \tfracw1{\piu }\Gaar{\ga^{S,a}}\Gaar{\ga^{S,a}\mn1}
		\fracw{\Gaar{1\mn\al^{S,b}}\;\Gaar{\be^{S,b}}}
				{\Gaar{1\mn\al^{S,a}}\; \Gaar{\be^{S,a}}}.
	}
(Again, the parameters $\al^{S,\cdot},\be^{S,\cdot}$
and $\ga^{S,a}$ are understood as evaluated at the respective values
of $l$ and at positive magic frequencies $\om \eq +\omag+nl$.)
We observe that \eqref{zzz_AdS_invar_complex_isometries_1691_j_cond}
fulfills conditions \eqref{zzz_AdS_invar_complex_isometries_681_m}
for commuting with the boosts.
Because of $\et^{S,b}_{\om \vc l m_l} \equiv 0$,
amplitude equivalence fixes only the factor
$j^{S,ba}$ of the complex structure
while not fixing $j^{S,aa}$ and $j^{S,ab}$
(the same happens for Minkowski spacetime).
Therefore, (as for Minkowski spacetime) we choose an anti-diagonal
complex structure: $j^{S,aa}_{\om l} \equiv 0$,
which induces $j^{S,ab} = -1/j^{S,ba}$.
This makes $J_\ro$ map $a$-modes to $b$-modes and vice versa.
This is the most direct implementation of the property that $J_\ro$
maps solutions well-defined on the whole rod region
to solutions that are well-defined only near the boundary hypercylinder.

Further, amplitude equivalence fixes $j^{S,ba}$
only for the discrete set of magic frequencies $\pm \omag+nl$,
it remains to fix $j^{S,ba}$ also for non-magic frequencies.
Let us recall the properties $j^{S,ba}$ still has to fulfill.
The first is frequency symmetry
$j^{S,ba}_{\om, l} \eq j^{S,ba}_{-\om, l}$.
The second is for $J_\ro$ to commute with the boost generators
also for non-magic frequencies, which amounts to
\eqref{zzz_AdS_invar_complex_isometries_681_m}.
We observe that this condition only relates factors $j^{S,ba}$
with a discrete difference in frequency $\om$ and
angular momentum $l$. 
This means that after fixing $j^{S,ba}$ for one frequency $\om_0$
and angular momentum $l_0$,
this condition then does not fix $j^{S,ba}$
for all other $\om$ and $l$, but only for those
that are at discrete steps from $\om_0$ and $l_0$.
The third requirement is that the flat limit of
$g_\ro$ reproduces the Minkowski real inner product $g_r$,
which is treated in Section \ref{complex_struct_AdS_KG:_flatlim_rod}.
We now define two candidate versions for $j^{S,ba}_{\om l}$
which both commute with boosts
\eqref{zzz_AdS_invar_complex_isometries_681_m},
called $\al$-version and $\be$-version:
\bal{\label{zzz_AdS_invar_complex_isometries_1783_j_al}
	j^{(\al)}_{\om l}
	& \eqn \fracw{\Gaar{\ga^{S,a}}\,\Gaar{\ga^{S,a}\mn1}}{\piu}
		\fracw{\Gaar{\al^{S,b}}}{\Gaar{\al^{S,a}}}
		\frac{\Gaar{1\mn\be^{S,a}}}{\Gaar{1\mn\be^{S,b}}
		}
		&
	j^{(\be)}_{\om l} 
	& \eqn \fracw{\Gaar{\ga^{S,a}}\,\Gaar{\ga^{S,a}\mn1}}{\piu}
		\fracw{\Gaar{\be^{S,b}}}{\Gaar{\be^{S,a}}}
		\frac{\Gaar{1\mn\al^{S,a}}}{\Gaar{1\mn\al^{S,b}}}.
	}
We recall that in each line the parameters $\al^{S,\cdotw}$,
$\be^{S,\cdotw}$ and $\ga^{S,a}$ are calculated from $\om$ and $l$.
Switching the sign of $\om$ corresponds to
interchanging $\al$ and $\be$-parameters, and thus
$j^{(\al)}_{\om, l}\eq j^{(\be)}_{-\om, l}$.
The only possibility for $j^{S,ba}$ to fulfill
\eqref{zzz_AdS_invar_complex_isometries_1691_j_cond}
with frequency symmetry is setting
\bal{\label{zzz_AdS_invar_complex_isometries_1784_magic_+}
	j^{S,ba}_{\omag+nl, l} 
	\eq j^{(\be)}_{\omag+nl, l}
	& \eq \tfracw1{\piu} \Gaar{\ga^{S,a}}\,\Gaar{\ga^{S,a}\mn1}
		\fracw{\Gaar{\be^{S,b}}}{\Gaar{\be^{S,a}}}
		\fracw{\Gaar{1\mn\al^{S,a}}}{\Gaar{1\mn\al^{S,b}}}
		\\
	\label{zzz_AdS_invar_complex_isometries_1784_magic_-}
	j^{S,ba}_{-\omag+nl, l} 
	\eq j^{(\al)}_{-\omag+nl, l}
	& \eq \tfracw1{\piu} \Gaar{\ga^{S,a}}\,\Gaar{\ga^{S,a}\mn1}
		\fracw{\Gaar{\al^{S,b}}}{\Gaar{\al^{S,a}}}
		\fracw{\Gaar{1\mn\be^{S,a}}}{\Gaar{1\mn\be^{S,b}}}.
	}
We recall that in \eqref{zzz_AdS_invar_complex_isometries_1784_magic_+}
the parameters $\al,\be$
are calculated from $+\omag+nl$, while in
\eqref{zzz_AdS_invar_complex_isometries_1784_magic_-}
they derive from $-\omag+nl$.
It is actually quite nontrivial that there exists a choice
for $J_\ro$ which induces amplitude equivalence,
because the factors related here have rather different origins:
The factor appearing in amplitude equivalence condition
\eqref{zzz_AdS_invar_complex_isometries_1691_j_cond}
stems from integrating a global solution over an equal-time
hyperplane $\Si_t$, while the factors in the boost conditions
\eqref{zzz_AdS_invar_complex_isometries_681_m}
stem from boost compatibility of the complex structure $J_\ro$
for more general solutions near a hypercylinder $\Si_\ro$.

\subsection{Two-branches choice $J\htx{two}_\ro$}
\label{complex_struct_AdS_KG:_two_branched}

We can define a first version for $j^{S,ba}$,
which we call simply $j\htx{two}$, by using
\eqref{zzz_AdS_invar_complex_isometries_1784_magic_+}
for all positive frequencies,
while using \eqref{zzz_AdS_invar_complex_isometries_1784_magic_-}
for all negative ones.
That is, we use the $\be$-version for
the ''positive branch'' $\om>0$ of the frequency axis
and the $\al$-version for the ''negative branch'' $\om<0$.
This results in a frequency symmetric expression:
\bals{j\htx{two}_{\om l} 
	& \eq \fracw{\Gaar{\tfrac12(2l\pn d)}\,	\Gaar{\tfrac12(2l\pn d\mn2)}}
					{\piu}
		\fracw{\Gaar{\tfrac12(\timp\pn\abs{\om}\mn l\mn d\pn2)}}
				{\Gaar{\tfrac12(\timp\pn\abs{\om}\pn l)}}
		\fracw{\Gaar{-\tfrac12(\timp\mn\abs{\om}\pn l\mn2)}}
				{\Gaar{-\tfrac12(\timp\mn\abs{\om}\mn l\mn d)}}.
	}
The choice $j\htx{two}$ commutes with time-translations
and rotations (that is: with isometries mapping the
hypercylinder to itself) and induces amplitude equivalence.
Alas, this choice violates the conditions
\eqref{zzz_AdS_invar_complex_isometries_681_m}
for commuting with the boosts.
This breaking occurs for the boosts generators
only for frequencies with $\abs\om < 1$,
because for these frequencies the action of boosts generators
creates frequencies $\om\pm1$ which cross the gluing point $\om=0$.
However, if we proceed to finite boosts by taking the exponential,
then there appear not only frequency shifts of $\pm1$,
but of any integer. Then, commutation with boosts 
becomes lost for the \emph{whole} frequency range.
Moreover, this version of $j\htx{two}_{\om l}$ vanishes at 
some frequencies and becomes singular for others.
Therefore, in the next subsection we also construct a second choice
for the complex structure $J_\ro$.

Having above considered $j^{S,ba}_{\om l}$,
we now deal with fixing $j^{S,ab}_{\om l}$,
which in turn fixes $j^{S,aa}_{\om l}$
through $\biglrr{j^{S,aa}_{\om l}}^2 
\eq - j^{S,ab}_{\om l} j^{S,ba}_{\om l} \mn 1$.
As discussed in Section \ref{complex_struct_AdS_KG:_flatlim_rod},
requiring the flat limit of the real inner product $g_\ro$ of AdS
to reproduce the Minkowski real inner product $g_r$ implies an 
anti-diagonal $J_\ro$, that is: $j^{S,aa}_{\om l} \equiv 0$,
which in turn implies $j^{S,ab}_{\om l} =-1/j^{S,ba}_{\om l}$.
With this, fixing $j^{S,ba}_{\om l}$ determines $J_\ro$ completely.

\subsection{Interlaced choice $J\htx{iso}_\ro$}
\label{complex_struct_AdS_KG:_interlaced}

Now we aim to construct another version of $j^{S,ba}_{\om l}$,
which we call simply $j\htx{iso}_{\om l}$,
that commutes with the boosts for all frequencies while respecting weak amplitude equivalence at the same time.
To this end we use the fact that the boost conditions
\eqref{zzz_AdS_invar_complex_isometries_681_m}
only relate factors $j^{S,ba}$ with integer frequency differences.
If we consider only $j^{S,ba}_{\om l}$ with $l$ fixed,
then these conditions relate only factors
whose frequency difference is an \emph{even} integer
(since we have to apply them twice to get back the original $l$).
Our starting point are again the conditions
\eqref{zzz_AdS_invar_complex_isometries_1784_magic_+} and
\eqref{zzz_AdS_invar_complex_isometries_1784_magic_-},
that is, $j^{S,ba}_{\omag+nl, l} \eq j^{(\be)}_{\omag+nl, l}$
and $	j^{S,ba}_{-\omag+nl, l} \eq j^{(\al)}_{-\omag+nl, l}$.
Thus, for the positive magic frequencies $+\omag+nl$
we choose the $\be$-version,
and for negative magic frequencies $-\omag+nl$ the $\al$-version.
This \emph{does not} violate the boost conditions
\eqref{zzz_AdS_invar_complex_isometries_681_m}
as long as positive and negative magic frequencies
are not separated by an even integer gap.
Since from the outset we only considered
non-integer $\nu$, this is an additional condition,
which implies that now we only consider values of 
both $\timp$ and $\nu$ that are neither integer nor half-integer.

Let us keep $l$ fixed for the moment, for example at $l=0$.
The above boost compatibility conditions
then induce the $\be$-version for all frequencies
at an even integer distance 
from the positive magic frequencies $+\omag+n{,0}$,
and the $\al$-version for all frequencies
at an even integer distance
from the negative magic frequencies $-\omag+n{,0}$.
Apart from this discrete set of frequencies
we are at liberty to choose our $j^{S,ba}_{\om l}$
(as long as we respect frequency symmetry and boost conditions).
We thus obtain a pattern of interlaced frequency intervals, 
on some of which we choose $j^{(\al)}_{\om l}$
while on others we choose $j^{(\be)}_{\om l}$.
We denote the result by $j\htx{iso}_{\om l}$.

We would like our $j\htx{iso}$
to avoid zeros and also singularities if possible.
For some fixed $l$, the $\al$-version vanishes if
the frequency $\om$ is either magic $\om \eq +\omag+nl$,
or if it is one of those which we call zero frequencies
$\om \eq +\omag0nl \eq -\timp \pn d \pn 2n \pn l$.
The $\be$-version vanishes if
the frequency $\om$ is either negative-magic $\om \eq -\omag+nl$,
or ''negative''-zero $\om \eq -\omag0nl \eq +\timp \mn d \mn 2n \mn l$.
For some fixed $l$, the $\al$-version becomes singular
if the frequency $\om$ is either ''sing'' 
$\om =+\om\htx{sing}_{nl} \eq \pn\timp \pn 2n \mn l\mn d \pn 2$
or ''ular'' $\om =+\om\htx{ular}_{nl} \eq -\timp +2n - l + 2$.
The $\be$-version becomes singular if the frequency $\om$
is either ''negative-sing'' 
$\om =-\om\htx{sing}_{nl} \eq \mn\timp \mn 2n \pn l\pn d \mn 2$
or ''negative-ular'' 
$\om =-\om\htx{ular}_{nl} \eq +\timp -2n + l - 2$.
Unfortunately, we find that the ''negative''-zero
frequencies $-\omag0{n_2}{,l}$ are at an even
integer distance from the ''sing'' frequencies
$\om\htx{sing}_{n_1,l}$, and the negative magic frequencies
are at even integer distance from the ''ular'' frequencies.
This implies, that we \emph{cannot} get a $j\htx{iso}$
which gives us amplitude equivalence
while avoiding zeros and singularities.

\begin{figure}[H]
	\centering
	\igx[width=0.48\linewidth]{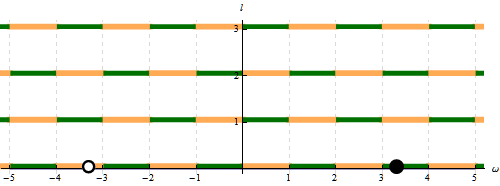}
	\caption{Interlaced complex structure $J_\ro\htx{iso}$:
				intervals in $(\om,l)$-space with $\al$ and
				$\be$-version of $j^{S,ba}_{\om l}$.
				}
	\label{J_al_be}
\end{figure}

So far we have assigned the $\al$ and $\be$-version
to a discrete set of frequencies related to the magic frequencies.
We now extend this choice to all $l\elof\naturals_0$
and $\om\elof \reals$ as sketched in Figure \ref{J_al_be}.
Therein, we have $\om$ on the horizontal axis and $l$ on the vertical.
Intervals on which we choose the $\al$-version
appear in orange (lighter gray in monochrome),
and intervals with the $\be$-version are dark green (darker gray).
The extension of $j\htx{iso}_{\om l}$ must fulfill
three properties: first, include the magic frequencies as in
\eqref{zzz_AdS_invar_complex_isometries_1784_magic_+} and
\eqref{zzz_AdS_invar_complex_isometries_1784_magic_-}.
Second, be frequency-symmetric: $j\htx{iso}_{\om l}
\eq j\htx{iso}_{-\om, l}$. Third, the pattern of
interlaced intervals where we choose the $\al$ and $\be$-versions
must be translation-invariant for steps of 2 in $\om$-direction
and $l$-direction.
This is necessary in order to comply with the boost
conditions \eqref{zzz_AdS_invar_complex_isometries_681_m},
which relate $j^{S,ba}_{\om l}$ to $j^{S,ba}_{\om\pm2, l}$
and $j^{S,ba}_{\om, l\pm2}$.
The simplest solution to these three properties is
for $l=0$ to associate the interval
$(\,\lfloor\timp\rfloor,\; \lceil \timp\rceil\,]$ to the $\be$-version.
We use the standard notation of $\lfloor x \rfloor$
for the floor function (largest integer $\leq x$), and
$\lceil x \rceil$ for ceiling (smallest integer $\geq x$).
This choice already determines all other intervals
through the boost conditions: the $\al$ and $\be$-version
alternate both horizontally ($\om$-direction) and vertically
($l$-direction), see again Figure \ref{J_al_be}.
In the $(\om,l)$-plane let us denote by $I_\be$ the set of intervals
associated to the $\be$-version as described above,
and by $I_\al$ the set of intervals associated to the $\al$-version.
Then, our interlaced choice $J\htx{iso}_\ro$ writes as
\bal{\label{5892_Ji_ro}
	j\htx{iso}_{\om l} & \eq
		\bcas{j^{(\al)}_{\om l} & (\om,l)\elof I_\al
				\\
			j^{(\be)}_{\om l} & (\om,l)\elof I_\be
			},
	}
wherein $j^{(\al)}_{\om l}$ and $j^{(\be)}_{\om l}$
are those of \eqref{zzz_AdS_invar_complex_isometries_1783_j_al}.
For this choice, two different patterns emerge:
for ''Case $\al$'': $\timp \elof (d\pn2n,\,d+2n\pn1)$
with $n\elof\naturals_0$ we have the unit interval
$\om\elof(0,1)$ for $l=0$ associated to the $\al$-version,
as shown in Figure \ref{J_al_be}
while for ''Case $\be$'': $\timp \elof (d\pn2n\pn1,\,d+2n\pn2)$
we have it associated to the $\be$-version (not shown here).
The label of the case thus refers to which version occupies the
unit interval $\om\elof(0,1)$ for $l=0$.
(We recall that $d$ is odd, and that we only consider values of $\timp$
 that are neither integer nor half-integer.)
In Figure \ref{J_al_be},
the position of $\om\eq\pn\timp$ is marked by
a black disk, and that of $\om\eq\mn\timp$ by a black circle.
For $d=3$ with $\RAdS=1$, the example in Figure \ref{J_al_be}
arises from Klein-Gordon mass $m=1$ giving $\timp\approx3.3$.
For any $\timp$, our choice implies that for $l=0$
the black disk of $\om\eq\pn\timp$ sits on a green (dark gray)
interval of the $\be$-version, and hence the circle of $\om\eq\mn\timp$
on an orange (light gray) interval of the $\al$-version.

We thus have fixed completely the element $j\htx{iso}_{\om l}$
through interlacing intervals on which we choose
the $\al$ respectively $\be$-version.
While not very elegant, this is physically motivated:
it makes our complex structure
commute \emph{with all} isometry actions, while also respecting weak amplitude
equivalence.
As for $J_\ro\htx{two}$, 
we choose an anti-diagonal $J_\ro\htx{iso}$,
that is: $j^{S,aa}_{\om l} \equiv 0$,
which in turn implies $j^{S,ab}_{\om l} =-1/j^{S,ba}_{\om l}$.

\subsection{Sign of the real inner product $g_\ro$}
\label{complex_struct_AdS_KG:_sign_g}

The amplitudes \eqref{eq:freeampl} are determined
by the real inner product \eqref{eq:ip}, 
which is in turn determined by the complex structure $J$
introduced in Section~\ref{sec:smatgbf}.
In \eqref{AdS_structures_180_n_nu}
we already wrote down the real inner product 
$g_t$ induced by the standard choice $J_t$ for equal-time hypersurfaces.
Then, amplitude equivalence fixes $g_\ro$ and thus $J_\ro$
for the hypercylinder,
but only for the magic frequencies $\omag+nl$.
Any anti-diagonal choice like $J_\ro\htx{iso}$ or
$J_\ro\htx{two}$ then further fixes the real inner product
\emph{for all} frequencies $\om$ to take the form
\bal{g_\ro(\et,\ze) = 2 \om_\ro(\et,J_\ro\ze) 
	& \eq 2\piu \RAdS^{d\mn1} \intomsumvclm 
		\biiglrr{\et^{a}_{\om \vc l m_l}\,
					(J_\ro\ze)^{b}_{-\om,\vc l,-m_l}
				-\et^{b}_{\om \vc l m_l}\,
				(J_\ro\ze)^{a}_{-\om,\vc l,-m_l}
				}\,
		(2l\pn d\mn2)
		\notag
		\\
	\label{AdS_invar_complex_isometries_24}
	& \eq 2\piu \RAdS^{d\mn1} \intomsumvclm 
		\biiglrr{\et^{a}_{\om \vc l m_l}\,
					\ze^{a}_{-\om,\vc l,-m_l}\,
					j^{S,ba}_{\om l}
				+\et^{b}_{\om \vc l m_l}\,
				\ze^{b}_{-\om,\vc l,-m_l}
				/ j^{S,ba}_{\om l}
				}\,
		(2l\pn d\mn2).
	}
For real solutions $\ph$ we have 
$\ph^{a}_{-\om,\vc l,-m_l} \eq \coco{\ph^{a}_{\om \vc l m_l}}$
and $\ph^{b}_{-\om,\vc l,-m_l} \eq \coco{\ph^{b}_{\om \vc l m_l}}$
and thus obtain
\bal{\label{zzz_AdS_invar_complex_isometries_2597_phph}
	g_\ro(\ph,\ph) 
	& \eq 2\piu \RAdS^{d\mn1} \intomsumvclm 
		\biiglrr{\absq{\ph^{a}_{\om \vc l m_l}}
					j^{S,ba}_{\om l}
				+\absq{\ph^{b}_{\om \vc l m_l}}
				/ j^{S,ba}_{\om l}
				}\,
		(2l\pn d\mn2).
	}
We can read off that the real inner product is positive (negative) 
for modes with $\om$ and $l$ such that $j^{S,ba}_{\om l}$
is positive (negative).
Let us have a look where this is the case.
Plotting the $\al$ and $\be$-version reveals that
the $\be$-version is positive for all $\om\geq (\timp\pn l)$
while the $\al$-version is positive for all $\om\leq-(\timp\pn l)$.
For all other frequencies, both versions alternate
between intervals with positive and negative sign.

Therefore, for the interlaced version $j\htx{iso}_{\om l}$
there is no simple rule telling us where it has positive
or negative sign. The only exception is that $j\htx{iso}_{\om l}$
is positive for all magic frequencies $\pm\omag+nl$.
For the two-branched version $j\htx{two}_{\om l}$
the situation is simpler: due to its definition,
$j\htx{two}_{\om l}$ is positive for all
frequencies with $\abs \om\geq (\timp\pn l)$, while it alternates
its sign for the remaining frequencies.
Hence $J_\ro\htx{two}$ makes $g_\ro$ positive definite
for all modes with $\abs \om\geq (\timp\pn l)$,
while $J_\ro\htx{iso}$ does so only for the Jacobi modes,
that is, the modes with magic frequencies
(with both $J_\ro$ letting $g_\ro$ alternate sign
 for the remaining frequencies).

Let us compare this to the complex structure $J_r$ and the induced
real inner product $g_r$ of a Minkowski hypercylinder given in \eqref{zzz_AdS_invar_complex_isometries_2902_Mink_txg_r}.
As for AdS, for Minkowski there is a complex structure $J_r\htx{iso}$
commuting with all isometries
while inducing a real inner product $g_r\htx{iso}$
that is not positive definite for evanescent modes
but only for the propagating ones.
Further, the Minkowski hypercylinder has a complex structure $J_r\htx{pos}$
inducing $g_r\htx{pos}$ which is positive definite for all modes \cite{Oe:holomorphic}.
For AdS, only the two-branched choice
$J_\ro\htx{two}$ induces a real inner product $g_\ro\htx{two}$
that comes close to this property:
above the mass threshold $\abs \om\geq (\timp\pn l)$
$g_\ro\htx{two}$ is positive definite.
In Section \ref{complex_struct_AdS_KG:_flatlim_rod}
we take this as a motivation for modifying
$J_\ro\htx{two}$ for the remaining frequencies
$\abs \om\leq (\timp\pn l)$ such that the induced $g_\ro$
becomes positive definite for all modes.

We conclude this subsection by giving explicit expressions
for the sign of the AdS real inner product $g_\ro$,
that is, determining where $j^{S,ba}_{\om l}$ is positive
and where negative. Combining the four relations
\eqref{zzz_AdS_invar_complex_isometries_443_ba}, we find that
\bal{
	j^{S,ba}_{\om\pn2,l}
	& \eq j^{S,ba}_{\om l} \ka^+_{\om l}
		&
	\ka^{1}_{\om l} 
	& \eq (\timp\pn\om\mn l\mn d\pn2)
		&
	\ka^{2}_{\om l} 
	& \eq (\timp\mn\om\pn l\mn2)
		\\
	\ka^+_{\om l}
	& \eq \fracw{\ka^{1}_{\om l}\,\ka^{2}_{\om l}}
						{\ka^{3}_{\om l}\,\ka^{4}_{\om l}}
		&
	\ka^{3}_{\om l} 
	& \eq (\timp\pn\om\pn l)
		&
	\ka^{4}_{\om l} 
	& \eq (\timp\mn\om\mn l\mn d).
		\notag
	}
These expressions relate the signs of $j^{S,ba}_{\om l}$
only for even integer frequency differences.
In particular, they are fulfilled by both
the $\al$-version $j^{(\al)}_{\om l}$ and
the $\be$-version $j^{(\be)}_{\om l}$.
Since $j^{(\be)}_{\om l} = j^{(\al)}_{\mn\om, l}$,
it is sufficient to consider the signs of the $\al$-version.
We denote by $\la^q_l$ the unique frequency
for which $\ka^q_{\om l}$ becomes zero (with $l$ fixed):
\bal{
	\la^1_l 
	& = -(\timp\mn l\mn d\pn2)
		&
	\la^2_l 
	& = +(\timp\pn l\mn2)
		&
	\la^3_l 
	& = -(\timp\pn l)
		&
	\la^4_l 
	& = +(\timp\mn l\mn d).
	}
Our starting point is that the $\al$-version $j^{(\al)}_{\om l}$
is positive for all $\om < \la^3_l+2 = -\timp-l + 2$, and 
we determine the sign for $\om > \la^3_l+2$ by counting sign changes.
We observe that the $\al$-version changes sign
at each singularity and at each zero.
The singularities are caused by the factors
$\ka^{3}_{\om l}$ and $\ka^{4}_{\om l}$
and appear to the right of $\la^{3}_{l}$ and
$\la^{4}_{l}$ in steps of $\De\om=2$.
The zeros come from $\ka^{1}_{\om l}$ and $\ka^{2}_{\om l}$
and appear to the right of $\la^{1}_{l}$ and
$\la^{2}_{l}$, also in steps of $\De\om=2$.
We use the usual notation of $\lfloor x \rfloor$
for the floor function (largest integer $\leq x$),
and $\te(x)$ for the Heaviside step function
which is 0 for all $x\leq0$ and 1 for $x>0$.
With $q=1,2,3,4$, each $\ka^{q}_{l}$
contributes $\si^q_{\om l}$ sign changes
between some frequency $\om > \la^q_l$ and $\la^q_l$:
\bal{
	\si^q_{\om l} \eq 
	\te(\om\mn\la^q_l)\; \lfloor (\om\mn\la^q_l)/2\rfloor.
	}
The total of sign changes is the sum of the four $\si$'s,
resulting in the following formula for the sign of
the $\al$-version:
\bal{\label{zzz_AdS_invar_complex_isometries_2794_si_al}
	\sign j^{(\al)}_{\om l}
	& \eq (-1)^{\si^\al_{\om l}}
		&
	\si^\al_{\om l} 
	& \eq \si^1_{\om l}+\si^2_{\om l}+\si^3_{\om l}+\si^4_{\om l}.
	}
Hence the signs of choices $j\htx{iso}_{\om l}$
and $j\htx{two}_{\om l}$ result to
be \eqref{zzz_AdS_invar_complex_isometries_2810_si_i},
which tells us whether the AdS hypercylinder's real inner product $g_\ro$
in \eqref{zzz_AdS_invar_complex_isometries_2597_phph}
is positive or negative for each mode $\mu\hbs{S,a}_{\om \vc l m_l}$
and $\mu\hbs{S,b}_{\om \vc l m_l}$:
\bal{\label{zzz_AdS_invar_complex_isometries_2810_si_i}
	\sign j\htx{iso}_{\om l}
	& \eq \bcas{(-1)^{\si^\al_{\om l}} & (\om,l)\elof I_\al
						\\
					(-1)^{\si^\al_{\mn\om,l}} & (\om,l)\elof I_\be
					}
		&
	\sign j\htx{two}_{\om l}
	& \eq \bcas{(-1)^{\si^\al_{\om l}} & \om < 0
						\\
					(-1)^{\si^\al_{\mn\om,l}} & \om >0
					}.
	}

\subsection{Flat limit of rod amplitudes and $J\htx{pos}_\ro$}
\label{complex_struct_AdS_KG:_flatlim_rod}

As above for the time-interval, a good choice of the complex structure
$J_\ro$ should induce a real inner product $g_\ro$
whose flat limit recovers the Minkowskian $g_r$
given in \eqref{zzz_AdS_invar_complex_isometries_2902_Mink_txg_r},
thereby making the flat limit of the AdS rod amplitude
the Minkowskian rod amplitude.
This points to the complex structure
$J_\ro$ of the AdS hypercylinder $\Si_\ro$
being anti-diagonal, that is: $j^ {S,aa}_{\om l} \equiv 0$.
The reason for this is that for the Minkowski hypercylinder
$\Si_r$ the complex structure $J_r$ is anti-diagonal:
$j^{aa}_{\tiom l} \equiv 0$. Hence the flat limit of
$j^{S,aa}_{\om l}$ must be zero, which indicates
that is is zero already before taking the limit.

For $J\htx{two}_\ro$ we now verify whether 
the flat limit of the induced $g_\ro$
recovers the Minkowskian $g_r$ in
\eqref{zzz_AdS_invar_complex_isometries_2902_Mink_txg_r}.
(This can be done for $J\htx{iso}_\ro$ as well,
but then only works for a discrete set of frequencies:
since $J\htx{iso}_\ro$ depends on the frequency $\om$,
while $J\htx{iso}_r$, see (\ref{eq:Jriso}), does not, there are issues
with overall signs which we could resolve only 
for a discrete set of frequencies.
Since this is not very satisfying, here we only present
the details for $J\htx{two}_\ro$.)
To this end we first calculate the flat limit 
of the $\al$ and $\be$-versions
$j^{(\al)}_{\om l}$ and $j^{(\be)}_{\om l}$.
The flat limit only affects the last two quotients of Gamma functions 
in \eqref{zzz_AdS_invar_complex_isometries_1784_magic_+}. 
Below we calculate the limit of the $\al$-version
for large $R$ (short for $\RAdS$), and setting $d=3$:
\bals{\frac{\Gaar{\al^{S,b}}}{\Gaar{\al^{S,a}}}
		\frac{\Gaar{1\mn\be^{S,a}}}{\Gaar{1\mn\be^{S,b}}}\!
	&= \fracw{\Gaar{-\tfrac12(\om\mn\timp\pn l\pn1)}}
				{\Gaar{-\tfrac12(\om\mn\timp\mn l)}}
		\fracw{\Gaar{-\tfrac12(\om\pn\timp\pn l\mn2)}}
				{\Gaar{-\tfrac12(\om\pn\timp\mn l\mn 3)}}
		\\
	 & \approx \!\biglrr{-\tfrac{1}2\biglrr{\om\mn\timp}}^{-l}\!
			\biglrr{-\tfrac{1}2\biglrr{\om\pn\timp}}^{-l}\!
		\frac{\Gaar{X^\al_-}}{\Gaar{X^\al_- +\tfrac12}}
		\frac{\Gaar{X^\al_+}}{\Gaar{X^\al_+ +\tfrac12}}
	= \biglrr{\tfracw{ R^2}4\biglrr{\tiom^2\mn m^2}}^{-l}
		\frac{\Gaar{X^\al_-}}{\Gaar{X^\al_- +\tfrac12}}
		\frac{\Gaar{X^\al_+}}{\Gaar{X^\al_+ +\tfrac12}},
	}
wherein $X^\al_- \eq -\tfrac12(\om\mn\timp)\mn\tfrac l2\mn\tfrac12$
and $	X^\al_+ \eq -\tfrac12(\om\pn\timp)\mn\tfrac l2\pn1$.
The calculation for the $\be$-version yields the same result,
but with $X^\be_- \eq +\tfrac12(\om\mn\timp)\mn\tfrac l2\pn1$
and $X^\be_+ \eq +\tfrac12(\om\pn\timp)\mn\tfrac l2\mn\tfrac12$.
In the flat limit, for all $X^{\al,\be}_\pm$ 
their absolute value $\abs{X}$ becomes very large.
Because Gamma alternates its sign for negative arguments,
for large $\abs X$ with $X$ positive or negative we get:
\bals{\fracw{\Gaar{X}}{\Gaar{X +\tfrac12}}
	\approx \abs{X}^{-1/2}\,\cdot\,
		\bcas{-1\quad & X \elof(-n+\tfrac12,-n+1)
					\qquad n\elof\naturals\,\!^+
					\\
				+1 \quad & \text{else}
				}
	}
With this we can complete the process of taking the flat limit
for the $\al$ and $\be$-version:
\bal{\label{zzz_AdS_invar_complex_isometries_2566_flatlim_jab_al}
	j^{(\al)}_{\om l}
	& \!\toflatw
	\tfracw1{\piu} \Gaar{l\pn\tfrac32}\,\Gaar{l\pn\tfrac12}\;
	\biiglrr{\tfrac R2 \tip_\tiom}^{-2l}\,
	\biiglrr{\tfrac R2 \tip\hreals_\tiom}^{-1}\;
	q^\al_-\, q^\al_+
		&
	q^\al_\mp 
	&\eqn \bcas{-1\quad & X^\al_\mp \elof (-n+\tfrac12,-n+1)
					\\
				+1 \quad & \text{else}
				}
		\\
	\label{zzz_AdS_invar_complex_isometries_2566_flatlim_jab_be}
	j^{(\be)}_{\om l}
	& \!\toflatw
	\tfracw1{\piu} \Gaar{l\pn\tfrac32}\,\Gaar{l\pn\tfrac12}\;
	\biiglrr{\tfrac R2 \tip_\tiom}^{-2l}\,
	\biiglrr{\tfrac R2 \tip\hreals_\tiom}^{-1}\;
	q^\be_-\, q^\be_+
		&
	q^\be_\mp 
	&\eqn \bcas{-1\quad & X^\be_\mp \elof (-n+\tfrac12,-n+1)
					\\
				+1 \quad & \text{else}
				}.
	}
Now we come back to the complex structure $J\htx{two}_\ro$.
We aim to reproduce in the flat limit the real inner product
$g_r$ for a Minkowski hypercylinder as induced by 
the Minkowski $J\htx{pos}_r$,
which is independent of frequency $\om$ and angular momentum $l$,
and induces a positive definite $g_r$.
For the AdS hypercylinder we start with the two-branched
complex structure $J\htx{two}_\ro$.
We already found that this choice does not commute
with AdS boosts. Hence we let go of this requirement
(keeping thus only commutation with time translations
and spatial rotations).
Therefore we are no longer tied to using
the $\al$ and $\be$-versions for all frequencies,
but only for the magic ones in order to 
keep our weak version of amplitude equivalence.
We now make use of this freedom, with the goal of modifying
$J\htx{two}_\ro$ such that it induces
a positive definite $g_\ro$ for all frequencies.

Like $J_t$ in Section \ref{complex_struct_AdS_KG:_flatlim_eqtime},
we need to construct a complex structure $J_\ro$ for AdS,
whose action ''commutes'' with the process
of taking the continuous flat limit:
\bal{\label{diag_J_ro_AdS_Mink}
	\bdia{\node{\ph\htx{AdS}}
				\arrow[2]{e,t}{J_\ro}
				\arrow[1]{s,l}{\text{cont. flat lim.}}
				\node{}
				\node{J_\ro \ph\htx{AdS}}
				\arrow[1]{s,r}{\text{cont. flat lim.}}
					\\
				\node{\ph\htx{Mink}}
				\arrow[2]{e,t}{J\htx{pos}_r}
				\node{}
				\node{J\htx{pos}_r \ph\htx{Mink}}
				}
	}
Our goal is to reproduce the action \eqref{zzz_Mink_J_r_124}
of the positive definite complex structure $J\htx{pos}_r$, given in (\ref{eq:Jrpos}),
of the Minkowski hypercylinder
as the continuous flat limit of $J_\ro\ph$. 
An anti-diagonal complex structure acts in the $S$-expansion as
\bals{\biglrr{J_\ro\ph} \artrOm 
		& \eq \intomsumvclm
		\biiglrc{(J_\ro\ph)^{S,a}_{\om \vc l m_l}\,\eu^{-\iu \om t}\;
			\spherharmonicar{m_l}{\vc l}\Om\;S^a_{\om l}\ar \ro	
			+ (J_\ro\ph)^{S,b}_{\om \vc l m_l}\,\eu^{-\iu \om t}\;
			\spherharmonicar{m_l}{\vc l}\Om\;S^b_{\om l}\ar \ro
			}
		\\
	& \eq \intomsumvclm
		\biiglrc{\frac{-1}{j^{S,ab}_{\om l}}
			\ph^{S,b}_{\om \vc l m_l}\,\eu^{-\iu \om t}\;
			\spherharmonicar{m_l}{\vc l}\Om\;S^a_{\om l}\ar \ro	
			+ j^{S,ab}_{\om l}
			\ph^{S,a}_{\om \vc l m_l}\,\eu^{-\iu \om t}\;
			\spherharmonicar{m_l}{\vc l}\Om\;S^b_{\om l}\ar \ro
			}.
	}
Switching to the flat $S$-representation
\eqref{AdS_KG_solutions_tube_flat_rep},
for $d=3$ this becomes
\bals{\biglrr{J_\ro\ph} \artrOm 
	& \eq \intomsumvclm \fracw{\tip\hreals_\tiom}{(4\piu)}
		\biiglrc{-T^{-1} \ph^{F,b}_{\om \vc l m_l}\,\eu^{-\iu \om t}\;
			\spherharmonicar{m_l}{\vc l}\Om\;
			\frac{(p\hreals_\om)^{l}}{(2l\pn1)!!}
			S^a_{\om l}\ar \ro	
			+ T\,\ph^{F,a}_{\om \vc l m_l}\,\eu^{-\iu \om t}\;
			\spherharmonicar{m_l}{\vc l}\Om\;
			\frac{(2l\mn1)!!}{(p\hreals_\om)^{l\pn1}}
			S^b_{\om l}\ar \ro	
			}.
	}
wherein $T = j^{S,ba}_{\om l}\,(p\hreals_\om)^{2l\pn1}/
((2l\pn1)!!\, (2l\mn1)!!)$. Taking the flat limit we obtain
\bals{\biglrr{J_\ro\ph} \artrOm 
	& \toflat \inttiomsumvclm \fracw{\tip\hreals_\tiom}{(4\piu)}
		\biiglrc{-T^{-1}
			\tiph^{F,b}_{\om \vc l m_l}\,\eu^{-\iu \tiom \ta}\;
			\spherharmonicar{m_l}{\vc l}\Om\;
			\check \jmath_{E l}\ar r
			+ T
			\tiph^{F,a}_{\om \vc l m_l}\,\eu^{-\iu \tiom \ta}\;
			\spherharmonicar{m_l}{\vc l}\Om\;
			\check n_{E l}\ar r
			}.
	}
Thus in order to reproduce the action \eqref{zzz_Mink_J_r_124}
of the Minkowski $J_r$,
we need to fix $j^{S,ba}_{\om l}$ such that
in the flat limit the factor $T$ becomes unity.
As found in Section \ref{complex_struct_AdS_KG:_sign_g},
the $\al$-version is positive for $\om \leq -(\timp\pn l)$
and the $\be$-version is positive for $\om \geq (\timp\pn l)$.
Hence together with
\eqref{zzz_AdS_invar_complex_isometries_2566_flatlim_jab_al} and
\eqref{zzz_AdS_invar_complex_isometries_2566_flatlim_jab_be}
we find that for these frequencies the flat limit
of both versions is
\bal{
	j^{(\al/\be)}_{\om l}
	\toflatw &
	\tfracw1{\piu} \Gaar{l\pn\tfrac32}\,\Gaar{l\pn\tfrac12}\;
	\biiglrr{\tfrac R2 \tip\hreals_\tiom}^{-2l-1}
		\\
	= & (2l\mn1)!!\,(2l\pn1)!!\;
	\biiglrr{R \tip\hreals_\tiom}^{-2l-1}.
	}
Therein we use that for odd $k$ we have
$k!! = \Gaar{\tfrac k2\pn1}\,2^{\frac{k\pn1}{2}}/\ruut\piu$,
and thus $\Gaar{l\pn\tfrac32}\Gaar{l\pn\tfrac12}\eq
(2l\mn1)!!\,(2l\pn1)!! 2^{-2l-1} \piu$.
That is: keeping the $\al$-version for $\om\leq -(\timp+l)$
and the $\be$-version for $\om\geq (\timp+l)$
makes the flat limit of $T$ unity for these frequencies.
For the remaining frequencies $|\om| \leq (\timp+l)$
the most obvious choice is then
\bal{
	j\htx{obv}_{\om l}
	= &
	\tfracw1{\piu} \Gaar{\ga^{S,a}}\,\Gaar{\ga^{S,a}\mn1}\;
	\biiglrr{\tfrac 12 p\hreals_\om}^{-2l-1}
		\\
	= & (2l\pn d\mn2)!!\,(2l\pn d\mn4)!!\;2^{3-d}
		\biiglrr{p\hreals_\om}^{-2l-1}.
	}
Then, for these frequencies with $d=3$
the factor $T$ becomes unity trivially.
This modified two-branched version 
shall be called positive definite version
and denoted by $J_\ro\htx{pos}$ with
\bal{\label{4469_J_pos}
	j\htx{pos}_{\om l} & \eq
		\bcas{j^{(\al)}_{\om l} & \om \leq -(\timp\pn l)
				\\
			j^{(\be)}_{\om l} & \om \geq +(\timp\pn l)
				\\
			j\htx{obv}_{\om l} & \text{else}
			}.
	}
We remark that $j\htx{pos}_{\om l} >0$ for all frequencies,
and thus induces a positive definite $g_\ro$.
Having thus verified that our new $J\htx{pos}_\ro$
commutes with the continuous flat limit,
we now study whether its induced real inner product $g_\ro$
is indeed positive definite.
We start with the real inner product
\eqref{AdS_invar_complex_isometries_24}
for an anti-diagonal $J_\ro$ with the solutions
written in their $S$-expansions.
Substituting the flat $S$-representation
\eqref{AdS_KG_solutions_tube_flat_rep} yields for $d=3$
\bal{g_\ro(\et,\ze) 
		& \eq \inttiomsumvclm 
		\fracw{\tip\hreals_\tiom}{8\piu}
		\biiglrc{\tiet^{F,a}_{\om \vc l m_l}\,
					\tize^{F,a}_{-\om,\vc l,-m_l}\,
					T
				+\tiet^{F,b}_{\om \vc l m_l}\,
				\tize^{F,b}_{-\om,\vc l,-m_l}
				T^{-1}
				},
	}
wherein again $T = j^{S,ba}_{\om l}\,(p\hreals_\om)^{2l\pn1}
/((2l\pn1)!!\, (2l\mn1)!!)$.
In order to reproduce the positive definite Minkowski $g_r$
\eqref{zzz_AdS_invar_complex_isometries_2902_Mink_txg_r},
we thus again need a $j^{S,ba}_{\om l}$ such that
in the flat limit the factor $T$ becomes unity.
Since our $j\htx{pos}_{\om l}$
is constructed precisely to fulfill this condition,
we have now verified that our complex structure
$J\htx{pos}_\ro$ in the flat limit indeed reproduces
the positive definite inner product $g_r$
of the Minkowski hypercylinder.

With this we have finished our construction of complex structures
$J_\ro$ for the AdS hypercylinder. The induced real inner product
$g_\ro=2\om_\ro(\cdot,\,J_\ro\cdot)$ then determines
the free amplitudes \eqref{eq:freeampl} and the amplitudes 
for interaction with a source field \eqref{eq:dampl}
for the AdS rod region.
These rod amplitudes inherit the properties of the complex structure
inducing them. Hence they are weakly equivalent
to the amplitudes of the time-interval region
and invariant under isometries preserving the hypercylinder
(time-translations and spatial rotations).
For the complex structure $J\htx{iso}$ the induced amplitudes
in addition are boost invariant, while the induced
inner product $g_\ro$ is not positive definite.
For $J\htx{pos}$ the induced amplitudes
are not boost invariant, but the induced
inner product $g_\ro$ is positive definite here.
Table \ref{table:_properties_J_ro} sums up all these properties.
Comparing it to Table \ref{table:_properties_J_r}
in Appendix \ref{class_KG_AdS:_sols_mink},
we can read off that the AdS $J_\ro\htx{iso}$
has the same properties as its Minkowski counterpart,
and the same holds for $J_\ro\htx{pos}$.
\setlength{\extrarowheight}{3pt}
\begin{table}[H]
	\centering
	\begin{tabular}{|c|c c|}
		\hline
		{\bfseries AdS} & $J_\ro\htx{iso}$ & $J_\ro\htx{pos}$
			\\
		\hhline{|=|==|}
		commute with time-translations & \checkmark & \checkmark 
			\\
		commute with spatial rotations & \checkmark & \checkmark 
			\\
		commute with boosts & \checkmark & - 
			\\
		\hline
		weak amplitude equivalence & \checkmark & \checkmark
			\\
		\hline
		induced real inner product $g_\ro$ 
			& indefinite &  positive-definite
			\\
		\hline
		flat limit & $J_r\htx{iso}$ & $J_r\htx{pos}$
			\\
		& $\htx{only for a discrete}\ltx{subset of frequencies}$
		&
			\\
		\hline
	\end{tabular}
	\caption{Properties of complex structures for AdS hypercylinder}%
	\label{table:_properties_J_ro}%
\end{table}

\section{Summary}
\label{summary_outlook}

The standard S-matrix is problematic in AdS spacetime. While it can be formulated (it is also reviewed in this paper), the asymptotic states are restricted to a set of modes with discrete frequencies. A generalized notion of S-matrix is more promising, where asymptotic states live on the hypercylinder boundary at infinite radius \cite{CDO:adsproc}. This is the subject of the present paper for the case of the real scalar Klein-Gordon field. As usual in field theory, in the quantization of the Klein-Gordon field there is an ambiguity which can be conveniently parametrized in terms of a \emph{complex structure}. In standard quantization in Minkowski spacetime this ambiguity is fixed by requiring the inner product to be positive definite and to be invariant under isometries. The main focus of the present paper is thus this complex structure in the case of AdS and for (asymptotic) fields on a hypercylinder geometry, induced by the boundary of AdS. Given this complex structure the quantization is fixed and the generalized S-matrix can be computed, see Section~\ref{sec:smatgbf}.

For an equal-time hypersurface in AdS there is a standard complex structure $J_t$, given in \eqref{complex_struct_AdS_KG:_ess_iso_J_t} of Section~\ref{complex_struct_AdS_KG:_time_interval}. The latter is fixed by positive definiteness and isometry invariance. This is analogous to standard quantization in Minkowski spacetime, but with the crucial difference that this works in AdS only for a discrete set of field modes with discrete frequencies, called \emph{magic frequencies} here. Consequently the standard S-matrix on AdS is restricted to these discrete modes.
For the AdS hypercylinder geometry (used for the generalized continuum-mode S-matrix) we find that there is no complex structure that is both isometry invariant and leads to a positive definite inner product for the whole continuum of modes. However, there are complex structures that partially satisfy these properties. Moreover, there are additional desirable properties for the complex structure that we take into account, see Section~\ref{sec:smcomplex}: One is an equivalence to the standard complex structure on equal-time hypersurfaces for the subset of modes with magic frequencies. We call this \emph{weak amplitude equivalence}. The other is the recovery of known complex structures on Minkowski spacetime in the limit that AdS becomes flat and the solutions of the Klein-Gordon equation become solutions on Minkowski spacetime. We call this the \emph{flat limit} for brevity.

We show that there is a class of complex structures on the AdS hypercylinder that is invariant under all isometries of AdS, see Section~\ref{complex_struct_AdS_KG:_essential_isometry}. Further imposing weak amplitude equivalence motivates an interlaced construction of a complex structure, $J_\ro\htx{iso}$ given in \eqref{5892_Ji_ro} of Section~\ref{complex_struct_AdS_KG:_interlaced}. This retains full isometry invariance and satisfies weak amplitude equivalence as well. A disadvantage is that it leads to an indefinite inner product on the space of modes and thus also on the space of quantum states. The space of quantum states is thus a Krein space rather than a Hilbert space. While this introduces a superselection rule, it does not spoil the probability interpretation of quantum theory \cite{Oe:freefermi}.

If instead we do not insist on full isometry invariance, but only on invariance with respect to isometries of the AdS hypercylinder (time-translations and spatial rotations), we obtain a complex structure $J_\ro\htx{pos}$ given in \eqref{4469_J_pos} of Section~\ref{complex_struct_AdS_KG:_flatlim_rod}, that yields a positive definite inner product. What is more, in the flat limit, this reproduces a complex structure on the Minkowski hypercylinder \cite{Oe:holomorphic} which is equivalent there to the standard quantization for propagating modes. Table~\ref{table:_properties_J_ro} in Section~\ref{complex_struct_AdS_KG} shows a summary of the properties of these complex structures. This is to be compared to complex structures in Minkowski spacetime, see Table~\ref{table:_properties_J_r} in Appendix~\ref{class_KG_AdS:_sols_mink}.

We have identified the key requirements and desirable properties for complex structures on the AdS hypercylinder, surveyed the space of complex structures that satisfy much of these and identified candidates that are particularly interesting. It remains to understand, from a physical perspective, the differences between these candidates and their respective induced scattering theories. In this direction we only point out here the intimate relationship between complex structures and Feynman propagators \cite{Lic:propquantgr,Oe:feynobs}. Much remains to be done in order to reach a satisfactory understanding of quantum field theory in AdS spacetime.

%
%
\begin{acknowledgments}%

	The authors are grateful 
	to Daniele Colosi
	for many stimulating and critical discussions.
	This work was supported by CONACyT scholarship 213531
	and UNAM-DGAPA-PAPIIT project grant IN100212.
\end{acknowledgments}%
%

%
\begin{appendix}
%
%
\section{Klein-Gordon theory on Minkowski spacetime}%
\label{class_KG_AdS:_sols_mink}%

On Minkowski spacetime the solutions allowed on time-interval
and rod regions and their boundaries are similar to the AdS ones.
To the hypergeometric $S^a$-modes on AdS correspond the
Bessel modes on Minkowski, which are always regular on the time axis.
For expanding a solution on a Minkowski time-interval region
we use the Bessel modes
\begin{equation}
 \mu\hbs j _{plm_l}\artrOm
\defeq \tfracw{2p}{\ruut{2\piu}} \eu^{-\iu E_p t}
\spherharmonicar{m_l}l\Om\,	j_{l}\ar{pr} ,
\end{equation}
wherein $E_p \defeq \ruut{p^2\pn m^2}$
and thus here $E_p^2>m^2$.
These modes decay for large radius, and are well-defined
on all of Minkowski spacetime (like the Jacobi modes on AdS).
The expansion of a complexified 
solution $\ph$ on a Minkowski time-interval is then
\bal{\label{class_KG_AdS:_sols_mink_eqtime}
	\ph \artrOm 
		\eq \intpsumlm \biiglrr{ \ph^+_{p l m_l} \mu\hbs j_{plm_l}\artrOm
					+\coco{\ph^-_{p l m_l}}\,\coco{\mu\hbs j_{plm_l}\artrOm}
					}.
	}
The complex structure $J_t$ of a Minkowski equal-time plane
is the standard one \cite{ash_mag:_qf_curved_spacetimes}: 
\begin{equation}
(J_t\ph)^\pm_{p l m_l}=-\iu\, \ph^\pm_{p l m_l} .
\end{equation}
The induced real inner product is,
\bal{\label{zzz_AdS_invar_complex_isometries_2642_Mink_t_txg}
	g_{t} \biglrr{\et,\,\ze}
	& \defeq 2\omega_t \biglrr{\et,\,J_t\ze} \eq \intpsumlm 2E_p\, \biiiglrr{
				 \coco{\et^-_{p l m_l}}\, \ze^+_{p l m_l}\,
				+ \et^+_{p l m_l}\,\coco{\ze^-_{p l m_l}}	
				} \;.
	}
For expanding a solution on a Minkowski rod region
we need modes that are regular on the time axis.
However, we do not need them regular on all of space,
and hence we can now allow also the Bessel modes with $E^2<m^2$,
which grow exponentially for large radius.
Here we define the Bessel modes as
$\mu\hbs a _{Elm_l}\artrOm
\defeq \fracws{{p\hreals_E }}{4\piu} \eu^{-\iu Et}
\spherharmonicar{m_l}l\Om\,	\check \jmath_{E l}\ar r$
and Neumann modes as $	\mu\hbs b _{Elm_l}\artrOm
\defeq \fracws{{p\hreals_E }}{4\piu}\eu^{-\iu Et}
\spherharmonicar{m_l}l\Om\,	\check n_{E l}\ar r$, wherein
$p\hreals_E\defeq\ruutabs{E^2\mn m^2}$ 
and the radial functions are
\bals{\check \jmath_{E l}\ar{r} & \eq
		\bcas{\spherbesselar l{{p\hreals_E }r} &\quad E^2>m^2
				\\
			\iu^{-l}\,\spherbesselar l{\iu {p\hreals_E }r} & \quad E^2<m^2
			}
		&
	\check n_{E l}\ar{r} & \eq
		\bcas{\spherneumannar l{{p\hreals_E }r} & \quad E^2>m^2
				\\
			\iu^{l\pn1}\,\spherneumannar l{\iu{p\hreals_E }r}&\quad E^2<m^2
			}.
	}
The expansion of a complexified
solution $\ph$ on a Minkowski rod is then
\bal{\ph \artrOm & \eq \intesumlm 
		\ph^a_{E l m_l} \mu\hbs a _{Elm_l}\artrOm.	
	}
For a solution near the boundary of a Minkowski rod
we can further allow modes that diverge on the time axis.
These are the above Neumann modes. Like the Bessel modes,
at large radius they decay for $E^2>m^2$ and grow exponentially
for $E^2<m^2$. The expansion of a complexified 
solution $\ph$ near a Minkowski rod's boundary is then
\bal{\label{class_KG_AdS:_sols_mink_hypcyl}
	\ph \artrOm & \eq \intesumlm \biiglrr{ 
			\ph^a_{E l m_l} \mu\hbs a _{Elm_l}\artrOm
			+ \ph^b_{E l m_l} \mu\hbs b _{Elm_l}\artrOm
			}.	
	}
For the complex structure $J_r$ on the Minkowski hypercylinder
we consider two choices. The first choice $J_r\htx{pos}$
acts as follows:
\begin{equation}
(J_r\htx{pos}\ph)^{a}_{\om l m_l}=- \ph^{b}_{\om l m_l}\quad\text{and}\quad
 (J_r\htx{pos}\ph)^{b}_{\om l m_l}=\ph^{a}_{\om l m_l}
\label{eq:Jrpos}
\end{equation}
In the GBF, this complex structure, restricted to propagating modes, has first been used implicitly
in the construction of the vacuum state in \cite{Oe:kgtl}.
Later in \cite{Oe:holomorphic}, $J_r\htx{pos}$ is given explicitly
in an equivalent form for complex linear combinations of
the modes used here. The advantage of $J_r\htx{pos}$
is that it induces a positive definite real inner product $g_r$
\emph{for all} frequencies $\om\in\reals$.
However, we later found that $J_r\htx{pos}$ does not commute
with the Minkowski boosts for $\abs{\om}<m$.
The second choice is a complex structure $J_r\htx{iso}$ that does commute with the boosts. It
acts as
\begin{equation}
(J_r\htx{iso}\ph)^{a}_{\om l m_l}=-(\pm1)^l \ph^{b}_{\om l m_l}\quad\text{and}\quad
(J_r\htx{iso}\ph)^{b}_{\om l m_l}=(\pm1)^l \ph^{a}_{\om l m_l},
\label{eq:Jriso}
\end{equation}
wherein the $+$ holds for $\abs{\om}>m$
and $-$ holds for $\abs{\om}<m$. 
In particular, both complex structures agree for propagating modes, i.e., for $\abs{\om}>m$. More details about this are to be published elsewhere.
The drawback of $J_r\htx{iso}$ is that the induced $g_r$
is negative definite for modes with $\abs\om<m$ for odd $l$,
see Table \ref{table:_properties_J_r}.
Setting $j_l \equiv 1$ for $J_r\htx{pos}$
while $j_l=(\pm1)^l$ for $J_r\htx{iso}$, we can write
the action of $J_r$ as
\bal{\label{zzz_Mink_J_r_124}
	(J_r\ph)\artrOm  
	\eq \intlim0\infty \dif\om \sumliml{l m_l}
			\fracw{p\hreals_\om}{(4\piu)} j_l
	& \biiglrc{- \ph^{b}_{\om l m_l}\;
			\mu\hbs{a}_{\om l m_l}\!\artroOm
			+\ph^{a}_{\om l m_l}\;
			\mu\hbs{b}_{\om l m_l}\!\artroOm
			}.
	}
The induced real inner product then becomes
\bal{\label{zzz_AdS_invar_complex_isometries_2902_Mink_txg_r}
	g_r(\et,\ze)
	& \eq \intesumlm \fracw{p\hreals_E}{8\piu} j_l\,
		 \biiiglrc{\et^a_{E l m_l}\ze^a_{\mn E,l,\mn m_l}\,
				+\et^b_{E l m_l}\ze^b_{\mn E,l,\mn m_l}
				}	\;.
	}
For $J_r\htx{pos}$, an equivalent $g_r$ is given in
\cite{Oe:holomorphic} for complex linear combinations of
the modes used here.
\setlength{\extrarowheight}{3pt}
\begin{table}[H]
	\centering
	\begin{tabular}{|c|c c|}
		\hline
		{\bfseries Minkowski} & $J_r\htx{iso}$ & $J_r\htx{pos}$
			\\
		\hhline{|=|==|}
		commute with time-translations & \checkmark & \checkmark 
			\\
		commute with spatial rotations & \checkmark & \checkmark 
			\\
		commute with boosts & \checkmark & - 
			\\
		\hline
		strong amplitude equivalence & \checkmark & \checkmark
			\\
		induced real inner product $g_r$ 
			& \;indefinite\; &  positive-definite
			\\
		\hline
	\end{tabular}
	\caption{Properties of complex structures 
					for Minkowski hypercylinder}%
	\label{table:_properties_J_r}%
\end{table}
%

%
%
\section{Flat limits of classical KG solutions on AdS}%
\label{class_KG_AdS:_flatlim_sols}%

After the rescaled coordinates $r=\ro \RAdS$ and $t=\ta\RAdS$
we now introduce the rescaled frequency $\tiom=\om/\RAdS$
and momentum $\tip\hreals_\tiom=p\hreals_\om/\RAdS$,
wherein $p\hreals_\om = \ruutabs{\om^2-m^2\RAdS^2}$
and thus $\tip\hreals_\tiom = \ruutabs{\tiom^2-m^2}$.
As derived in \cite{Doh:classads}, for $d=3$
we have the following flat limits of the radial functions:
\bal{\label{zzz_AdS_large_R_limit_50}
	\tfrac{(p\hreals_\om)^l}{(2l\pn d\mn2)!!}\, S^a_{\om l}\ar \ro \toflatw
	& \;\check \jmath_{\tiom l}\ar{r}
		&
	\tfrac{(2l\pn d\mn4)!!}{(p\hreals_\om)^{l\pn d\mn2}}\,S^b_{\om l}\ar \ro \toflatw
	& \;\check n_{\tiom l}\ar{r} \;.
	}
Since $\Jacar+nl \ro$ is a special case of $S^a_{\om l}$
its flat limit for $d=3$ is
\bal{\label{zzz_AdS_large_R_limit_51_5}
	\tfrac{(p\hreals_\om)^l}{(2l\pn d\mn2)!!}\, \Jacar+ nl \ro \toflat
	& \; \spherbesselar l{\tip_\tiom r} \;.
	}
Further, the discrete sum over $n$ for Jacobi modes becomes
for any function $f$
\bal{\label{zzz_AdS_large_R_limit_51_57}
	\sumlim{n=0}\infty f(\omag+nl) \toflatw
	\tfracw{\RAdS}2 \intlim 0\infty \!\dif\tip\, \fracws{\tip}{\tiom_\tip}
			f(\RAdS\tiom_\tip),
	}
while with $\RAdS^{-1}\ti\ph_{\tiom\vc l m_l} 
\eq \ph_{\om \vc l m_l}$ the frequency integral becomes
\bals{\phar{t,\ro,\Om} 
		\eq \intomsumvclm \biiglrr{\ph_{\om \vc l m_l}\,
		f\ar{t,\ro,\Om,\om,\vc l, m_l} + \comcon}
		\eq \inttiomsumvclm \biiglrr{\ti\ph_{\tiom\vc l m_l}\,
		f\ar{\tau/\RAdS,\,r/\RAdS,\Om,\tiom\RAdS,\vc l,m_l} + \comcon}.
	}
Rescaling the momentum representation 
(where the label F stands for flat,
and $\tip$ short for $\tip_{nl}$)
\bal{\label{class_flat_eqtim_4734}
	\ph^{\pm}_{n \vc l m_l}
	& 	\eq \ph^{\txF,\pm}_{n \vc l m_l}\,
		\tfrac{4\omag+nl}{\ruut{2\piu}\RAdS}
		\tfracw{(p\hreals_{nl})^l}{(2l\pn d\mn2)!!}
		&
	\ph^{\txF,\pm}_{n \vc l m_l}
	& \eq \RAdS^{-1}\,\tiph^{\txF,\pm}_{\tip \vc l m_l},
	}
gives the flat Jacobi expansion of an AdS solution on a time-interval
\bal{\label{AdS_KG_solutions_202_flat}
	\ph \artroOm & \eq \sumliml{n \vc l m_l}
		\tfrac{4\omag+nl}{\ruut{2\piu}\RAdS}
		\tfracw{(p\hreals_{nl})^l}{(2l\pn d\mn2)!!}
 		\biiglrr{\ph^{\txF,+}_{n \vc l m_l}\;
			\mu\hbs+_{n\vc l m_l}\artroOm
			+ \coco{\ph^{\txF,-}_{n \vc l m_l}}\;
			\coco{\mu\hbs+_{n\vc l m_l}\artroOm}\,
			} \;.
	}
With the above ingredients for $d=3$ we then find that
the flat limit of the flat Jacobi expansion
becomes the Minkowski time-interval expansion
\eqref{class_KG_AdS:_sols_mink_eqtime}: 
\bal{\label{AdS_KG_solutions_slice_flatlim}
	\ph \artrOm \toflatw
		& \intlim{0}{\infty}\,\vol{\tip}\!\sumliml{l,m_l}\,
		\tfrac{2\tip}{2\piu}\,\spherbesselar l{\tip r}
		\biiglrr{\tiph^{\txF,+}_{\tip l m_l}\,\eu^{-\iu \tiom_\tip \ta}\,
			\spherharmonicar{m_l}l\Om
			+ \coco{\tiph^{\txF,-}_{\tip l m_l}}\,
			\eu^{\iu \tiom_\tip \ta}\,
			\coco{\spherharmonicar{m_l}l\Om}
			}\;.
	}
We can also rescale the momentum representation
of solutions on AdS rods respectively near its boundary:
\bal{\label{AdS_KG_solutions_tube_flat_rep}
	\ph^{S,a}_{\om \vcl m_l}
	& \eq \ph^{\txF,a}_{\om \vc l m_l}\,
		\tfracw{\tip\hreals_\tiom}{(4\piu)}
		\tfracw{(p\hreals_\om)^l}{(2l\pn d\mn2)!!}
		&
	\ph^{\txF,a}_{\om \vc l m_l}
	& \eq \RAdS^{\mn1} \tiph^{\txF,a}_{\tiom \vc l m_l}
		\\
	\ph^{S,b}_{\om \vcl m_l} 
	& \eq \ph^{\txF,b}_{\om \vc l m_l}\,
		\tfracw{\tip\hreals_\tiom}{(4\piu)}
		\tfracw{(2l\pn d\mn4)!!}{(p\hreals_\om)^{l\pn d\mn2}}
		&
	\ph^{\txF,b}_{\om \vc l m_l}
	& \eq \RAdS^{\mn1} \tiph^{\txF,b}_{\tiom \vc l m_l}.
		\notag
	}
which gives us the flat $S$-expansion of a solution near an
AdS rod's boundary:
\bal{\label{kg_ads_solutions_520_flat}
	\ph \artrOm  & \eq \sma{\intomsumvclm}
		\tfracw{\tip\hreals_\tiom}{(4\piu)}
		\biiglrr{\ph^{\txF,a}_{\om \vc l m_l}\,
		\tfracw{(p\hreals_\om)^l}{(2l\pn d\mn2)!!}\;
			\mu\hbs{S,a}_{\om \vc l m_l}\artroOm
			+\ph^{\txF,b}_{\om \vc l m_l}\,
		\tfracw{(2l\pn d\mn4)!!}{(p\hreals_\om)^{l\pn d\mn2)}}
			\mu\hbs{S,b}_{\om \vc l m_l}\artroOm
			}.
	}
Again, for $d=3$ we then find that
the flat limit of the flat $S$-expansion
becomes the expansion \eqref{class_KG_AdS:_sols_mink_hypcyl}
of a solution near a Minkowski rod's boundary: 
\bal{\label{zzz_AdS_completeness_solutions_280}
	\ph\artrOm \toflatw
	& \sma{\int\!\vol\tiom\sumliml{l,m_l}} \fracws{\tip\hreals_\tiom}{4\piu}
		\biiglrr{ \tiph^{\txF,a}_{\tiom l m_l} \eu^{-\iu \tiom\ta}
			\spherharmonicar{m_l}{l}\Om\,	\check \jmath_{\tiom l}\ar r
			+  \tiph^{\txF,b}_{\tiom l m_l} \eu^{-\iu \tiom\ta}
			\spherharmonicar{m_l}{l}\Om\,	\check n_{\tiom l}\ar r
			}\;.
	}
This completes the correspondence through the flat limit
between solutions on AdS and Minkowski regions.
%
%
\section{Isometry actions on AdS Klein-Gordon solutions}%
\label{class_KG_AdS:_iso_actions}%

In order to make the complex structure commute with
the actions of isometry generators on the solutions,
we use the actions computed in \cite{Doh:classads}.
Let $k_{\De t}$ denote a finite time translation,
$\ph$ a solution, and $k_{\De t}\actson\ph$ denotes
the new solution obtained by $k_{\De t}$ acting on $\ph$.
For a solution $\ph$ on an AdS time-interval,
the new momentum representation then becomes
\bal{\label{zzz_AdS_isometries_time_81}
	\biglrr{k_{\De t}\actson\ph}^+_{n\vc l m_l} 
	& \eq \eu^{\iu\omag\pm nl \De t}\;\ph^+_{n\vc l m_l}
		&
	\coco{\biglrr{k_{\De t}\actson\ph}^-_{n\vc l m_l}} 
	& \eq \eu^{\mn\iu\omag\pm nl \De t}\;\coco{\ph^-_{n\vc l m_l}}\;.
	}
For a solution $\ph$ near the boundary of an AdS rod,
the new momentum representation becomes
\bal{\label{zzz_AdS_isometries_time_153}
	\biglrr{k_{\De t}\actson\ph}^{S,a}_{\om\vc l m_l} 
	&\eq \eu^{\iu\om\De t}\,\ph^{S,a}_{\om\vc l m_l}
		&
	\biglrr{k_{\De t}\actson\ph}^{S,b}_{\om\vc l m_l} 
	&\eq \eu^{\iu\om\De t}\,\ph^{S,b}_{\om\vc l m_l}.
	}
Let now $\opR\arvc\al$ denote a finite rotation
about angles $\vc\al$ with respect to some fixed set of axes,
and let $\opR\arvc\al\actson\ph$ the solution
obtained by letting it act on $\ph$.
For a solution $\ph$ on an AdS time-interval,
the momentum representation then becomes
(with $D$ denoting elements of Wigner's $D$-matrix)
\bal{\label{zzz_AdS_isometries_symplectic_rot_100}
	\biglrr{\opR\arvc\al\actson\ph}^+_{nl\,\tivc l\, m_l}
	&\eqn \sumliml{\tivc l',\,m'_l}	\ph^+_{nl\,\tivc l'\, m'_l}
		\bglrr{D^l_{\tivc l,\tivc l'}\arvc \al}_{m_l m'_l}
		&
	\coco{\biglrr{\opR\arvc\al\actson\ph}^-_{nl\,\tivc l\, m_l}\,}
	&\eqn \sumliml{\tivc l',\,m'_l}	\coco{\ph^-_{nl\,\tivc l'\, m'_l}\,}\,
		\coco{\bglrr{D^l_{\tivc l,\tivc l'}\arvc \al}_{m_l m'_l}}\;,
	}
For a solution $\ph$ near the boundary of an AdS rod,
the momentum representation becomes
\bal{\label{zzz_AdS_isometries_symplectic_rot_515}
	\biglrr{\opR\arvc\al\actson\ph}^{S,a}_{\om l\,\tivc l\, m_l}
	&\eqn \sumliml{\tivc l',\,m'_l}	\ph^{S,a}_{\om l\,\tivc l'\, m'_l}
		\bglrr{D^l_{\tivc l,\tivc l'}\arvc \al}_{m_l m'_l}
		&
	\biglrr{\opR\arvc\al\actson\ph}^{S,b}_{\om l\,\tivc l\, m_l}
	&\eqn \sumliml{\tivc l',\,m'_l}	\ph^{S,b}_{\om l\,\tivc l'\, m'_l}
		\bglrr{D^l_{\tivc l,\tivc l'}\arvc \al}_{m_l m'_l}
	}
We can only give the actions for infinitesimal boosts.
The action of the boost generator $K\ls{0d}$
on a solution on an AdS time-interval writes
\balsplit{\label{zzz_AdS_isometries_boosts_171}
	\biglrr{K\ls{0d}\actson\ph}^+\ls{n\vc l m\ls l} \!
	& \eqn \tfrac{\iu}{2}
		z\hs{(\pn)0-}\ls{n,l\pn1,\til}\,\ph^+\ls{n,l\pn1,\tivc l,m\ls l}
		\pn \tfrac{\iu}{2}
		z\hs{(\pn)-+}\ls{n\pn1,l\mn1,\til}\,\ph^+\ls{n\pn1,l\mn1,\tivc l,m\ls l}
		\pn \tfrac{\iu}{2}
		\tiz\hs{(\pn)+-}\ls{n\mn1,l\pn1,\til}\,\ph^+\ls{n\mn1,l\pn1,\tivc l,m\ls l}
		\pn \tfrac{\iu}{2}
		\tiz\hs{(\pn)0+}\ls{n,l\mn1,\til}\,\ph^+\ls{n,l\mn1,\tivc l,m\ls l}
		\\
	\coco{\biglrr{K\ls{0d}\actson\ph}^-\ls{n\vc l m\ls l}}\!
	& \eqn \mn \tfrac{\iu}{2}
		z\hs{(\pn)0-}\ls{n,l\pn1,\til}\,\coco{\ph^-\ls{n,l\pn1,\tivc l,m\ls l}}
		\mn \tfrac{\iu}{2}
		z\hs{(\pn)-+}\ls{n\pn1,l\mn1,\til}\,\coco{\ph^-\ls{n\pn1,l\mn1,\tivc l,m\ls l}}
		\mn \tfrac{\iu}{2}
		\tiz\hs{(\pn)+-}\ls{n\mn1,l\pn1,\til}\,\coco{\ph^-\ls{n\mn1,l\pn1,\tivc l,m\ls l}}
		\mn \tfrac{\iu}{2}
		\tiz\hs{(\pn)0+}\ls{n,l\mn1,\til}\,\coco{\ph^-\ls{n,l\mn1,\tivc l,m\ls l}}.
	}
For a solution near an AdS rod's boundary,
the action of the boost generator $K\ls{0d}$ is
\balsplit{\label{zzz_AdS_isometries_boosts_5201}
	\biglrr{K\ls{0d}\actson\ph}^{S,a}_{\om\vc l m_l}
	& \!\eqn \tfrac{\iu}2
			\tiz\hs{(S,a)+-}\ls{\om\mn1,l\pn1,\til}\,\ph^{S,a}\ls{\om\mn1,l\pn1,\tivc l,m_l}
		\pn \tfrac{\iu}2
			\tiz\hs{(S,a)++}\ls{\om\mn1,l\mn1,\til}\,\ph^{S,a}\ls{\om\mn1,l\mn1,\tivc l,m_l}
		\pn \tfrac{\iu}2
			z\hs{(S,a)--}\ls{\om\pn1,l\pn1,\til}\,\ph^{S,a}\ls{\om\pn1,l\pn1,\tivc l,m_l}
		\pn \tfrac{\iu}2
			z\hs{(S,a)-+}\ls{\om\pn1,l\mn1,\til}\,\ph^{S,a}\ls{\om\pn1,l\mn1,\tivc l,m_l}
		\\
	\biglrr{K\ls{0d}\actson\ph}^{S,b}_{\om\vc l m_l}
	& \!\eqn	\tfrac{\iu}2
			\tiz\hs{(S,b)+-}\ls{\om\mn1,l\pn1,\til}\,\ph^{S,b}\ls{\om\mn1,l\pn1,\tivc l,m_l}
		\pn \tfrac{\iu}2
			\tiz\hs{(S,b)++}\ls{\om\mn1,l\mn1,\til}\,\ph^{S,b}\ls{\om\mn1,l\mn1,\tivc l,m_l}
		\pn \tfrac{\iu}2
			z\hs{(S,b)--}\ls{\om\pn1,l\pn1,\til}\,\ph^{S,b}\ls{\om\pn1,l\pn1,\tivc l,m_l}
		\pn \tfrac{\iu}2
			z\hs{(S,b)-+}\ls{\om\pn1,l\mn1,\til}\,\ph^{S,b}\ls{\om\pn1,l\mn1,\tivc l,m_l}.
	}
These actions are already sufficient for our purposes, because
the actions of $K\ls{d\pn1,d}$ are the same up to the $\iu$'s
becoming $\pm1$, and the actions of the other boosts
come from commutators of these boosts and rotations,
see again \cite{Doh:classads} for details.
The $z$-factors of the boosts write for the time-interval
\bal{	\label{zzz_AdS_recurrence_jacobi_99}
	\tiz\hs{(\pn)+-}_{nl\til} & \eq +(2l\pn d\mn2)\,
		\ki\hb{d\mn1}_-\ar{l,\til}
		&
	\tiz\hs{(\pn)0+}_{nl\til} & \eq -2(n\pn l\pn\Timp)
		\fracw{(n\pn l\pn \tfrac d2)}{(l\pn \tfrac d2)}
		\, \ki\hb{d\mn1}_+\ar{l,\til}
		\\
	z\hs{(\pn)0-}_{nl\til} & \eq -(2l\pn d\mn2)\, \ki\hb{d\mn1}_-\ar{l,\til}
		&
	z\hs{(\pn)-+}_{nl\til} & \eq +2n\fracw{(n\pn\nu)}{(l\pn \fracsss d2)}
		\, \ki\hb{d\mn1}_+\ar{l,\til} \;,
		\notag
	}
and for the rod's boundary
(the hyperspherical $\ki$-factors are also given in
\cite{Doh:classads}, however, their inner form 
is not needed for the computations of the complex structure).
\bal{	\label{zzz_AdS_recurrence_hypergeo_99}
	\tiz\hs{(\!S,a\!) \pn\,\mn}_{\om l \til} 
	& \eq \eq (2l\pn d\mn2)\, \ki\hb{d\mn1}_-\ar{l,\til}
		&
	\tiz\hs{(\!S,a\!) \pn\,\pn}_{\om l \til} 
	& \eq (\timp\mn\om\mn l\mn d)
		\fracw{(\timp\pn\om\pn l)}{(2l\pn d)} \ki\hb{d\mn1}_+\ar{l,\til}
		\\ 
	\tiz\hs{(\!S,b\!)\pn\,\mn}_{\om l \til} 
	& \eq -(\timp\mn\om\pn l\mn2)
		\fracw{(\timp\pn\om\mn l\mn d\pn2)}{(2l\pn d\mn4)}
		\ki\hb{d\mn1}_-\ar{l,\til}
		&
	\tiz\hs{(\!S,b\!)\pn\,\pn}_{\om l \til} 
	& \eq -(2l\pn d\mn2)\, \ki\hb{d\mn1}_+\ar{l,\til}
		\notag
		\\ 
	z\hs{(\!S,a\!) \mn\,\mn}_{\om l \til} 
	& \eq -(2l\pn d\mn2)\, \ki\hb{d\mn1}_-\ar{l,\til}
		&
	z\hs{(\!S,a\!) \mn\,\pn}_{\om l \til} 
	& \eq -(\timp\pn\om\mn l\mn d)\fracw{(\timp\mn\om\pn l)}{(2l\pn d)} 
		\ki\hb{d\mn1}_+\ar{l,\til}
		\notag
		\\ 
	z\hs{(\!S,b\!) \mn\,\mn}_{\om l \til} 
	& \eq (\timp\pn\om\pn l\mn2)
		\fracw{(\timp\mn\om\mn l\mn d\pn2)}{(2l\pn d\mn4)}
		\ki\hb{d\mn1}_-\ar{l,\til}
		&
	z\hs{(\!S,b\!) \mn\,\pn}_{\om l \til} 
	& \eq (2l\pn d\mn2)\, \ki\hb{d\mn1}_+\ar{l,\til} \;.
		\notag
	}
%
%
\section{Fixing $J_\ro$ for AdS: essentials and isometries}
\label{zzz_complex_struct_AdS_KG:_essential_isometry}

We use the $S$-expansion \eqref{kg_ads_solutions_520}
for solutions near an AdS hypercylinder $\Si_\ro$.
$J_\ro$ must map solutions to solutions,
and since all solutions can be expanded in this way,
the new solution $J_\ro\ph$ must write as an $S$-expansion, too:
\bal{\label{complex_struct_AdS_KG:_essential_time_rot:_5096_J_phi}
	\biglrr{J_\ro\ph} \artrOm & \eq \intomsumvclm
		\biiglrr{\biglrr{J_\ro\ph}^{S,a}_{\om \vc l m_l}\,
			\mu\hb{S,a}_{\om\vc l m_l}\artroOm
			+ \biglrr{J_\ro\ph}^{S,b}_{\om \vc l m_l}\,
			\mu\hb{S,b}_{\om\vc l m_l}\artroOm.
			}
		}
That is, we express the action of the complex structure
on a solution completely in momentum space as in
$J_\ro\maps\ph^{S,a}_{\om \vc l m_l}
\to (J_\ro\ph)^{S,a}_{\om \vc l m_l}$.
Requiring only $\reals$-linearity for real solutions,
the most general form of $J_\ro$ is
\bals{\biglrr{J_\ro\ph}^{S,a}_{\om \vc l m_l}
	\eqn \intn\dif\!{\om'}\sumliml{\vc l', m'_l}
		& \biiiglrr{\;
			j^{S,aa}\biiglrr{\,\!^{\om\; \vc l\; m_l}_{\om' \vcl' m'_l}}\,
			\ph^{S,a}_{\om' \vc l' m'_l}
			+ j^{S,ab}\biiglrr{\,\!^{\om\; \vc l\; m_l}_{\om' \vcl' m'_l}}\,
			\ph^{S,b}_{\om' \vc l' m'_l}
			+ \tij^{S,aa}\biiglrr{\,\!^{\om\; \vc l\; m_l}_{\om' \vcl' m'_l}}\,
			\coco{\ph^{S,a}_{\om' \vc l' m'_l}}
			+ \tij^{S,ab}\biiglrr{\,\!^{\om\; \vc l\; m_l}_{\om' \vcl' m'_l}}\,
			\coco{\ph^{S,b}_{\om' \vc l' m'_l}}\;
			}
		\\
	\biglrr{J_\ro\ph}^{S,b}_{\om \vc l m_l}
	\eqn \intn\dif\!{\om'}\sumliml{\vc l', m'_l}
		& \biiiglrr{\;
			j^{S,ba}\biiglrr{\,\!^{\om\; \vc l\; m_l}_{\om' \vcl' m'_l}}\,
			\ph^{S,a}_{\om' \vc l' m'_l}
			+ j^{S,bb}\biiglrr{\,\!^{\om\; \vc l\; m_l}_{\om' \vcl' m'_l}}\,
			\ph^{S,b}_{\om' \vc l' m'_l}
			+ \tij^{S,ba}\biiglrr{\,\!^{\om\; \vc l\; m_l}_{\om' \vcl' m'_l}}\,
			\coco{\ph^{S,a}_{\om' \vc l' m'_l}}
			+ \tij^{S,bb}\biiglrr{\,\!^{\om\; \vc l\; m_l}_{\om' \vcl' m'_l}}\,
			\coco{\ph^{S,b}_{\om' \vc l' m'_l}}\;
			}.
	}
Here, the integral kernels $j^{S,\cdotw\cdotw}$ are complex functions
of two sets of momenta, and completely determine the complex structure.
It turned out more effective to first require $J_\ro$
to commute with time translation and rotations, 
and to impose the essential properties only after doing this.
Imposing thus commutation with finite time translations $k_{\De t}$
as in $J_\ro(k_{\De t}\actson\ph)=k_{\De t}\actson(J_\ro\ph)$,
and rotations $\opR\arvc\al$ as in
$J_\ro(\opR\arvc\al\actson\ph)=\opR\arvc\al\actson(J_\ro\ph)$,
using actions \eqref{zzz_AdS_isometries_time_153}
respectively \eqref{zzz_AdS_isometries_symplectic_rot_515},
we can get rid of the integrals above.
After this straightforward calculation
we obtain a much simpler form of $J_\ro$:
\bal{\label{zzz_AdS_invar_complex_isometries_genJ_tube_571_J}
	\biglrr{J_\ro\ph}^{S,a}_{\om \vc l m_l}
	& \eq j^{S,aa}_{\om l}\,\ph^{S,a}_{\om \vc l m_l}
			+ j^{S,ab}_{\om l}\,\ph^{S,b}_{\om \vc l m_l}
		&
	\biglrr{J_\ro\ph}^{S,b}_{\om \vc l m_l}
	& \eq j^{S,ba}_{\om l}\,\ph^{S,a}_{\om \vc l m_l}
			+ j^{S,bb}_{\om l}\,\ph^{S,b}_{\om \vc l m_l}.
	}
This is $\complex$-linear for complex solutions,
which thus we could have imposed right from the start.
The complex structure is now determined by the complex
matrix elements $j^{S,\cdot\cdot}_{\om l}$,
which only depend on frequency $\om$ and total angular momentum $l$.
We recall that real solutions $\ph$ are those with
$\ph^{S,a}_{\mn\om,\vcl,\mn m_l} \eq \coco{\ph^{S,a}_{\om \vcl m_l} }$
while 
$\ph^{S,b}_{\mn\om,\vcl,\mn m_l} \eq \coco{\ph^{S,b}_{\om \vcl m_l} }$.
Plugging form \eqref{zzz_AdS_invar_complex_isometries_genJ_tube_571_J}
into expansion
\eqref{complex_struct_AdS_KG:_essential_time_rot:_5096_J_phi},
we can read off, that the condition that J must turn real solutions into real solutions becomes:
\bal{\label{zzz_AdS_invar_complex_isometries_Jgen_8}
	j^{S,aa}_{\mn\om,l}
	& \eqn \cocow{j^{S,aa}_{\om l}}
		&
	j^{S,ab}_{\mn\om,l}
	& \eqn \cocow{j^{S,ab}_{\om l}}
		&
	j^{S,ba}_{\mn\om,l}
	& \eqn \cocow{j^{S,ba}_{\om l}}
		&
	j^{S,bb}_{\mn\om,l}
	& \eqn \cocow{j^{S,bb}_{\om l}}.
	}
It is also straightforward to check that
$J^2 \eq \mn\One$ translates into the conditions
$
	\mn 1 
	\eqn (j^{S,aa}_{\om l})^2\!
		+ j^{S,ab}_{\om l}\, j^{S,ba}_{\om l}
	\eqn (j^{S,bb}_{\om l})^2\!
		+ j^{S,ab}_{\om l}\, j^{S,ba}_{\om l}
$
and 
$	0  \eqn j^{S,ab}_{\om l}
		\biiglrr{j^{S,aa}_{\om l}\! + j^{S,bb}_{\om l} }
	\eqn j^{S,ba}_{\om l}
		\biglrr{j^{S,aa}_{\om l}\! + j^{S,bb}_{\om l} }
$.
Requiring compatibility with the symplectic structure
\eqref{zzz_AdS_structures_421} implies the conditions
%
$	1 
	\eqn j^{S,aa}_{\om l}\, \cocow{j^{S,bb}_{\om l} }
		- j^{S,ba}_{\om l}\, \cocow{j^{S,ab}_{\om l} }
$ and
$j^{S,aa}_{\om l}\,\cocow{j^{S,ba}_{\om l} }
\eqn \cocow{j^{S,aa}_{\om l} }\,j^{S,ba}_{\om l}
$ with
$j^{S,bb}_{\om l}\,\cocow{j^{S,ab}_{\om l} }
\eqn \cocow{j^{S,bb}_{\om l} }\,j^{S,ab}_{\om l}
$.
We further want that the real inner product $g_\ro$
induced by $J_\ro$ does not vanish for any single mode:
$g_\ro(\mu^{S,a}_{\om\vcl m_l},\mu^{S,a}_{\om\vcl m_l})\neq0$ and 
$g_\ro(\mu^{S,b}_{\om\vcl m_l},\mu^{S,b}_{\om\vcl m_l})\neq0$,
which implies $j^{S,ab}_{\om l}\neq0\neq j^{S,ba}_{\om l}$.
This implies $j^{S,bb}_{\om l}\eq-j^{S,aa}_{\om l}$.
Combining all above conditions implies that all 
$j^{S,\cdotw\cdotw}_{\om l}$
must be real and frequency-symmetric:
$j^{S,\cdotw\cdotw}_{\om l}=j^{S,\cdotw\cdotw}_{\mn\om, l}$.
This establishes the following complex structure:
\bal{\label{zzz_AdS_invar_complex_isometries_Jgen_X13}
	\biglrr{J_\ro\ph}^{S,a}_{\om \vc l m_l}
	& \eqn j^{S,aa}_{\om l}\,\ph^{S,a}_{\om \vc l m_l}
			\pn j^{S,ab}_{\om l}\,\ph^{S,b}_{\om \vc l m_l}
		&	
	\biglrr{J_\ro\ph}^{S,b}_{\om \vc l m_l}
	& \eqn j^{S,ba}_{\om l}\,\ph^{S,a}_{\om \vc l m_l}
			\mn j^{S,aa}_{\om l}\,\ph^{S,b}_{\om \vc l m_l}
		&
	\biglrr{j^{S,aa}_{\om l}}^2
	& \eqn - j^{S,ab}_{\om l} j^{S,ba}_{\om l} \mn 1.
	}
%
Next, we impose also commutation with the boost
actions \eqref{zzz_AdS_isometries_boosts_5201}.
First we compare the pair of actions
\bals{\biiglrr{J_\ro \biglrr{K_{0d}\actson\ph} }^{S,a}_{\om\vc l m_l}
	& \!= j^{S,aa}_{\om l}
		\biiiglrr{\!\tfrac{\iu}2 \tiz\hs{(S,a)+-}_{\om\mn1,l\pn1,\til}\,
			\ph^{S,a}_{\om\mn1,l\pn1,\tivc l,m_l}
			\!\pn \tfrac{\iu}2 \tiz\hs{(S,a)++}_{\om\mn1,l\mn1,\til}\,
			\ph^{S,a}_{\om\mn1,l\mn1,\tivc l,m_l}
			\!\pn \tfrac{\iu}2 z\hs{(S,a)--}_{\om\pn1,l\pn1,\til}\,
			\ph^{S,a}_{\om\pn1,l\pn1,\tivc l,m_l}
			\!\pn \tfrac{\iu}2 z\hs{(S,a)-+}_{\om\pn1,l\mn1,\til}\,
			\ph^{S,a}_{\om\pn1,l\mn1,\tivc l,m_l}
			}
		\notag
		\\
	& + j^{S,ab}_{\om l}
		\biiiglrr{\!\tfrac{\iu}2 \tiz\hs{(S,b)+-}_{\om\mn1,l\pn1,\til}\,
			\ph^{S,b}_{\om\mn1,l\pn1,\tivc l,m_l}
			\!\pn \tfrac{\iu}2 \tiz\hs{(S,b)++}_{\om\mn1,l\mn1,\til}\,
			\ph^{S,b}_{\om\mn1,l\mn1,\tivc l,m_l}
			\!\pn \tfrac{\iu}2 z\hs{(S,b)--}_{\om\pn1,l\pn1,\til}\,
			\ph^{S,b}_{\om\pn1,l\pn1,\tivc l,m_l}
			\!\pn \tfrac{\iu}2 z\hs{(S,b)-+}_{\om\pn1,l\mn1,\til}\,
			\ph^{S,b}_{\om\pn1,l\mn1,\tivc l,m_l}
			}
	}
\bals{\biiglrr{K_{0d}\actson\biglrr{J_\ro \ph} }^{S,a}_{\om\vc l m_l}
	& \eq \tfrac{\iu}2 j^{S,aa}_{\om\mn1, l\pn1}
			\tiz\hs{(S,a)+-}_{\om\mn1,l\pn1,\til}\,
			\ph^{S,a}_{\om\mn1,l\pn1,\tivc l,m_l}
		+ \tfrac{\iu}2 j^{S,aa}_{\om\mn1, l\mn1}
			\tiz\hs{(S,a)++}_{\om\mn1,l\mn1,\til}\,
			\ph^{S,a}_{\om\mn1,l\mn1,\tivc l,m_l}
		+\tfrac{\iu}2 j^{S,aa}_{\om\pn1, l\pn1}
			z\hs{(S,a)--}_{\om\pn1,l\pn1,\til}\,
			\ph^{S,a}_{\om\pn1,l\pn1,\tivc l,m_l}
		\notag
		\\
	&\quad 
		+\tfrac{\iu}2 j^{S,aa}_{\om\pn1, l\mn1}
			z\hs{(S,a)-+}_{\om\pn1,l\mn1,\til}\,
			\ph^{S,a}_{\om\pn1,l\mn1,\tivc l,m_l}
		+ \tfrac{\iu}2 j^{S,ab}_{\om\mn1, l\pn1}
			\tiz\hs{(S,a)+-}_{\om\mn1,l\pn1,\til}\,
			\ph^{S,b}_{\om\mn1,l\pn1,\tivc l,m_l}
		+ \tfrac{\iu}2 j^{S,ab}_{\om\mn1, l\mn1}
			\tiz\hs{(S,a)++}_{\om\mn1,l\mn1,\til}\,
			\ph^{S,b}_{\om\mn1,l\mn1,\tivc l,m_l}
		\notag
		\\
	&\quad 
		+\tfrac{\iu}2 j^{S,ab}_{\om\pn1, l\pn1}
			z\hs{(S,a)--}_{\om\pn1,l\pn1,\til}\,
			\ph^{S,b}_{\om\pn1,l\pn1,\tivc l,m_l}
		+\tfrac{\iu}2 j^{S,ab}_{\om\pn1, l\mn1}
			z\hs{(S,a)-+}_{\om\pn1,l\mn1,\til}\,
			\ph^{S,b}_{\om\pn1,l\mn1,\tivc l,m_l}
	}
We can read off that $J$ and $K_{0d}$ commute, 
if the following equalities become fulfilled:
\bal{\label{zzz_AdS_invar_complex_isometries_443_aa}
	j^{S,aa}_{\om l} \eqn j^{S,aa}_{\om\mn1,l\pn1} \eqn j^{S,aa}_{\om\mn1,l\mn1}
						\eqn j^{S,aa}_{\om\pn1,l\pn1} \eqn j^{S,aa}_{\om\pn1,l\mn1}
	}
together with (see \eqref{zzz_AdS_recurrence_hypergeo_99}
for the $z$-factors)
\bal{\label{zzz_AdS_invar_complex_isometries_443}
	j^{S,ab}_{\om\mn1,l\pn1}\, \tiz\hs{(S,a)+-}_{\om\mn1,l\pn1,\til}
	& \eq j^{S,ab}_{\om l}\,\tiz\hs{(S,b)+-}_{\om\mn1,l\pn1,\til}
		&
	j^{S,ab}_{\om\mn1,l\mn1}\, \tiz\hs{(S,a)++}_{\om\mn1,l\mn1,\til}
	& \eq j^{S,ab}_{\om l}\, \tiz\hs{(S,b)++}_{\om\mn1,l\mn1,\til}
		\\
	j^{S,ab}_{\om\pn1,l\pn1}\, z\hs{(S,a)--}_{\om\pn1,l\pn1,\til}
	& \eq j^{S,ab}_{\om l}\, z\hs{(S,b)--}_{\om\pn1,l\pn1,\til}
		&
	j^{S,ab}_{\om\pn1,l\mn1}\, z\hs{(S,a)-+}_{\om\pn1,l\mn1,\til}
	& \eq j^{S,ab}_{\om l}\, z\hs{(S,b)+-}_{\om\pn1,l\mn1,\til}
		\notag
	}
Comparing in the same way the pairs of actions
$\biglrr{J_\ro (K_{0d}\actson\ph) }^{S,b}_{\om\vc l m_l}
=\biglrr{K_{0d}\actson(J_\ro \ph) }^{S,b}_{\om\vc l m_l}$
and $	\biglrr{J_\ro (K_{d\pn1,d}\actson\ph) }^{S,a}_{\om\vc l m_l}
= \biglrr{K_{d\pn1,d}\actson(J_\ro \ph) }^{S,a}_{\om\vc l m_l}$
and $	\biglrr{J_\ro(K_{d\pn1,d}\actson\ph) }^{S,b}_{\om\vc l m_l}
=\biglrr{K_{d\pn1,d}\actson(J_\ro \ph) }^{S,b}_{\om\vc l m_l}$,
we get the additional conditions
\bal{\label{zzz_AdS_invar_complex_isometries_443_ba}
	j^{S,ba}_{\om\mn1,l\pn1}\, \tiz\hs{(S,b)+-}_{\om\mn1,l\pn1,\til}
	& \eq j^{S,ba}_{\om l}\,\tiz\hs{(S,a)+-}_{\om\mn1,l\pn1,\til}
		&
	j^{S,ba}_{\om\mn1,l\mn1}\, \tiz\hs{(S,b)++}_{\om\mn1,l\mn1,\til}
	& \eq j^{S,ba}_{\om l}\, \tiz\hs{(S,a)++}_{\om\mn1,l\mn1,\til}
		\\
	j^{S,ba}_{\om\pn1,l\pn1}\, z\hs{(S,b)--}_{\om\pn1,l\pn1,\til}
	& \eq j^{S,ba}_{\om l}\, z\hs{(S,a)--}_{\om\pn1,l\pn1,\til}
		&
	j^{S,ba}_{\om\pn1,l\mn1}\, z\hs{(S,b)-+}_{\om\pn1,l\mn1,\til}
	& \eq j^{S,ba}_{\om l}\, z\hs{(S,a)+-}_{\om\pn1,l\mn1,\til}.
		\notag
	}
This means, that if we can find any solution $j^{S,ab}$
of \eqref{zzz_AdS_invar_complex_isometries_443},
then setting $j^{S,ba} = -1/j^{S,ab}$ yields a solution
of \eqref{zzz_AdS_invar_complex_isometries_443_ba}.
Therefore it is sufficient to study the four conditions
\eqref{zzz_AdS_invar_complex_isometries_443_ba}.
Plugging the values of the $z$-factors
\eqref{zzz_AdS_recurrence_hypergeo_99} into them,
we see that shifting $^{\om\to\om\pn1}_{\,l\to\, l\mn1}$
in condition one reproduces the second,
and $^{\om\to\om\pn1}_{\,l\to\, l\pn1}$ in the third
reproduces the fourth.
Thus only two conditions remain which write
\bal{\label{zzz_AdS_invar_complex_isometries_681_m}
	j^{S,ba}_{\om\mn1,l\pn1}
	& \eqn -j^{S,ba}_{\om l}\,
		\fracwss{(2l\pn d)\,(2l\pn d\mn2)}
				{(\Timp \pn\om \mn l \mn d)\,(\Timp \mn\om \pn l)}
		&
	j^{S,ba}_{\om\pn1,l\pn1}
	& \eqn - j^{S,ba}_{\om l}\,
		\fracwss{(2l\pn d)\,(2l\pn d\mn2)}
				{(\Timp \mn\om \mn l \mn d)\,(\Timp \pn\om \pn l)}.
	}
Further, recalling $\biglrr{j^{S,aa}_{\om l}}^2
\eqn - j^{S,ab}_{\om l} j^{S,ba}_{\om l} \mn 1$
from \eqref{zzz_AdS_invar_complex_isometries_Jgen_X13},
we find that conditions \eqref{zzz_AdS_invar_complex_isometries_443_aa}
are met automatically, given that
\eqref{zzz_AdS_invar_complex_isometries_443} and
\eqref{zzz_AdS_invar_complex_isometries_443_ba} are fulfilled.
Hence \eqref{zzz_AdS_invar_complex_isometries_681_m}
indeed remain the only conditions to be solved.
%
%
%
\end{appendix}
%

%
\bibliographystyle{stdnodoi}
\bibliography{stdrefsb,zzzzz_references}

\begin{thebibliography}{10}
\providecommand{\url}[1]{\texttt{#1}}
\providecommand{\urlprefix}{URL }
\providecommand{\selectlanguage}[1]{\relax}
\providecommand{\eprint}[2][]{\url{#2}}

\bibitem{Gid:smatrixadscft}
S.~B. Giddings, \textit{Boundary S Matrix and the Anti-de~Sitter Space to
  Conformal Field Theory Dictionary}, Phys. Rev. Lett. \textbf{83} (1999)
  2707--2710, \eprint{hep-th/9903048}.

\bibitem{Oe:timelike}
R.~Oeckl, \textit{States on timelike hypersurfaces in quantum field theory},
  Phys. Lett. \textbf{B 622} (2005) 172--177, \eprint{hep-th/0505267}.

\bibitem{Oe:boundary}
R.~Oeckl, \textit{A ``general boundary'' formulation for quantum mechanics and
  quantum gravity}, Phys. Lett. \textbf{B 575} (2003) 318--324,
  \eprint{hep-th/0306025}.

\bibitem{Oe:GBQFT}
R.~Oeckl, \textit{General boundary quantum field theory: Foundations and
  probability interpretation}, Adv. Theor. Math. Phys. \textbf{12} (2008)
  319--352, \eprint{hep-th/0509122}.

\bibitem{Oe:kgtl}
R.~Oeckl, \textit{General boundary quantum field theory: Timelike hypersurfaces
  in Klein-Gordon theory}, Phys. Rev. \textbf{D 73} (2006) 065017,
  \eprint{hep-th/0509123}.

\bibitem{CoOe:spsmatrix}
D.~Colosi, R.~Oeckl, \textit{S-matrix at spatial infinity}, Phys. Lett.
  \textbf{B 665} (2008) 310--313, \eprint{0710.5203}.

\bibitem{CoOe:smatrixgbf}
D.~Colosi, R.~Oeckl, \textit{Spatially asymptotic S-matrix from general
  boundary formulation}, Phys. Rev. \textbf{D 78} (2008) 025020,
  \eprint{0802.2274}.

\bibitem{Col:desitterletter}
D.~Colosi, \textit{S-matrix in de~Sitter spacetime from general boundary
  quantum field theory}, \eprint{0910.2756}.

\bibitem{Col:desitterpaper}
D.~Colosi, \textit{General boundary quantum field theory in de~Sitter
  spacetime}, \eprint{1010.1209}.

\bibitem{CoDo:smatrixcsp}
D.~Colosi, M.~Dohse, \textit{On the structure of the S-matrix in general
  boundary quantum field theory in curved space}, \eprint{1011.2243}.

\bibitem{CDO:adsproc}
D.~Colosi, M.~Dohse, R.~Oeckl, \textit{S-Matrix for AdS from General Boundary
  QFT}, J. Phys.: Conf. Ser. \textbf{360} (2012) 012012, \eprint{1112.2225}.

\bibitem{Oe:holomorphic}
R.~Oeckl, \textit{Holomorphic Quantization of Linear Field Theory in the
  General Boundary Formulation}, SIGMA \textbf{8} (2012) 050, 31 pages,
  \eprint{1009.5615v3}.

\bibitem{Doh:classads}
M.~Dohse, \textit{Classical Klein-Gordon solutions, symplectic structures, and
  isometry actions on AdS spacetimes}, J. Geom. Phys. \textbf{70} (2013)
  130--156, \eprint{1212.2945}.

\bibitem{KiTu:symplectic}
J.~Kijowski, W.~M. Tulczyjew, \textit{A Symplectic Framework for Field
  Theories}, Springer, Berlin, 1979.

\bibitem{Oe:freefermi}
R.~Oeckl, \textit{Free Fermi and Bose Fields in TQFT and GBF}, SIGMA \textbf{9}
  (2013) 028, 46 pages, \eprint{1208.5038v2}.

\bibitem{ItZu:qft}
C.~Itzykson, J.-B. Zuber, \textit{Quantum Field Theory}, McGraw-Hill, New York,
  1980.

\bibitem{Oe:affine}
R.~Oeckl, \textit{Affine holomorphic quantization}, J. Geom. Phys. \textbf{62}
  (2012) 1373--1396, \eprint{1104.5527v3}.

\bibitem{Oe:feynobs}
R.~Oeckl, \textit{Schrödinger-Feynman quantization and composition of
  observables in general boundary quantum field theory}, to appear in Adv.
  Theor. Math. Phys., \eprint{1201.1877v1}.

\bibitem{Lic:propquantgr}
A.~Lichnerowicz, \textit{Propagateurs et quantification en relativit{\'e}
  g{\'e}n{\'e}rale}, Relativistic Theories of Gravitation (Warsaw, 1962), (ed.
  L.~Infeld), Pergamon Press, Oxford, 1964, pp. 177--188.

\bibitem{CoRa:qftrindler}
D.~Colosi, D.~R\"atzel, \textit{Quantum Field Theory on Timelike Hypersurfaces
  in Rindler Space}, Phys. Rev. \textbf{D 87} (2013) 125001,
  \eprint{1303.5873}.

\bibitem{bagidlaw:_what_cft_tell_about_ads}
V.~Balasubramanian, S.~Giddings, A.~Lawrence, \textit{What do CFTs tell us
  about Anti-de Sitter spacetimes?}, JHEP \textbf{9903} (1999) 001.

\bibitem{ash_mag:_qf_curved_spacetimes}
A.~Ashtekar, A.~Magnon, \textit{Quantum fields in curved space-times},
  Proc.R.Soc.Lond.A \textbf{346} (1975) p.375--394.

\end{thebibliography}
%
%
\end{document}